\documentclass[aps,prd,twocolumn,superscriptaddress,amssymb,nofootinbib]{revtex4}

\usepackage{color}
\usepackage{epsfig}
\usepackage{graphicx}
\usepackage{epstopdf}

\newcommand{\nc}{\newcommand}
\nc{\beq}{\begin{equation}}   \nc{\eeq}{\end{equation}}
\nc{\bea}{\begin{eqnarray}}   \nc{\eea}{\end{eqnarray}}
\nc{\baa}{\begin{array}}      \nc{\eaa}{\end{array}}
\nc{\bit}{\begin{itemize}}    \nc{\eit}{\end{itemize}}
\nc{\ben}{\begin{enumerate}}  \nc{\een}{\end{enumerate}}
\nc{\bce}{\begin{center}}     \nc{\ece}{\end{center}}
\def\beqa{\begin{eqnarray}}
\def\eeqa{\end{eqnarray}}
\def\bed{\begin{description}}
\def\eed{\end{description}}

\def\vev#1{\langle #1 \rangle}

\def\susy{{\rm SUSY}}

\def\ymax{y_{\rm max}}
\def\nmax{n_{\rm max}}
\def\hsm{h_{\rm SM}}
\def\mhsm{m_{\hsm}}

\def\hh{H}

\def\hp{h^+}
\def\hm{h^-}
\def\hpm{h^{\pm}}

\def\mhp{m_{\hp}}

\def\tanb{\tan\beta}
 
\def\mt{m_t}
\def\mb{m_b}
\def\mz{m_Z}

\def\mz{m_Z}

\def\mh{m_h}
\def\hi{h_1}
\def\hii{h_2}
\def\ai{a_1}
\def\mhi{m_{h_1}}
\def\mhii{m_{h_2}}   
\def\mai{m_{a_1}}
\def\mueff{\mu_{\rm eff}}
\def\mtau{m_\tau} 
\def\noi{\noindent}

\def\what{\widehat}

\def\lam{\lambda}  
\def\kap{\kappa}   
\def\alam{A_\lambda}
\def\akap{A_\kappa}

\def\mhusq{m_{H_u}^2}
\def\mhdsq{m_{H_d}^2}
\def\mssq{m_S^2} 
\def\mqsq{m_Q^2}
\def\musq{m_U^2} 
\def\mdsq{m_D^2}   
\def\mlsq{m_L^2}
\def\mesq{m_E^2}  

\def\cvisq{C_V^2(\hi)}
\def\cviisq{C_V^2(\hii)}
\def\cta{\cos\theta_A}
\def\ctasq{\cos^2\theta_A}

\def\tauptaum{\tau^+\tau^-}
\def\mups{m_\Upsilon}
\def\brups{\br(\Upsilon\to \gam\ai)}
\def\caibb{C_{\ai b\anti b}}
\def\cmax{|C_{\ai b\anti  b}|^{\rm max}}

\def\mh{m_h}
\def\hi{h_1}
\def\hii{h_2}
\def\ai{a_1}
\def\mhi{m_{h_1}}
\def\mhii{m_{h_2}}
\def\mai{m_{a_1}}
\def\mueff{\mu_{\rm eff}}
\def\mtau{m_\tau}
\def\noi{\noindent}

\def\lam{\lambda}
\def\kap{\kappa}
\def\alam{A_\lambda}

\def\akap{A_\kappa}

\def\mh{m_h}
\def\mhp{m_{\hp}}

\def\mhusq{m_{H_u}^2}
\def\mhdsq{m_{H_d}^2}
\def\mssq{m_S^2}
\def\mqsq{m_Q^2}
\def\musq{m_U^2}
\def\mdsq{m_D^2}
\def\mlsq{m_L^2}
\def\mesq{m_E^2}

\def\h{h}
\def\mh{m_{h}}

\def\what{\widehat}

\def\hbar{\overline h}
\def\lam{\lambda}

\def\mz{m_Z}

\def\hi{h_i^0}
\def\mhi{m_{\hi}}

\def\h{h}
\def\mh{m_{\h}}

\def\lam{\lambda}

\def\hpm{h^{\pm}}

\def\what{\widehat}
\def\tauptaum{\tau^+\tau^-}

\def\lsim{\mathrel{\raise.3ex\hbox{$<$\kern-.75em\lower1ex\hbox{$\sim$}}}}
\def\gsim{\mathrel{\raise.3ex\hbox{$>$\kern-.75em\lower1ex\hbox{$\sim$}}}}
\def\ifmath#1{\relax\ifmmode #1\else $#1$\fi}

\def\vev#1{\langle #1 \rangle}
\def\lam{\lambda}

\def\mhi{m_{h_1}}

\def\etc{{\em etc.}}

\def\eg{{\it e.g.}}
\def\etal{{\it et al.}}

\def\msusy{m_{\rm SUSY}}

\def\susy{{\rm SUSY}}

\def\eg{{\it e.g.}}
\def\etal{{\it et al.}}

\def\hsm{h_{\rm SM}}
\def\mhsm{m_{\hsm}}

\def\hh{H}

\def\hp{h^+}
\def\hm{h^-}
\def\hpm{h^{\pm}}

\def\mhp{m_{\hp}}

\def\tanb{\tan\beta}

\def\mt{m_t}
\def\mb{m_b}
\def\mz{m_Z}

\def\lam{\lambda}
\def\br{B}
\def\tauptaum{\tau^+\tau^-}

\def\gam{\gamma}

\def\etal{{\it et al.}}

\def\anti{\overline}
\def\epem{e^+e^-}
\def\mupmum{\mu^+\mu^-}

\def\wstarp{W^{+\,(\star)}}
\def\mupmum{\mu^+\mu^-}

\def\ie{{\it i.e.}}
\def\eg{{\it e.g.}}

\def\anti{\overline}

\def\ai{a_1}
\def\aii{a_2}
\def\mai{m_{\ai}}

\def\gev{~{\rm GeV}}

\def\mt{m_t}
\def\mb{m_b}

\def\hi{\h_1}
\def\hii{\h_2}
\def\hiii{\h_3}
\def\mhi{m_{\hi}}
\def\mhii{m_{\hii}}

\def\mhi{m_{h_1}}

\def\eg{{\it e.g.}}
\def\etal{{\it et al.}}

\def\tanb{\tan\beta}

\begin{document}

\title{Many Light Higgs Bosons in the NMSSM}

\author{Radovan Derm\' \i\v sek}
\email[]{dermisek@indiana.edu}

\affiliation{Physics Department, Indiana University, Bloomington, IN 47405}

\author{John F. Gunion}
\email[]{gunion@physics.ucdavis.edu}

\affiliation{Department of Physics, University of California at Davis, Davis, CA 95616}



\date{\today}

\begin{abstract}
  The next-to-minimal supersymmetric model with a light doublet-like
  CP-odd Higgs boson and small $\tan \beta$ can satisfy all
  experimental limits on Higgs bosons even with light superpartners.
  In these scenarios, the two lightest CP-even Higgs bosons, $\hi$ and
  $\hii$, and the charged Higgs boson, $\hp$, can all be light enough
  to be produced at LEP and yet have decays that have not been looked
  for or are poorly constrained by existing collider experiments. The
  channel $\hi\to \ai\ai$ with $\ai\to \tau^+\tau^-$ or $2j$ is still
  awaiting LEP constraints for $\mhi>86\gev$ or $82\gev$, respectively.
  LEP data may also contain $\epem\to \hii\ai$ events where $\hii\to
  Z\ai$ is the dominant decay, a channel that was never examined.
  Decays of the charged Higgs bosons are often dominated by $H^\pm \to
  W^{\pm (\star)} \ai$ with $\ai \to gg,c \bar c, \tau^+ \tau^-$. This
  is a channel that has so far been ignored in the search for $t\to
  \hp b$ decays at the Tevatron. A specialized analysis might reveal a
  signal. The light $\ai$ might be within the reach of $B$ factories
  via $\Upsilon\to \gam \ai$ decays.  We study typical mass ranges and
  branching ratios of Higgs bosons in this scenario and compare these
  scenarios where the $\ai$ has a large doublet component to the more
  general scenarios with arbitrary singlet component for the $\ai$.
\end{abstract}


\maketitle






\section{Introduction}

Discovery of Higgs bosons and exploration of their properties is the
key to understanding electroweak symmetry breaking and a major
step in uncovering the ultimate theory of particle physics.  The Higgs
boson is the last missing piece of the standard model (SM).  In
theories beyond the SM the Higgs sector is typically more complicated,
\eg\ in the minimal supersymmetric model (MSSM) there are two Higgs
doublets which lead to five Higgs bosons in the spectrum: light and
heavy CP-even Higgses, $h$ and $H$, a CP-odd Higgs, $A$, and a pair
of charged Higgs bosons, $H^\pm$. In the next-to-minimal
supersymmetric model (NMSSM) which contains an additional singlet 
superfield with complex component scalar field
there are three CP-even Higgses, $h_{1,2,3}$, two CP-odd Higgses,
$a_{1,2}$ and a pair of charged Higgs bosons; and there are many
simple models with an even more complicated Higgs sector.  

Since searches for Higgs bosons rely on detection of their decay
products, it is crucial to understand the way the Higgs bosons decay.
Although it is usually the case that one of the Higgs bosons has
couplings to the $W,Z$ bosons and to fermions that are close to those
of the SM Higgs, it is not necessarily true that such a Higgs decays
in the way the SM Higgs does~\cite{Chang:2008cw}. A significant model
dependence of decay modes applies to other Higgses as well.

It has been recently argued that supersymmetric models in the region
of parameter space for which $\tanb$ is small and, in addition, there
is a light doublet-like CP-odd Higgs predict that all the Higgses
resulting from the two Higgs doublets ($h$, $H$, $A$ and $H^\pm$)
could have been produced already at LEP or the Tevatron, but would
have escaped detection because the decay modes have either not been
searched for (or the searches have been incomplete) or are ones to
which the experiments are not sensitive.  Although this scenario is
ruled out in the MSSM, it is only marginally disallowed for $m_A <
2m_b$ and $\tan \beta \lesssim 2.5$ and thus can possibly be viable in
simple extensions of the MSSM~\cite{Dermisek:2008id}. The reason is
that for $m_A \ll m_W$ and $\tan \beta \simeq 1$ the light CP-even
Higgs boson becomes SM-like, and although it is massless at the tree
level, its mass will receive a contribution from superpartners and the
tree level relation between the light CP-even and CP-odd Higgses, $m_h
< m_A$, is dramatically changed by SUSY corrections. Even for modest
superpartner masses the light CP-even Higgs boson will be heavier than
$2m_A$. In particular, for superpartner masses between 300 GeV and 1
TeV and $\tanb\simeq 1$, one finds $m_h \simeq 40 - 60 $ GeV and the
$h \to AA$ decay mode is open and generically dominant.

Since the $h$ has SM-like $WW,ZZ$ couplings, $e^+ e^- \to hA$ is
highly suppressed and the limits from the $Z$ width measurements can
be easily satisfied even for $m_h + m_A < m_Z$.  On the other hand,
the $\epem \to Zh$ cross section would be maximal. However,
for  $\tan \beta\sim 1$  and $m_A<2m_b$ the decay width of the $A$
is shared between $\tau^+ \tau^-$, $c \bar c$ and $gg$ final states and thus the
(dominant) $h\to AA$ decays  are spread over many different final states: $4\tau$,
$2\tau 2g$, $4g$, $4c$, $2g2c$, $2\tau 2c$ and $b\bar b$, the latter
being greatly suppressed relative to the SM expectation due to the
presence of the $h\to AA$ decays.  As a consequence, the LEP limits in
each channel separately are very substantially weakened. Of course, the decay mode
independent limit requires a Higgs with SM-like $ZZ$ coupling to be above 82
GeV~\cite{Abbiendi:2002qp}. It is this fact that rules out this scenario
in the MSSM, since there $m_h$ cannot be pushed above 82 GeV by
radiative corrections.

The rest of the Higgs spectrum is basically not constrained at all in
this scenario.  The heavy CP-even and the CP-odd Higgses could have
been produced at LEP in $e^+ e^- \to H A$ but they would have escaped
detection because $H$ dominantly decays to $ZA$ - a mode that has not
been searched for. Additional constraints are discussed in detail in
Ref.~\cite{Dermisek:2008id}.  The charged Higgs is also very little
constrained and up to $\sim 40 \%$ of top quarks produced at the
Tevatron could have decayed into charged Higgs and the $b$ quark since
the dominant decay mode for the charged Higgs $H^\pm \to W^{\pm (\star)}
A$ with $A \to c \bar c$, $gg$ or $\tau^+ \tau^-$ would not have been
separated from the the generic top sample~\footnote{We thank Ricardo
  Eusebi (CDF)  for a
  detailed discussion of the CDF and D0 analyses.}.  In addition, pair production
of a charged Higgs boson with the properties emerging in this scenario
and mass close to the mass of the $W$ boson could explain the $2.8
\sigma$ deviation from lepton universality in $W$ decays measured at
LEP~\cite{:2004qh} as discussed in~\cite{Dermisek:2008dq}.

The mass of the light CP-even Higgs is the only problematic part in
this scenario.  There are however various ways to increase the mass of
the SM-like Higgs boson in extensions of the MSSM. A simple
possibility is to consider singlet extensions of the MSSM containing a
$\lambda \what S \what H_u \what H_d$ term in the superpotential.  It
is known that this term itself contributes $\lambda^2 v^2 \sin^2 2
\beta$, where $v = 174\gev$, to the mass squared of the CP-even
Higgs~\cite{Ellis:1988er} and thus can easily push the Higgs mass
above the decay-mode independent limit of $82$ GeV. Note that this
contribution is maximized for $\tan \beta \simeq 1$. In addition, it
need not be the case that the light CP-even Higgs has full strength
$ZZ$ coupling, in which case the model-independent limit on $\mh$ is
reduced, while at the same time the $\hh$ which carries the rest of
the $ZZ$ coupling can have mass above the LEP kinematic reach and/or
decay to modes for which the LEP limit of $114\gev$ does not apply.
Thus, it is not surprising that in the NMSSM it is possible to find
scenarios in which the lightest CP-odd Higgs has mass below $2m_b$ and
the two lightest CP-even Higgs bosons and the charged Higgs would all
have been produced at LEP and yet escaped detection.

In this paper we study NMSSM scenarios with a light CP-odd Higgs boson
and small $\tan \beta$. We will in particular examine the subset of
these scenarios in which the light CP-odd Higgs boson is mainly
doublet-like ($\ai$-doublet-like scenarios) and will find that they
have many features in common with the MSSM scenarios discussed above,
except that they are not ruled out by Higgs searches --- they are
phenomenologically viable even with very light superpartners.  For the
subset of the $\ai$-doublet-like scenarios in which the $\hi$ has
nearly SM-like couplings, the $\hi$ can be as light as 82 GeV (the
decay-mode independent limit) by virtue of dominant decays $\hi\to
\ai\ai \to 2\tau 2c, 4\tau, 4c$, \etc.  There are also scenarios
for which the $\hi$ has reduced coupling to $ZZ$,
$g_{ZZ\hi}^2/g_{ZZ\hsm}^2 \simeq 0.5$.~\footnote{In general, in singlet extensions it is possible
  to alter the couplings of the Higgses to $Z$ and $W$ through mixing,
  see e.g.  Refs.~\cite{Barger:2006dh, Dermisek:2007ah} or to provide new
  Higgs decay modes~\cite{Chang:2008cw}.} In these latter cases, the
CP-even Higgs boson can have $\mhi$ as low as $\sim 55$ GeV.  All
these scenarios are similar to the scenario with a light singlet-like
CP-odd Higgs in the NMSSM~\cite{Dermisek:2005ar, Dermisek:2005gg,
  Dermisek:2006wr, Dermisek:2007yt, Dermisek:2006py} in that it is the
unexpected Higgs decays that allow one or more light Higgs bosons to
have escaped LEP detection.  The important difference is that the
scenario discussed in Refs.~\cite{Dermisek:2005ar, Dermisek:2005gg,
  Dermisek:2006wr, Dermisek:2007yt, Dermisek:2006py} is the usual
decoupled scenario as far as the two Higgs doublet part of the Higgs
spectrum is concerned (the CP-odd Higgs, the heavy CP-even Higgs and
the charged Higgs are heavy and approximately degenerate) and the
light CP-odd Higgs is supplied by the additional singlet. In contrast,
in the $\ai$-doublet-like low-$\tanb$ scenarios,
the CP-even and CP-odd Higgses coming from the
additional singlet are typically heavy and do not drastically alter
the two Higgs doublet part of the Higgs sector. The latter then looks
like the Higgs sector of the MSSM with somewhat modified mass
relations.

\section{Light doublet-like $a_1$ in the NMSSM }

As already mentioned the scenario with a light doublet-like CP-odd Higgs and
small $\tan \beta$ is phenomenologically viable  
in the simplest extension of the MSSM, the next-to-minimal
supersymmetric model
which adds only one singlet chiral
superfield, $\widehat{S}$. The very attractive nature
of the NMSSM extension of the MSSM on general grounds has been
discussed for many years \cite{allg}; in particular, it avoids
the need for the $\mu$ parameter of the MSSM
superpotential term $\mu \what H_u\what H_d$. The NMSSM particle content differs from the
MSSM by the addition of one CP-even and one CP-odd state in the   
neutral Higgs sector (assuming CP conservation), and one
additional neutralino.  We will follow the conventions of
\cite{Ellwanger:2004xm}.  Apart from the usual quark and lepton
Yukawa couplings, the scale invariant superpotential is
\beq \label{1.1}
\lambda \ \widehat{S}
\widehat{H}_u \widehat{H}_d + \frac{\kappa}{3} \ \widehat{S}^3
\eeq
\noi depending on two dimensionless
couplings $\lambda$, $\kappa$ beyond the MSSM.  [Hatted (unhatted)
capital letters denote superfields (scalar superfield
components).]  An effective $\mu$ term arises from the first term of
Eq.~(\ref{1.1}) when the scalar component of $\what S$ acquires a 
vacuum expectation value, $s\equiv\vev{\what S}$, yielding
\beq
\mueff=\lam s\,.
\eeq
The trilinear soft terms associated with the superpotential terms in
Eq.~(\ref{1.1}) are
\beq \label{1.2}
\lambda A_{\lambda} S H_u H_d + \frac{\kappa} {3} A_\kappa S^3 \,.
\eeq
The final input parameter is
\beq \label{1.3} \tan \beta = h_u/h_d\,,
\eeq
where $h_u\equiv
\vev {H_u}$, $h_d\equiv \vev{H_d}$.
The vevs $h_u$, $h_d$ and $s$, along with $\mz$, can be viewed as
determining the three \susy\ breaking masses squared for $H_u$, $H_d$
and $S$ (denoted $\mhusq$, $\mhdsq$ and $\mssq$)
through the three minimization equations of the scalar potential.
Thus, as compared to the three independent parameters needed in the
MSSM context (often chosen as $\mu$, $\tan \beta$ and $M_A$), the
Higgs sector of the NMSSM is described by the six parameters \beq
\label{6param} \lambda\ , \ \kappa\ , \ A_{\lambda} \ , \ A_{\kappa},
\ \tan \beta\ , \ \mu_\mathrm{eff}\ .  \eeq (We employ a convention in
which all parameters are evaluated at scale $\mz$ unless otherwise
stated.)  We will choose sign conventions for the fields such that
$\lambda$ and $\tan\beta$ are positive, while $\kappa$, $A_\lambda$,
$A_{\kappa}$ and $\mu_{\mathrm{eff}}$ should be allowed to have either
sign.  In addition, values must be input for the gaugino masses
($M_{1,2,3}$) and for the soft terms related to the (third generation)
squarks and sleptons ($\mqsq$, $\musq$, $\mdsq$, $\mlsq$, $\mesq$,
$A_t$, $A_b$ and $A_\tau$) that contribute to the radiative
corrections in the Higgs sector and to the Higgs decay widths.  For
small $\tanb$, the soft parameters which play the most prominent role
are $\mqsq$, $\musq$ and $A_t$.

A complete survey of the parameter space is difficult. To present
results in a manageable way, we fix $\mu$ and $\tan \beta$ together
with all soft SUSY breaking masses and scan over trilinear and
soft-trilinear couplings. We will plot results in various
two-dimensional planes. The input parameters of Eq.~(\ref{1.2}) are
scanned over the following regions with fixed steps: $\lambda \in
(0.001,0.6)$ using 60 steps of size $0.01$; $\kappa \in (-0.6, 0.6)$
using 120 steps of size $0.01$, with some refined scans for
$\kappa\in(-0.06,0.06)$ using 120 steps of size $0.001$; $A_\lambda
\in (-600 \, {\rm GeV}, 600 \, {\rm GeV} )$ using 200 steps of size
$6$ GeV; and finally $A_\kappa \in (-600 \, {\rm GeV}, 600 \, {\rm
  GeV} )$ using 200 steps of size $6$ GeV, with refined scans for this
same range with 1000 steps of size $1.2$ GeV.  Varying the fixed soft
SUSY breaking masses leads to smaller changes than does varying
$\tanb$. Thus, we will consider only a few choices of soft SUSY
breaking masses and will focus on the important changes that occur as
$\tanb$ is changed.

All scans are performed in the context of NMHDECAY. NMHDECAY checks a
long list of experimental constraints, especially those coming from
LEP data. It also checks various theoretical constraints on the model
incorporated, such as requiring that the vacuum be a true vacuum.
NMHDECAY also issues a warning if any of the couplings, $\lam$,
$\kap$, $h_t$ or $h_b$ become non perturbative ($h_t$ and $h_b$ are
the Yukawa couplings) after evolution to the GUT scale. We will
consider scenarios in which these become non-perturbative as well as
scenarios in which they remain perturbative. Aside from this, all
plotted points are consistent with all the NMHDECAY constraints.  

As stated in the introduction, we wish to focus on cases for which
$\mai<10\gev$. Such scenarios have the most unusual features.  In this
mass region, it is important to incorporate the constraints arising from
recently improved limits on $\brups$ with $\ai\to\tau\tau$ from
CLEO-III~\cite{cleoiii} as well as old CUSB-II limits~\cite{cusbii} on
$\brups$ where $\ai$ is only assumed to be visible.  These basically
place an upper limit on the $b\anti b \ai$ coupling defined by
\beq
{\cal L}_{\ai b\anti b}\equiv i C_{\ai b\anti b}{ig_2m_b\over2m_W}\anti b \gamma_5 b
\ai 
\eeq
in the region $\mai<\mups$. Further constraints on this coupling were
obtained at LEP by looking for $b\anti b \ai$ production with $\ai\to
\tau\tau$ and $\ai\to b\anti b$~\cite{Abbiendi:2001kp,delphi}. The
former channel is important in the $\mai<10\gev$ that we focus on.
The upper limits on $\caibb$ using the above inputs are given in
Fig.~1 of \cite{Gunion:2008dg}.  At any given $\tanb$, a limit $\cmax$ on
$|\caibb|$ converts to a limit on $|\cta|$ using $\caibb=\cta\tanb$, \ie\ 
$|\cta|\leq \cmax/\tanb$.  The resulting values for $|\cta|^{\rm max}$
appear in Fig.~3 of \cite{Gunion:2008dg}. It also turns out that Tevatron
limits on $pp\to \ai\to \mupmum$ provide some constraints on $\caibb$
in the region from $8\gev<\mai<9\gev$ that are stronger than those
from $\epem$ data~\cite{tevlimitpaper}. These too are incorporated.

A final addendum to NMHDECAY is to include  off-shell decays
involving an $\ai$ and a gauge boson in the final state. In
particular, $\hp\to W^* \ai$ and $\hii\to Z^*\ai$ virtual decays are
of occasional importance in the small $\mai$ region.

\subsection{Results for $\tan \beta =2$}

\begin{figure}
\includegraphics[height=0.4\textwidth,angle=90]{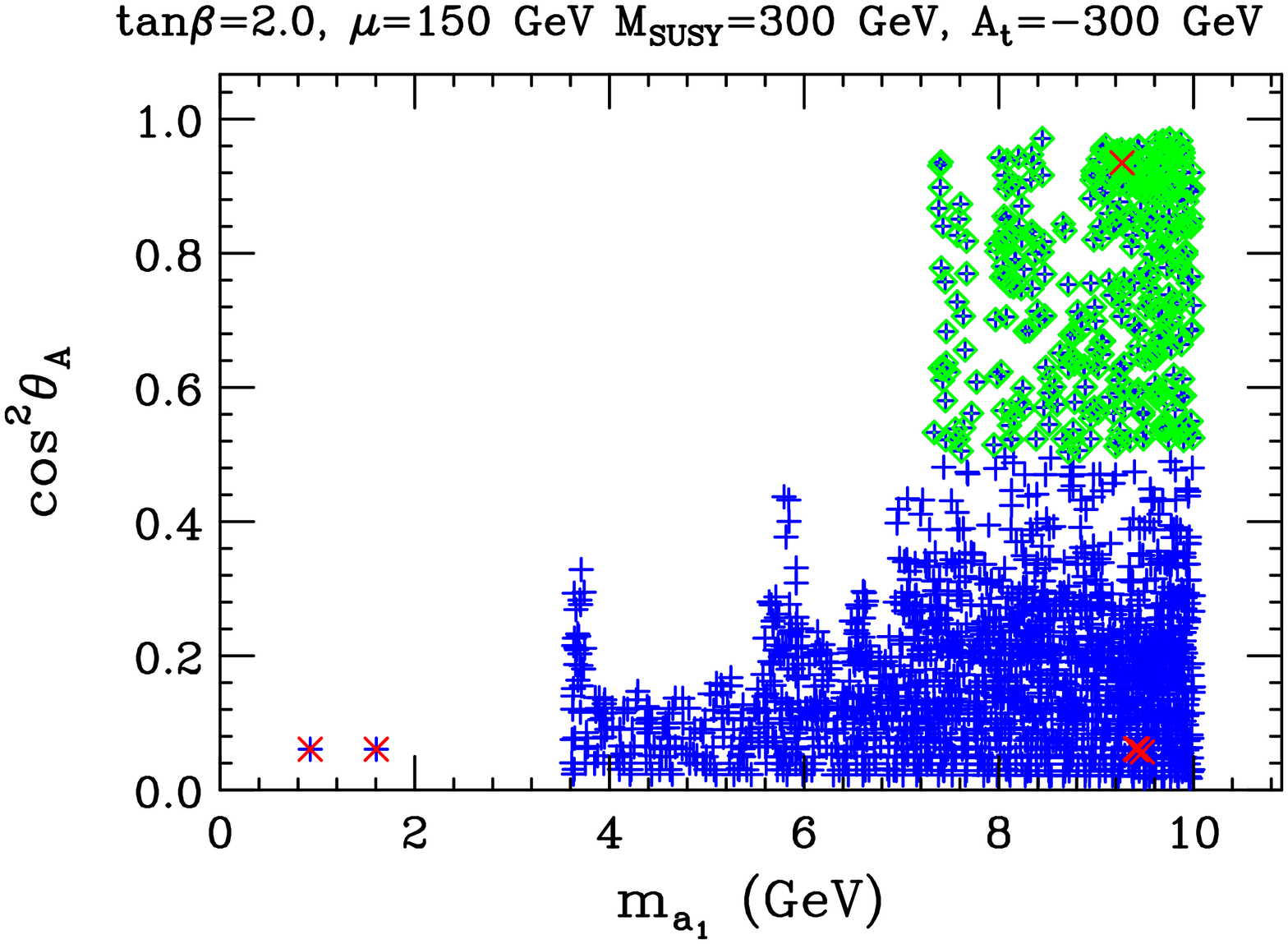}
\caption{$\cos^2\theta_A$ is plotted vs. $\mai$ for the $\tanb=2$,
  $\msusy=300\gev$, $A=-300\gev$ scenario.}
\label{ctasqvsmai_tb2}
\end{figure}

\begin{figure}
\includegraphics[height=0.4\textwidth,angle=90]{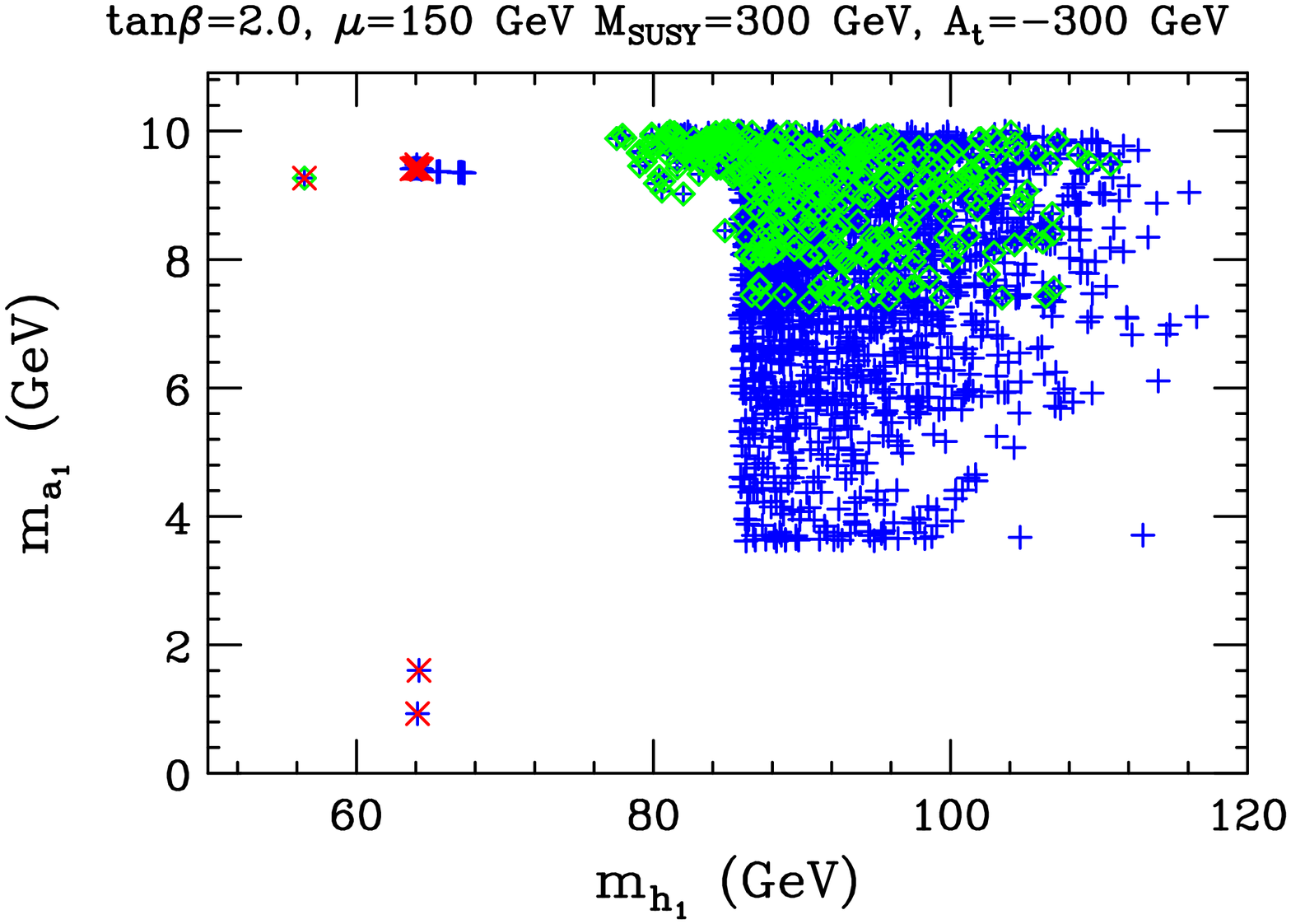}
\caption{$\mai$ is plotted vs. $\mhi$ for the $\tanb=2$,
  $\msusy=300\gev$, $A=-300\gev$ scenario.}
\label{maivsmhi_tb2}
\end{figure}

\begin{figure}
\includegraphics[height=0.4\textwidth,angle=90]{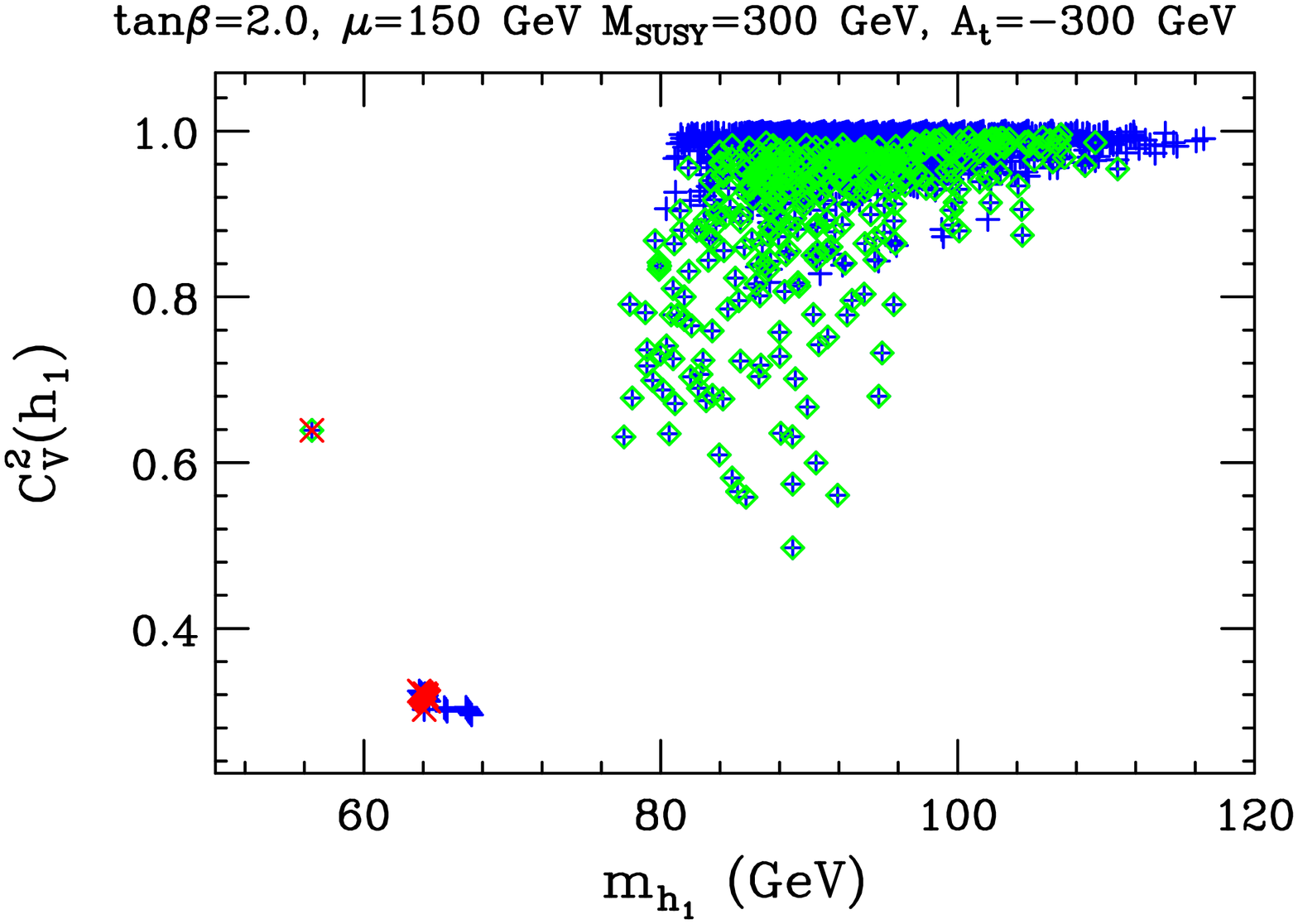}
\caption{$\cvisq$ is plotted vs. $\mhi$ for the $\tanb=2$,
  $\msusy=300\gev$, $A=-300\gev$ scenario.}
\label{cvisqvsmhi_tb2}
\end{figure}

\begin{figure}
\includegraphics[height=0.4\textwidth,angle=90]{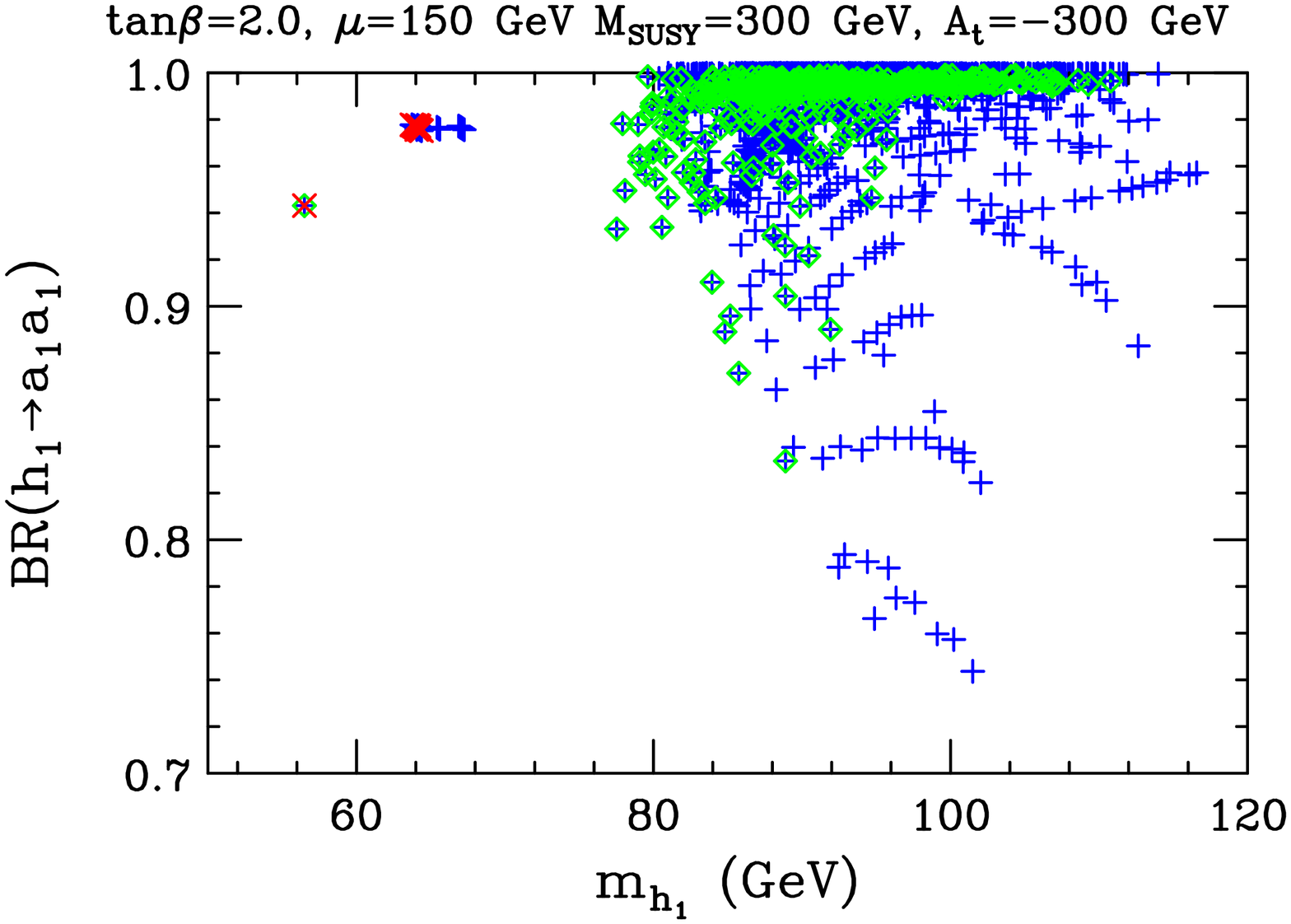}
\caption{$\br(\hi\to\ai\ai)$ is plotted vs. $\mhi$ for the $\tanb=2$,
  $\msusy=300\gev$, $A=-300\gev$ scenario.}
\label{brhiaiaivsmhi_tb2}
\end{figure}

\begin{figure}
\includegraphics[height=0.4\textwidth,angle=90]{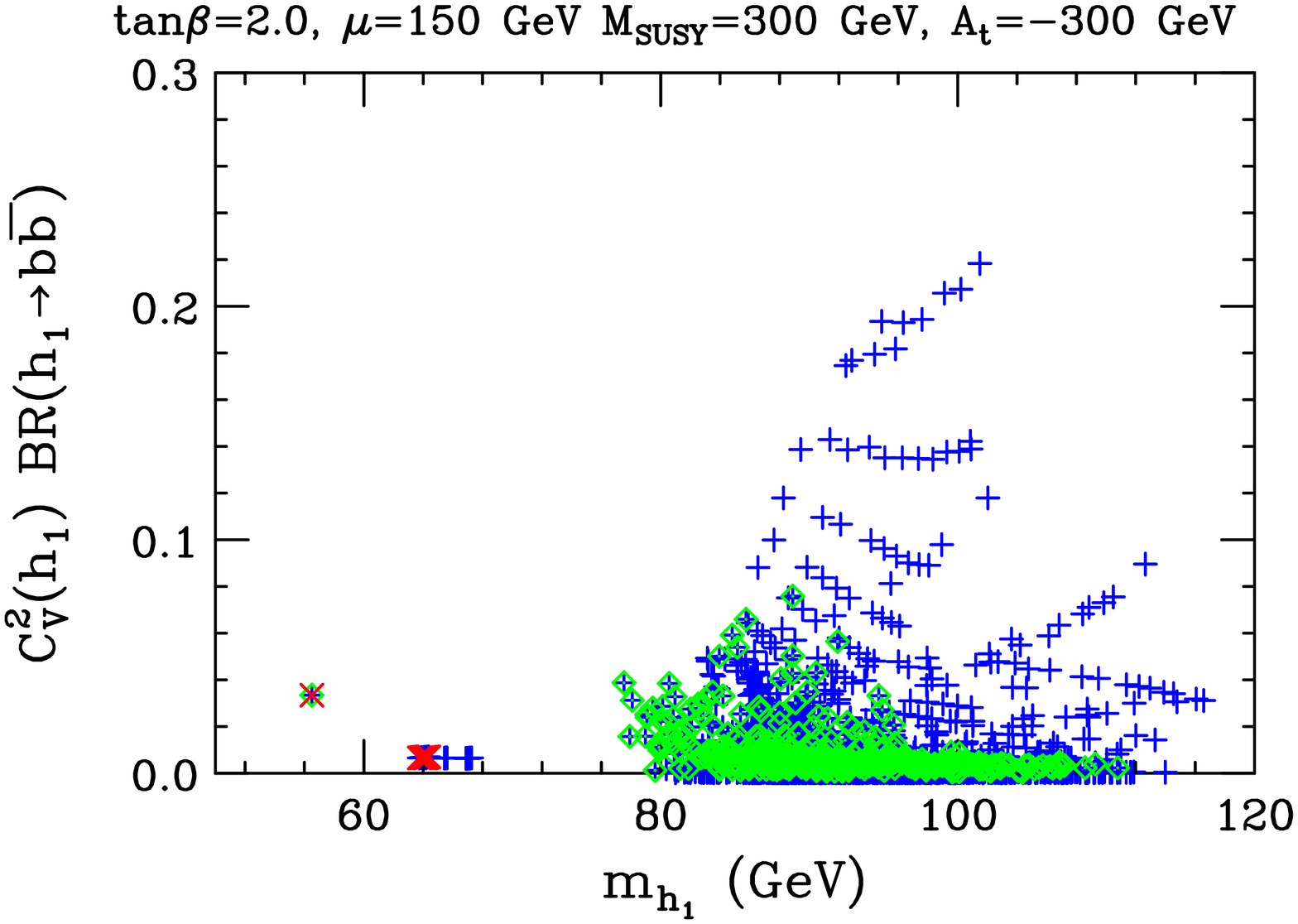}
\caption{$\cvisq\br(\hi\to b\anti b)$ is plotted vs. $\mhi$ for the $\tanb=2$,
  $\msusy=300\gev$, $A=-300\gev$ scenario.}
\label{gzzhisqbrhibbvsmhi_tb2}
\end{figure}

\begin{figure}
\includegraphics[height=0.4\textwidth,angle=90]{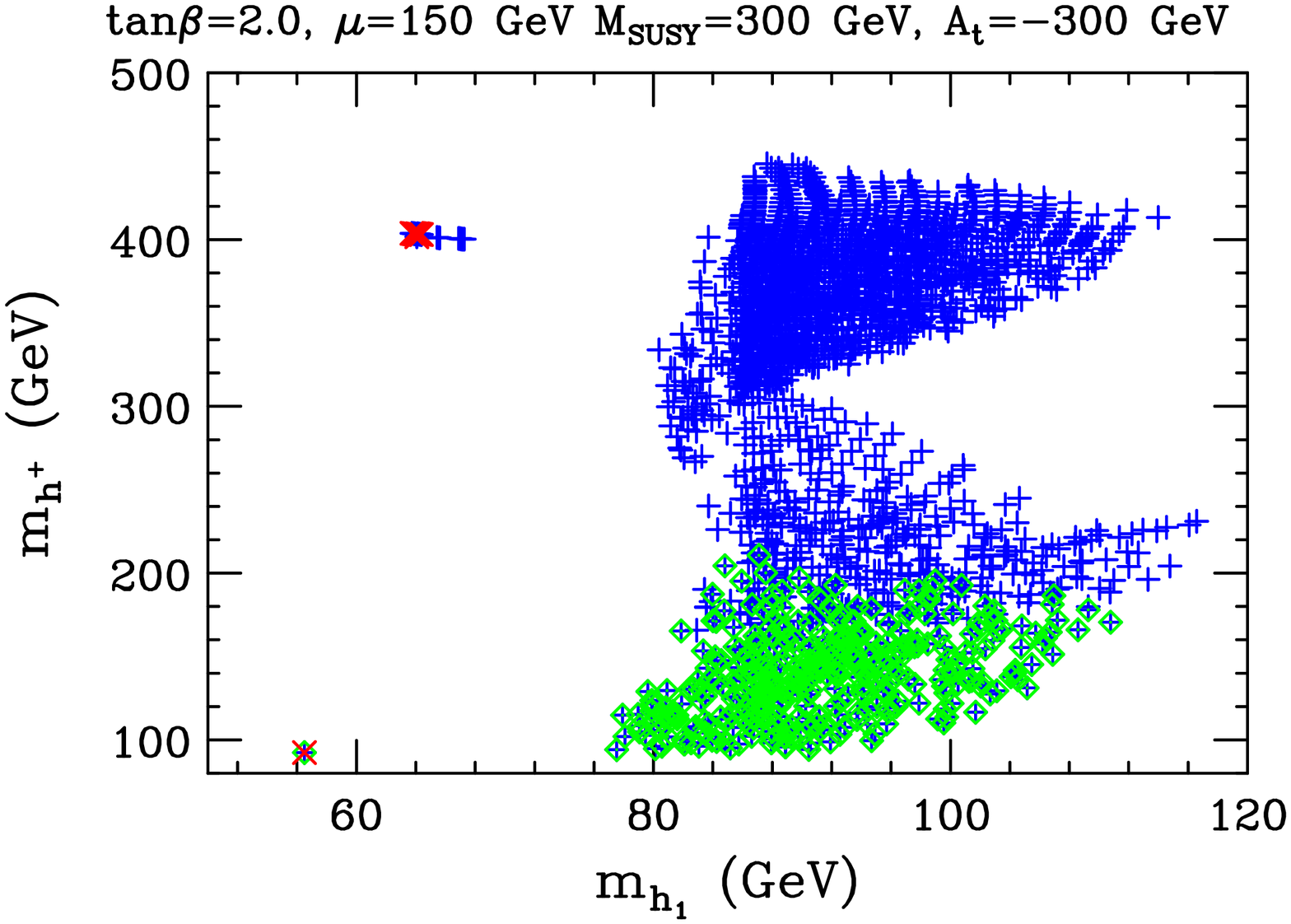}
\caption{$\mhp$ is plotted vs. $\mhi$ for the $\tanb=2$,
  $\msusy=300\gev$, $A=-300\gev$ scenario.}
\label{mhpvsmhi_tb2}
\end{figure}

\begin{figure}
\includegraphics[height=0.4\textwidth,angle=90]{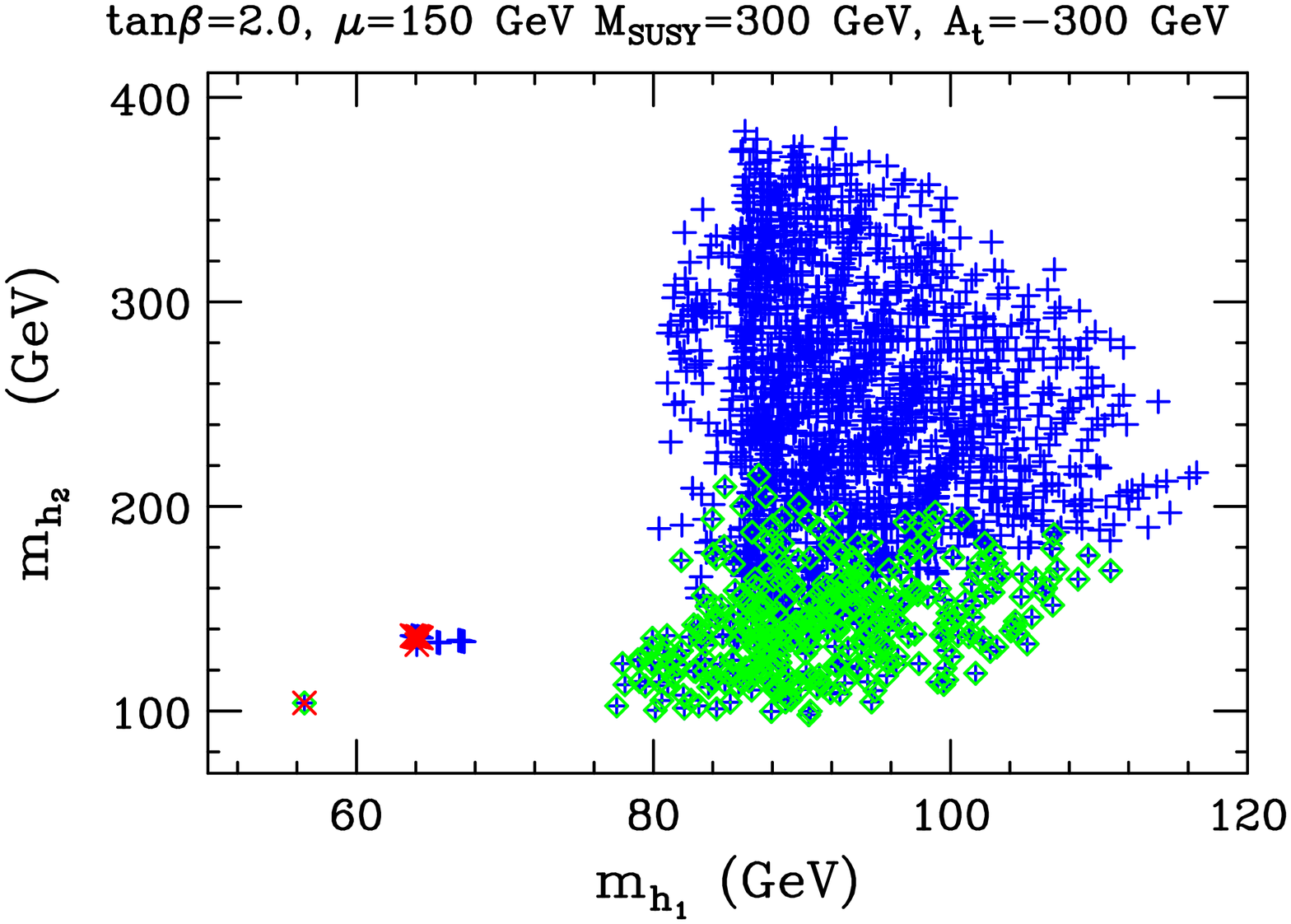}
\caption{$\mhii$ is plotted vs. $\mhi$ for the $\tanb=2$,
  $\msusy=300\gev$, $A=-300\gev$ scenario.}
\label{mhiivsmhi_tb2}
\end{figure}

\begin{figure}
\includegraphics[height=0.4\textwidth,angle=90]{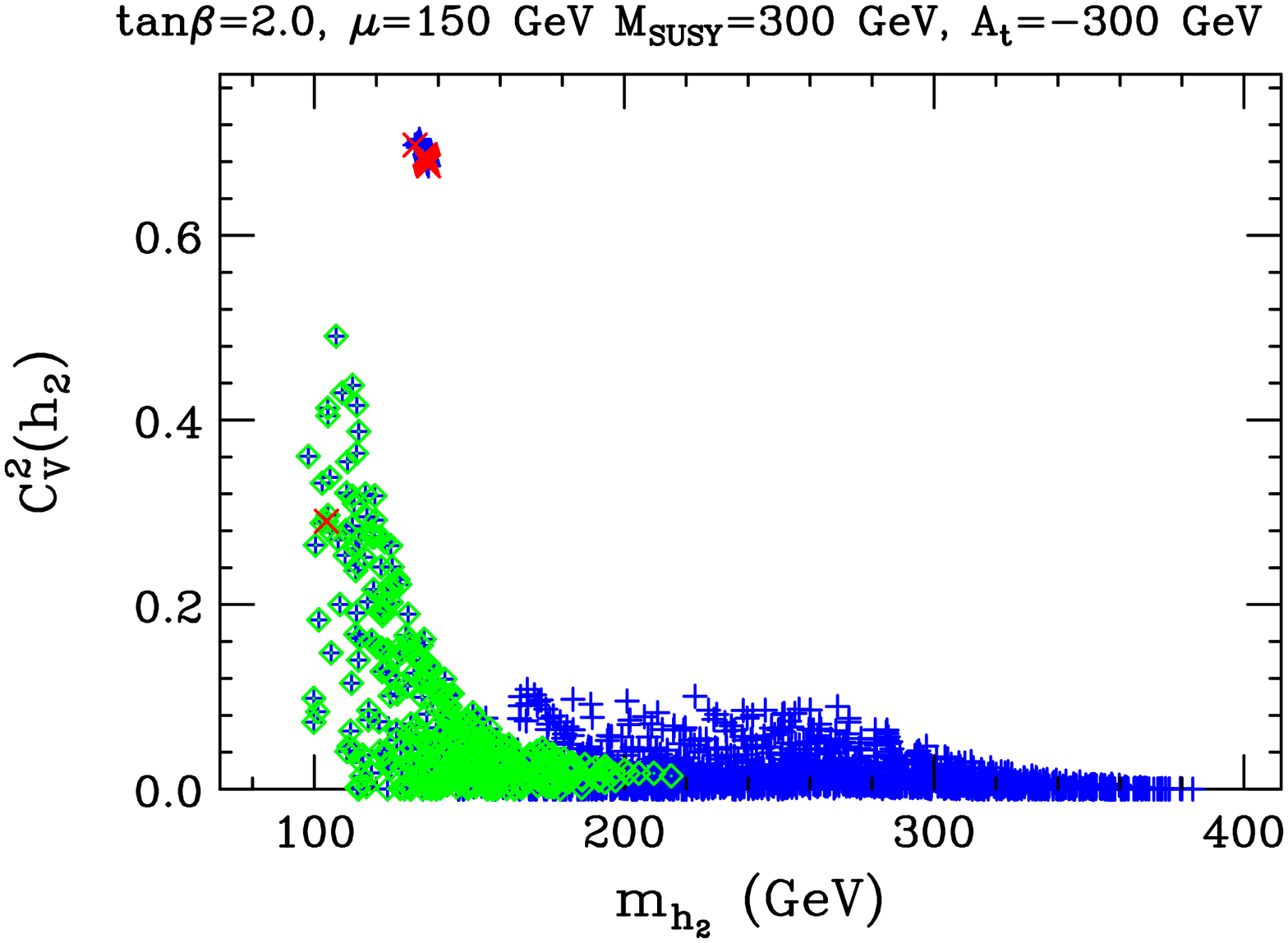}
\caption{$\cviisq$ is plotted vs. $\mhii$ for the $\tanb=2$,
  $\msusy=300\gev$, $A=-300\gev$ scenario.}
\label{cviisqvsmhii_tb2}
\end{figure}

\begin{figure}
\includegraphics[height=0.4\textwidth,angle=90]{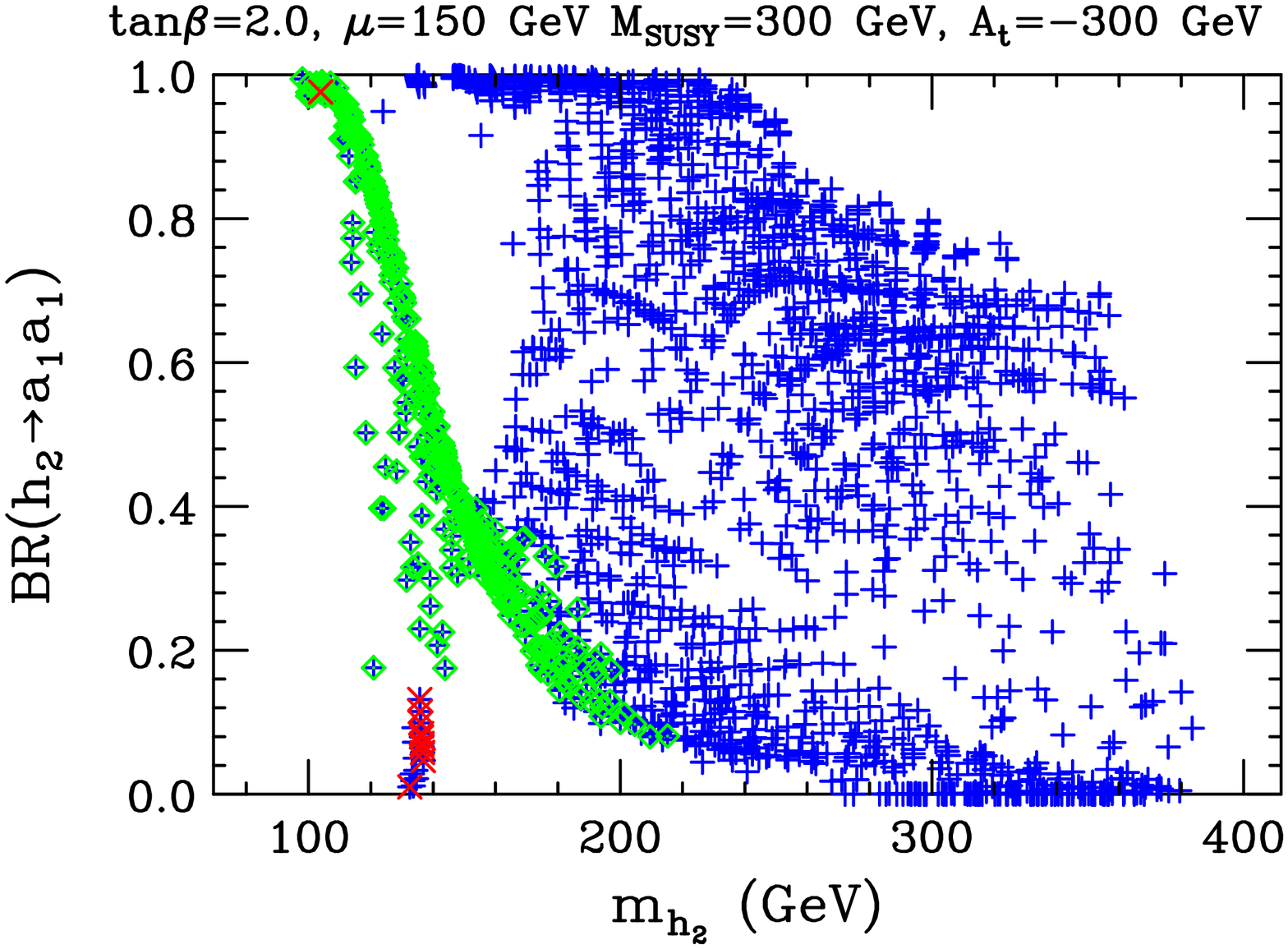}
\caption{$\br(\hii\to\ai\ai)$ is plotted vs. $\mhii$ for the $\tanb=2$,
  $\msusy=300\gev$, $A=-300\gev$ scenario.}
\label{brhiiaiaivsmhii_tb2}
\end{figure}

\begin{figure}
\includegraphics[height=0.4\textwidth,angle=90]{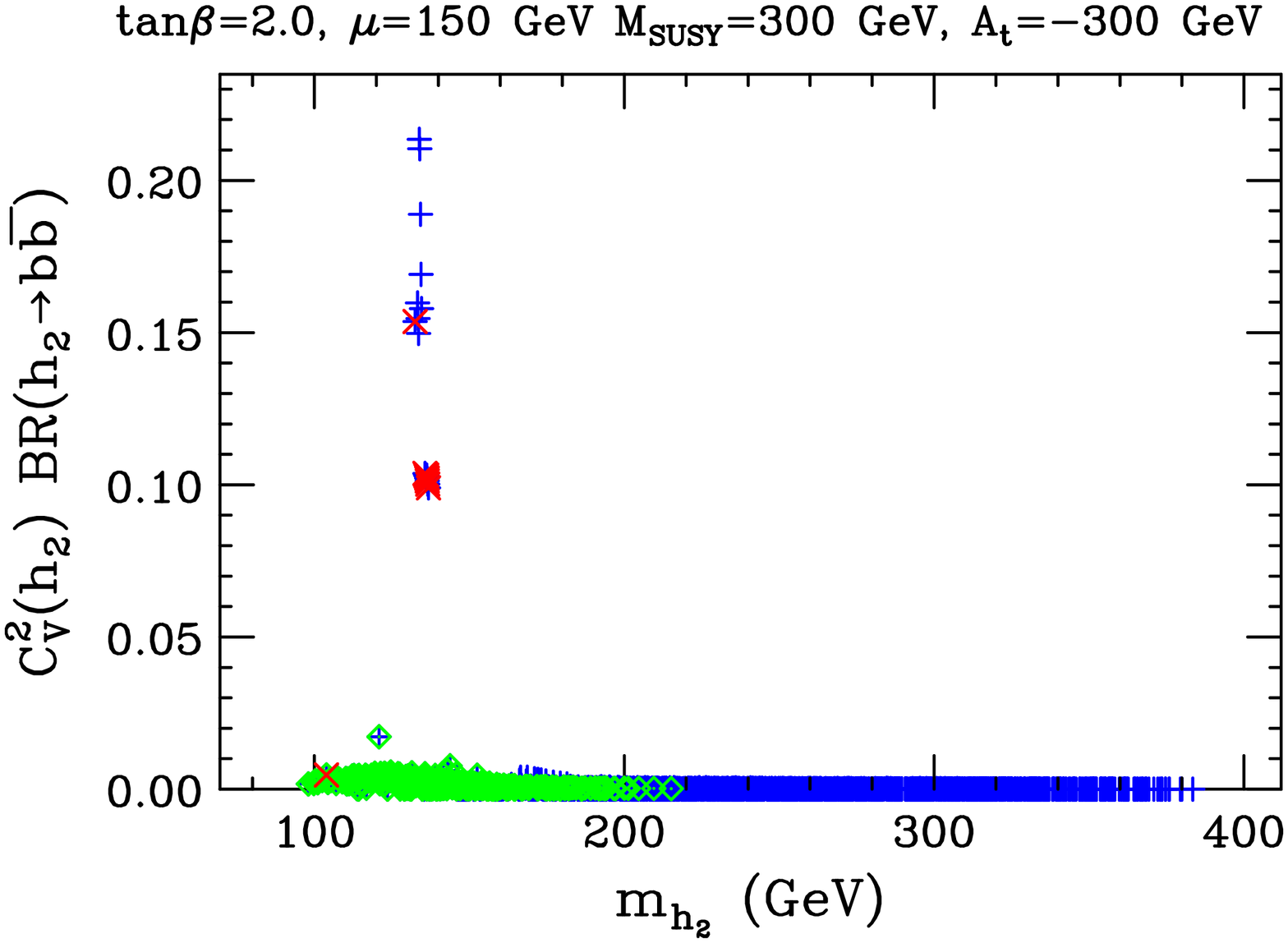}
\caption{$\cviisq\br(\hii\to b\anti b)$ is plotted vs. $\mhii$ for the $\tanb=2$,
  $\msusy=300\gev$, $A=-300\gev$ scenario.}
\label{gzzhiisqbrhiibbvsmhii_tb2}
\end{figure}

\begin{figure}
\includegraphics[height=0.4\textwidth,angle=90]{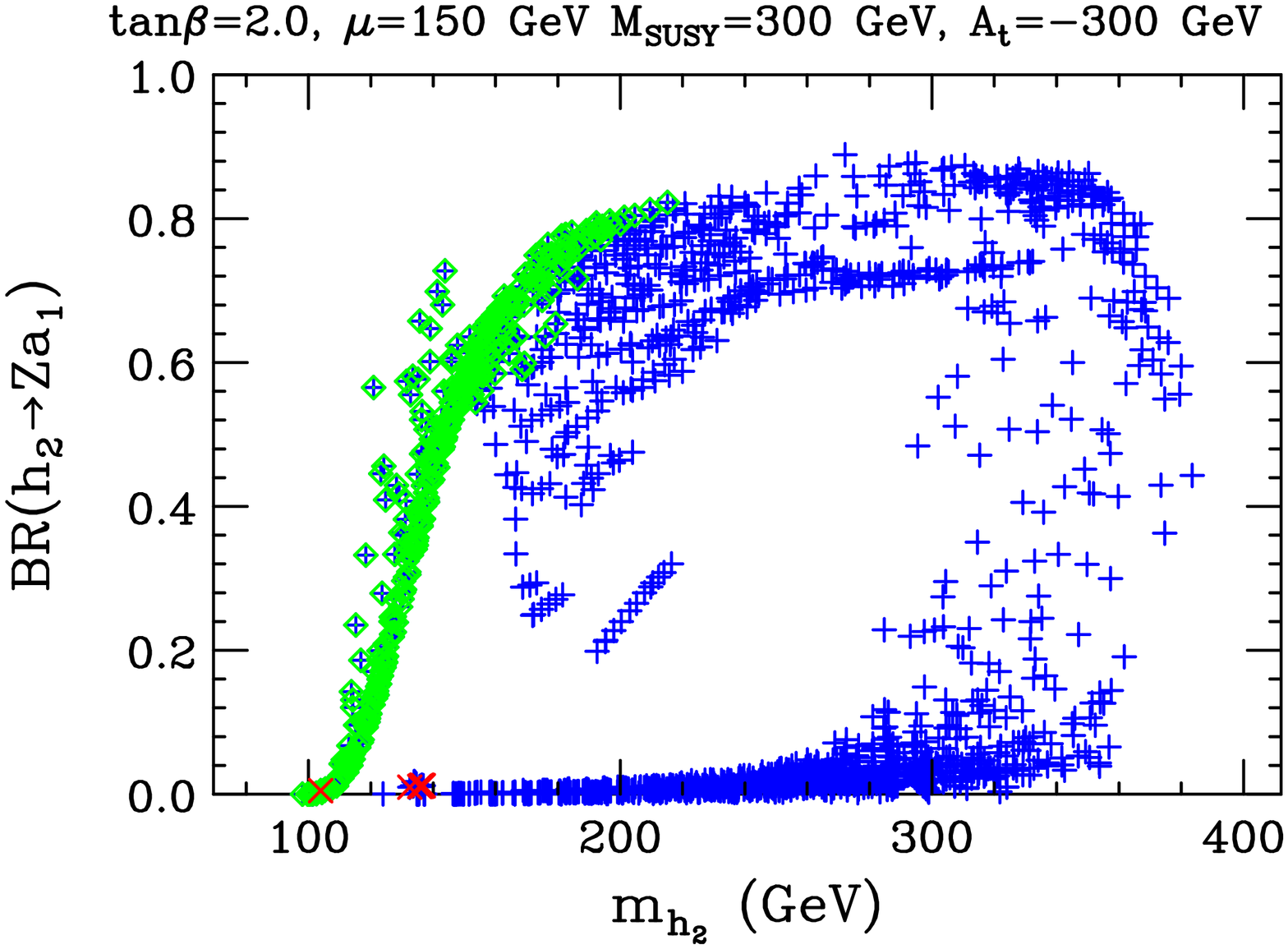}
\caption{$\br(\hii\to Z\ai)$ is plotted vs. $\mhii$ for the $\tanb=2$,
  $\msusy=300\gev$, $A=-300\gev$ scenario.}
\label{brhiizaivsmhii_tb2}
\end{figure}

\begin{figure}
\includegraphics[height=0.4\textwidth,angle=90]{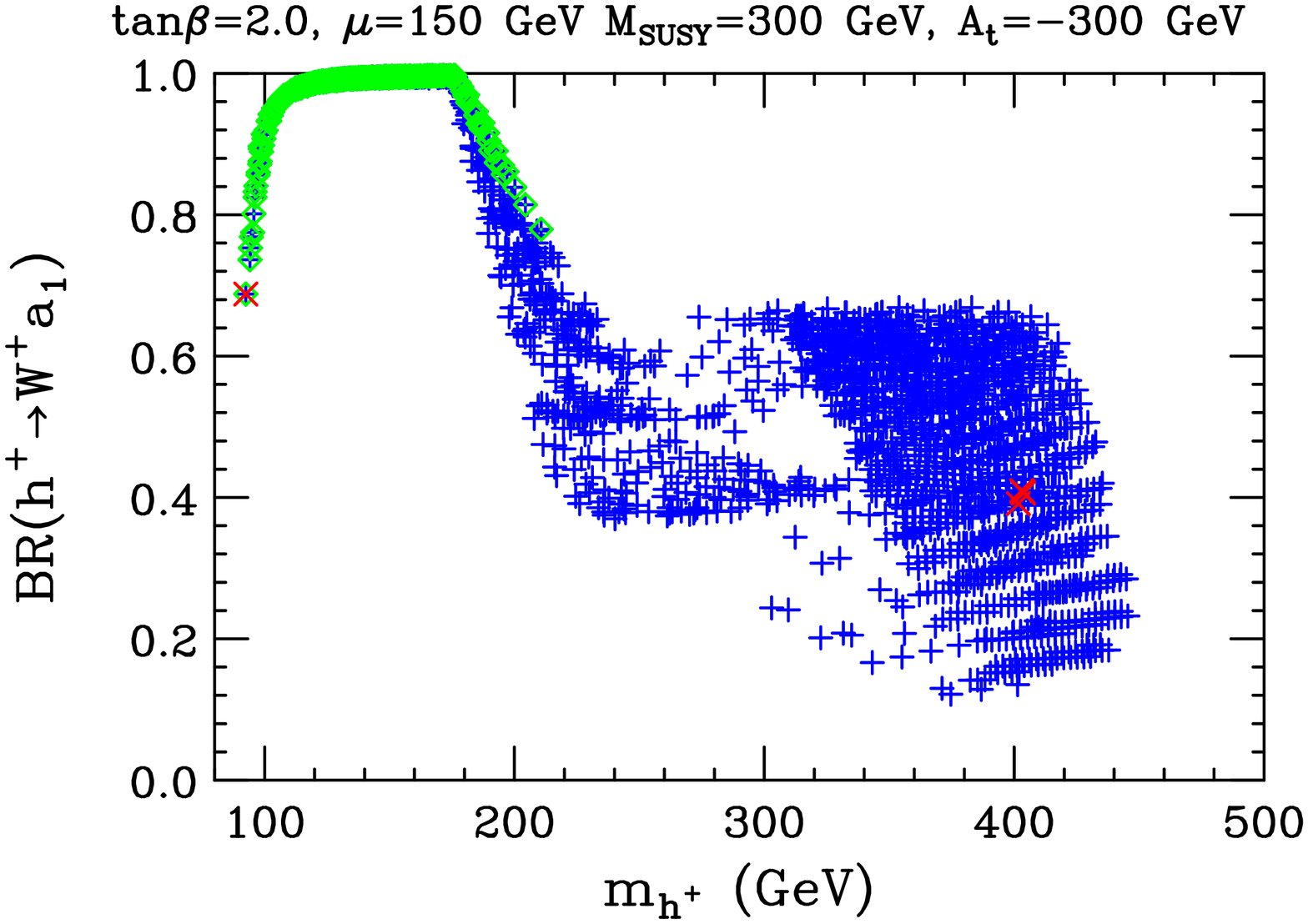}
\caption{$\br(\hp\to \wstarp\ai)$ is plotted vs. $\mhp$ for the $\tanb=2$,
  $\msusy=300\gev$, $A=-300\gev$ scenario.}
\label{brhpwai_tb2}
\end{figure}

\begin{figure}
\includegraphics[height=0.4\textwidth,angle=90]{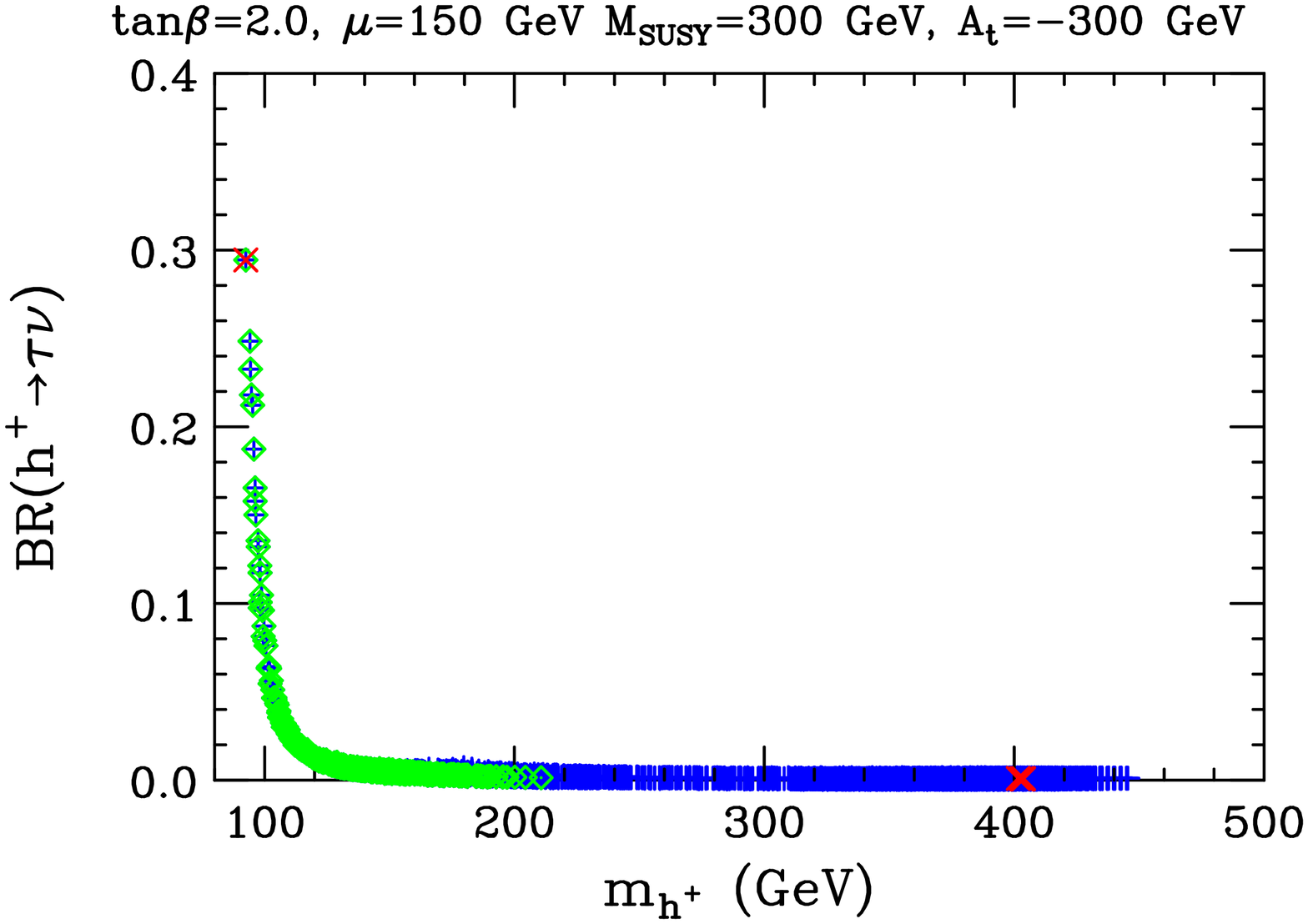}
\caption{$\br(\hp\to \tau^+\nu_\tau)$ is plotted vs. $\mhp$ for the $\tanb=2$,
  $\msusy=300\gev$, $A=-300\gev$ scenario.}
\label{brhptaunu_tb2}
\end{figure}

\begin{figure}
\includegraphics[height=0.4\textwidth,angle=90]{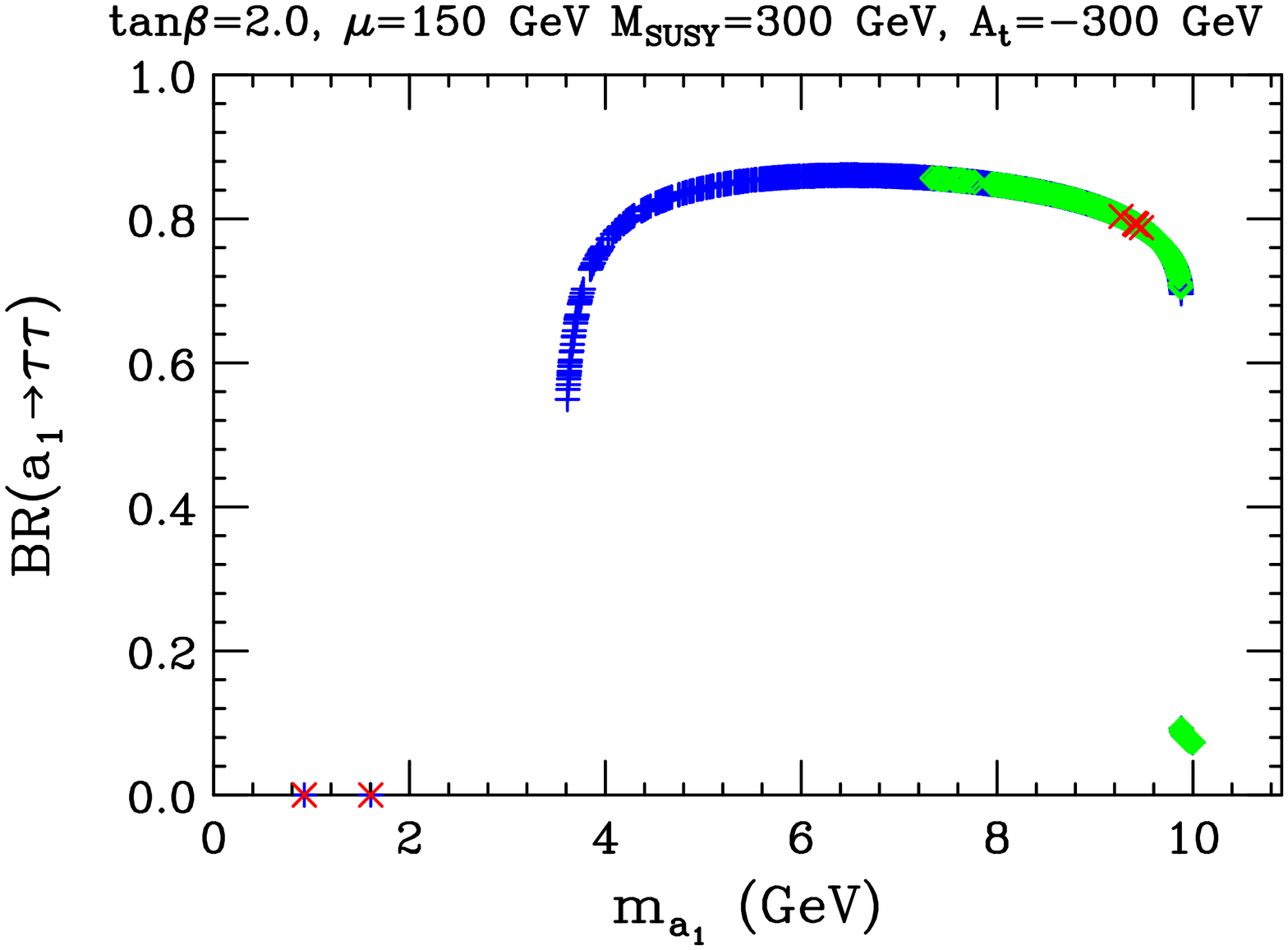}
\caption{$\br(\ai\to \tau^+\tau^-)$ is plotted vs. $\mai$ for the $\tanb=2$,
  $\msusy=300\gev$, $A=-300\gev$ scenario.}
\label{braitautau_tb2}
\end{figure}

\begin{figure}
\includegraphics[height=0.4\textwidth,angle=90]{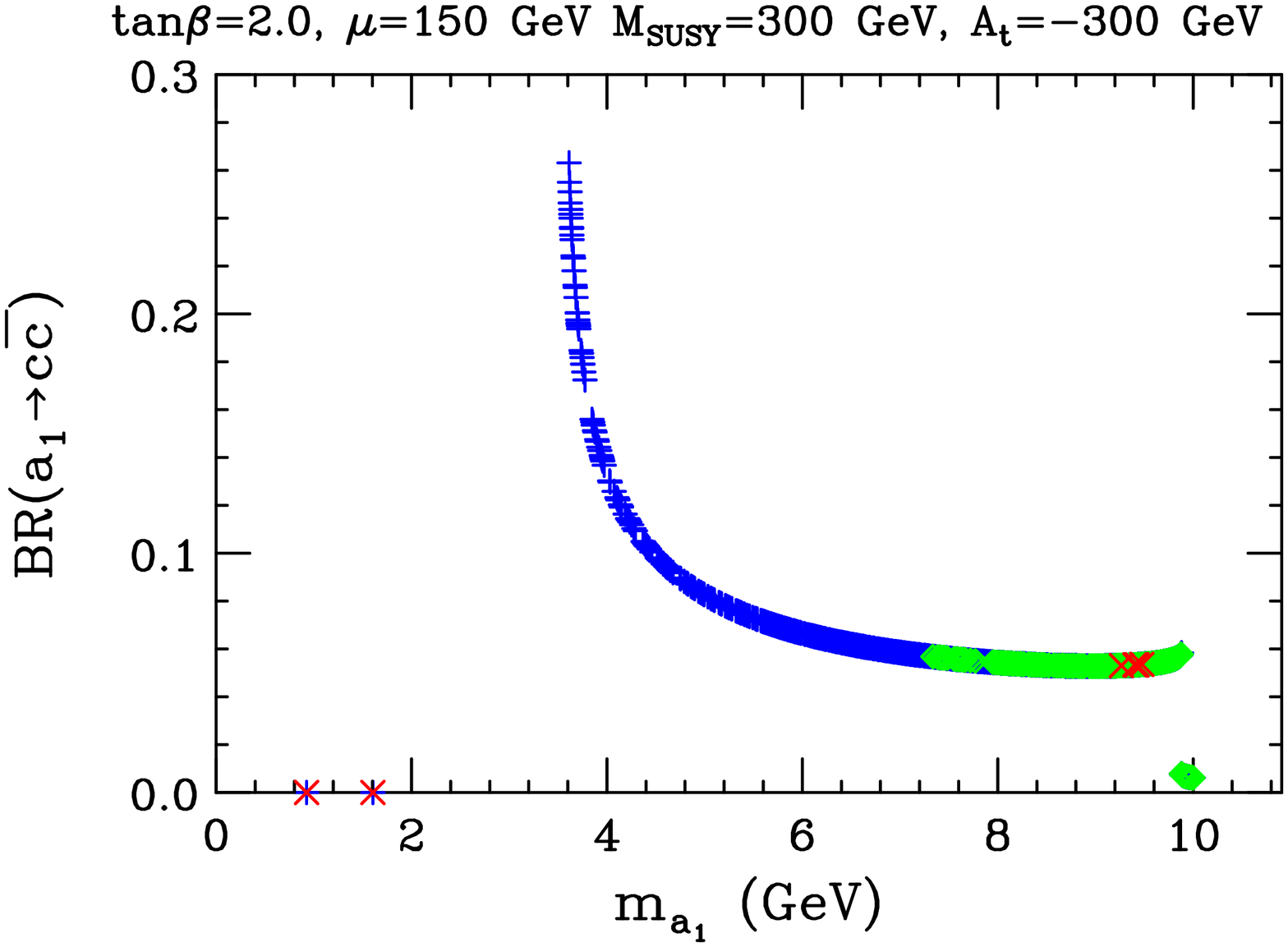}
\caption{$\br(\ai\to c\anti c)$ is plotted vs. $\mai$ for the $\tanb=2$,
  $\msusy=300\gev$, $A=-300\gev$ scenario.}
\label{braicc_tb2}
\end{figure}

\begin{figure}
\includegraphics[height=0.4\textwidth,angle=90]{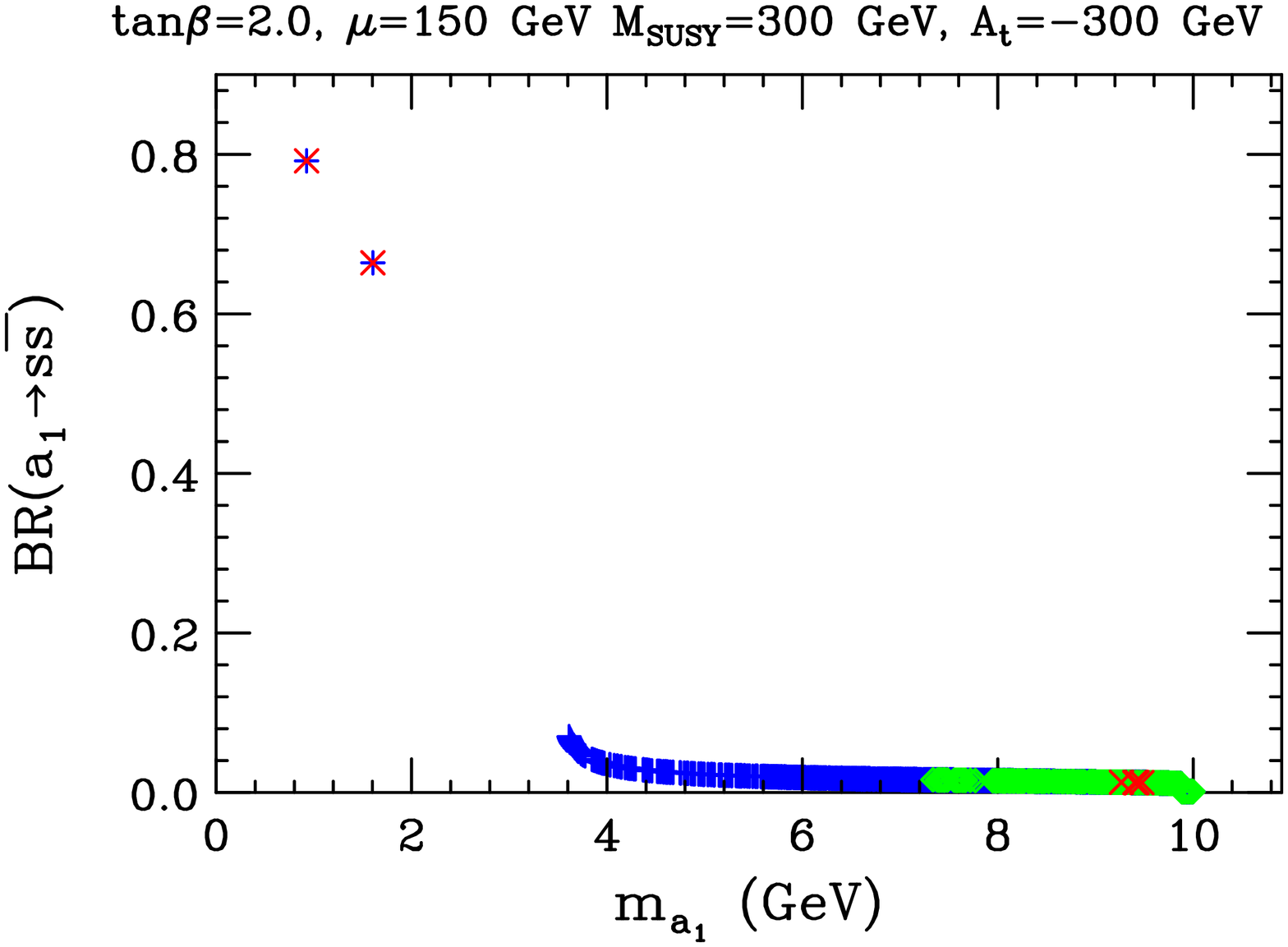}
\caption{$\br(\ai\to s\bar s)$ is plotted vs. $\mai$ for the $\tanb=2$,
  $\msusy=300\gev$, $A=-300\gev$ scenario.}
\label{braiss_tb2}
\end{figure}

\begin{figure}
\includegraphics[height=0.4\textwidth,angle=90]{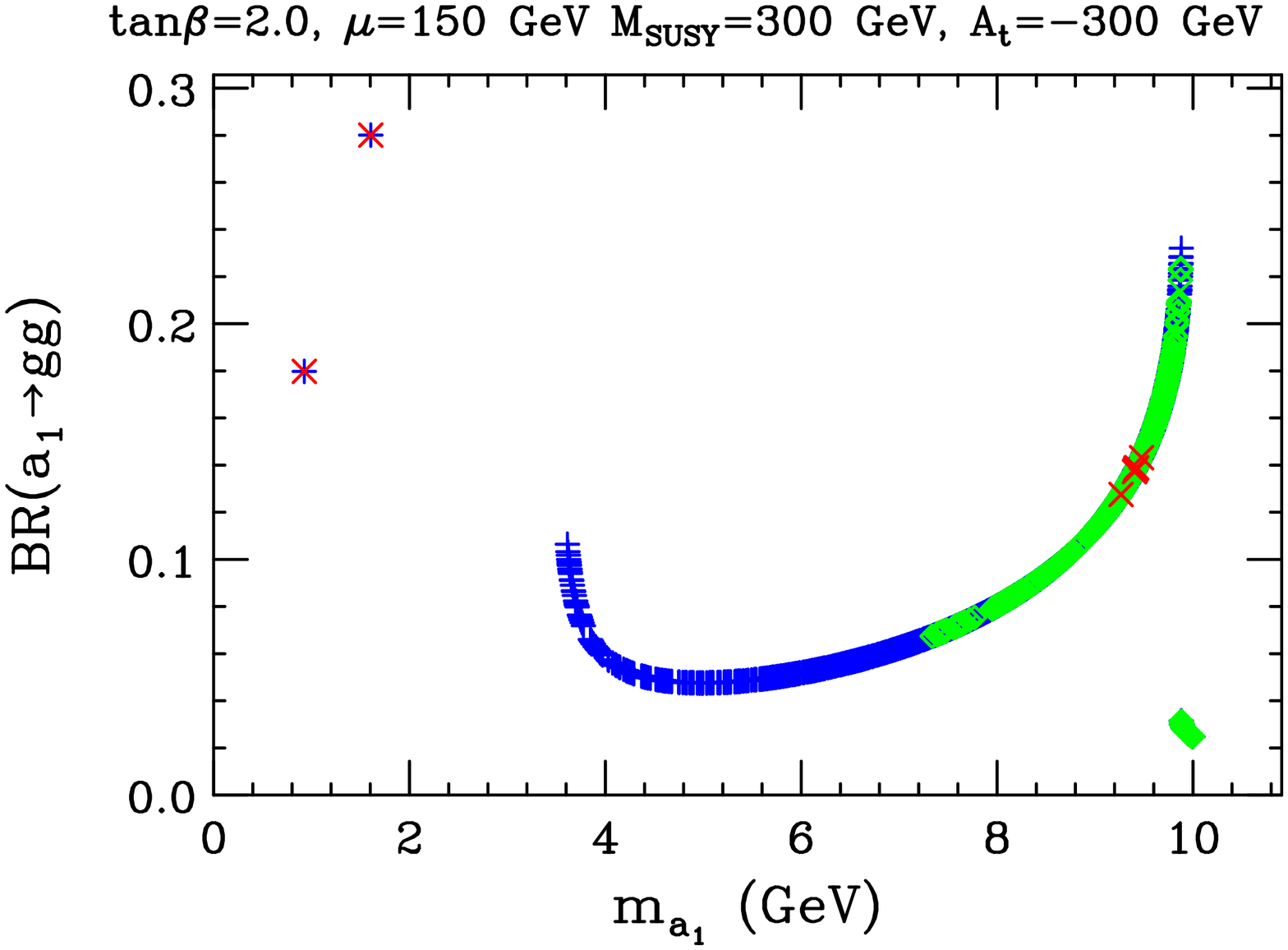}
\caption{$\br(\ai\to gg)$ is plotted vs. $\mai$ for the $\tanb=2$,
  $\msusy=300\gev$, $A=-300\gev$ scenario.}
\label{braigg_tb2}
\end{figure}

\begin{figure}
\includegraphics[height=0.4\textwidth,angle=90]{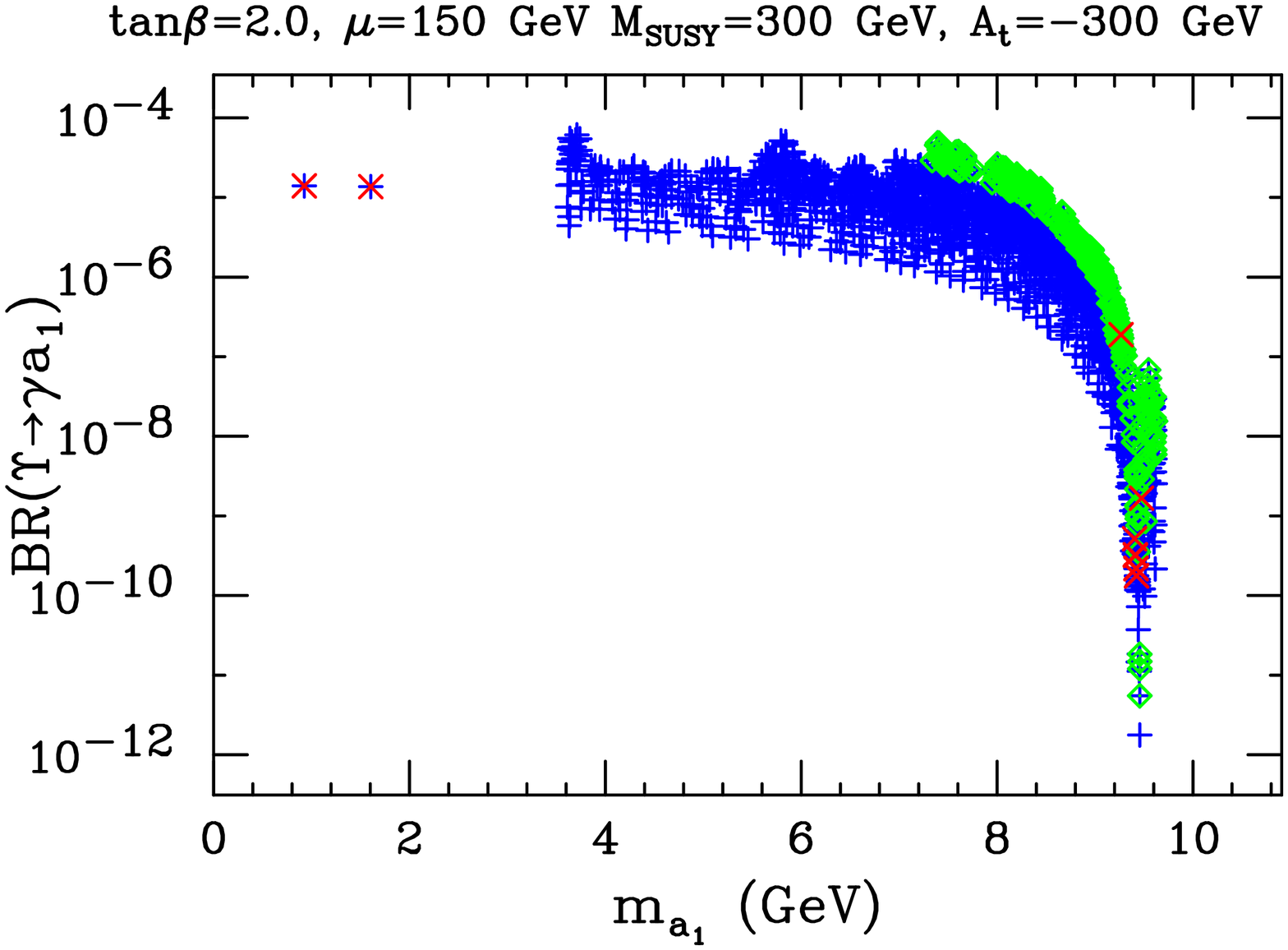}
\caption{$\brups$ is plotted vs. $\mai$ for the $\tanb=2$,
  $\msusy=300\gev$, $A=-300\gev$ scenario.}
\label{brupsvsmai_tb2}
\end{figure}

\begin{figure}
\includegraphics[height=0.4\textwidth,angle=90]{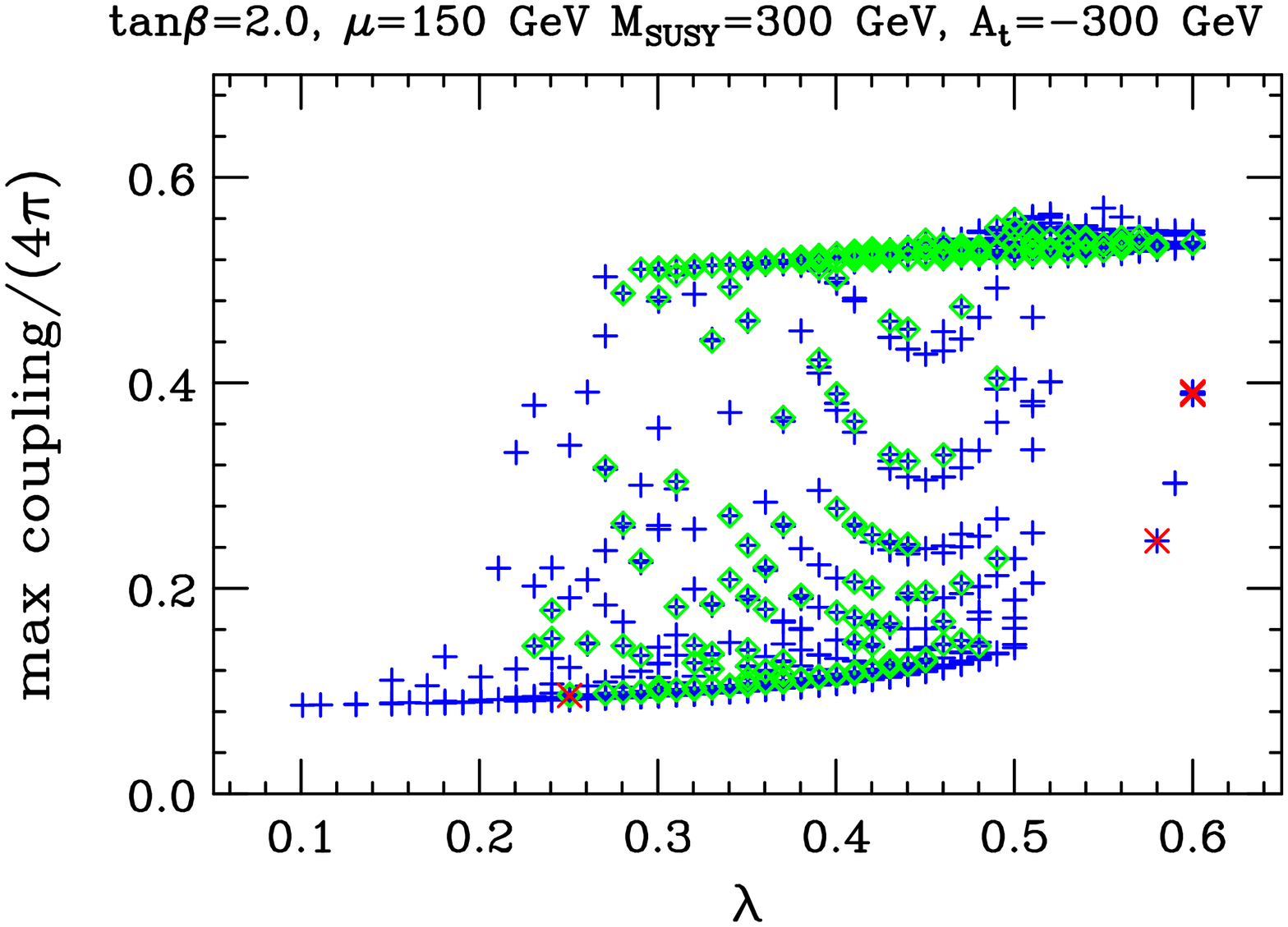}
\caption{$\ymax$ is plotted vs. $\lam$ for the $\tanb=2$,
  $\msusy=300\gev$, $A=-300\gev$ scenario.}
\label{ymaxvslam_tb2}
\end{figure}

\begin{figure}
\includegraphics[height=0.4\textwidth,angle=90]{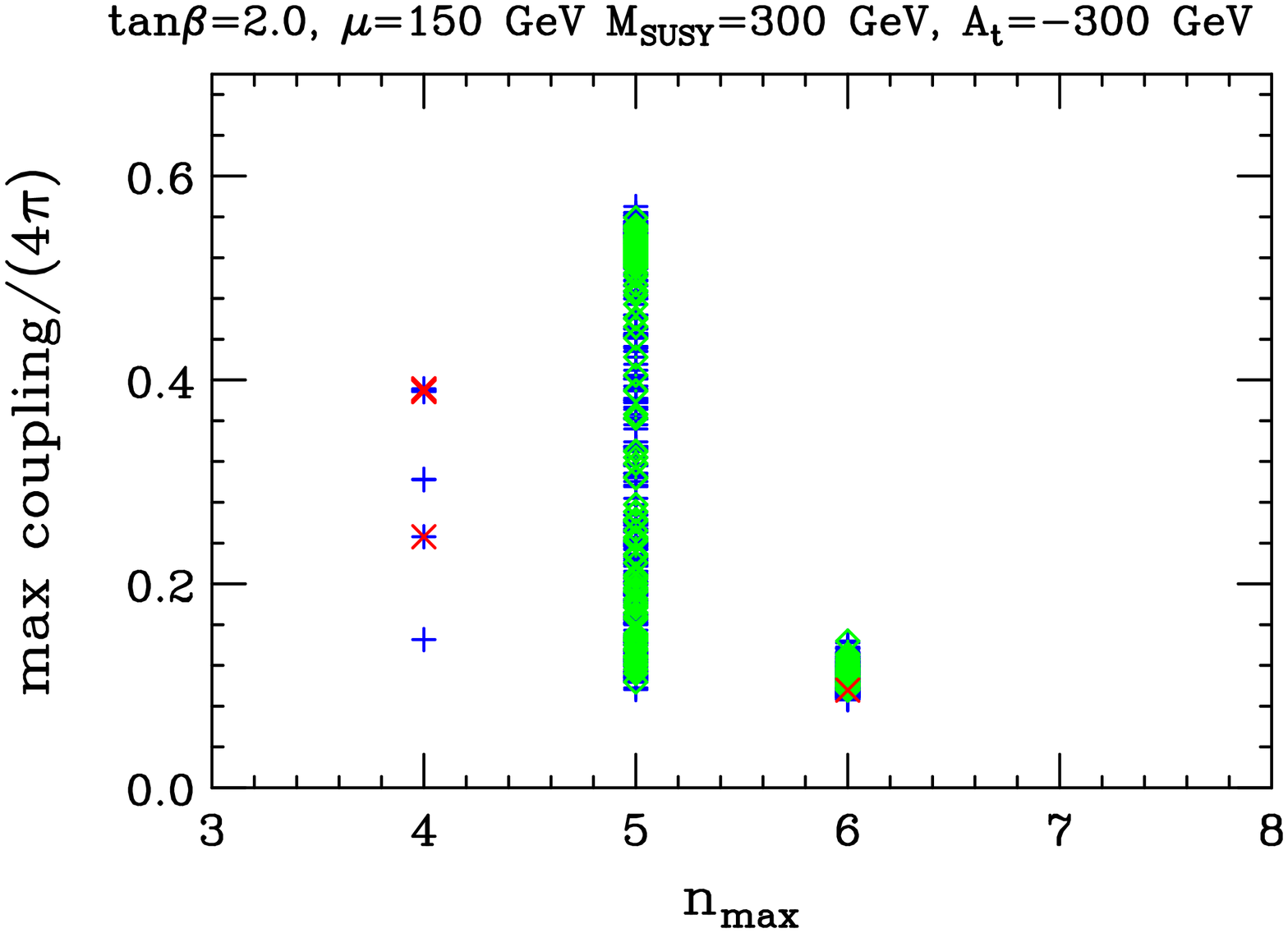}
\caption{$\ymax$ is plotted vs. $\nmax$ for the $\tanb=2$,
  $\msusy=300\gev$, $A=-300\gev$ scenario.}
\label{ymaxvsnmax_tb2}
\end{figure}

\begin{figure}
\includegraphics[height=0.4\textwidth,angle=90]{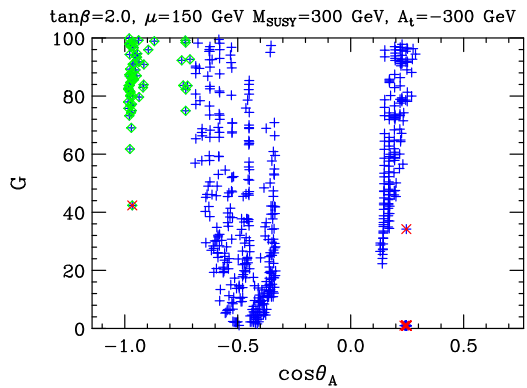}
\caption{$G$ is plotted vs. $\cta$ for the $\tanb=2$,
  $\msusy=300\gev$, $A=-300\gev$ scenario. The displayed points
  comprise only a small fraction of the total number of points
  appearing in previous figures.}
\label{gvscta_tb2}
\end{figure}

\begin{figure}
\includegraphics[height=0.4\textwidth,angle=90]{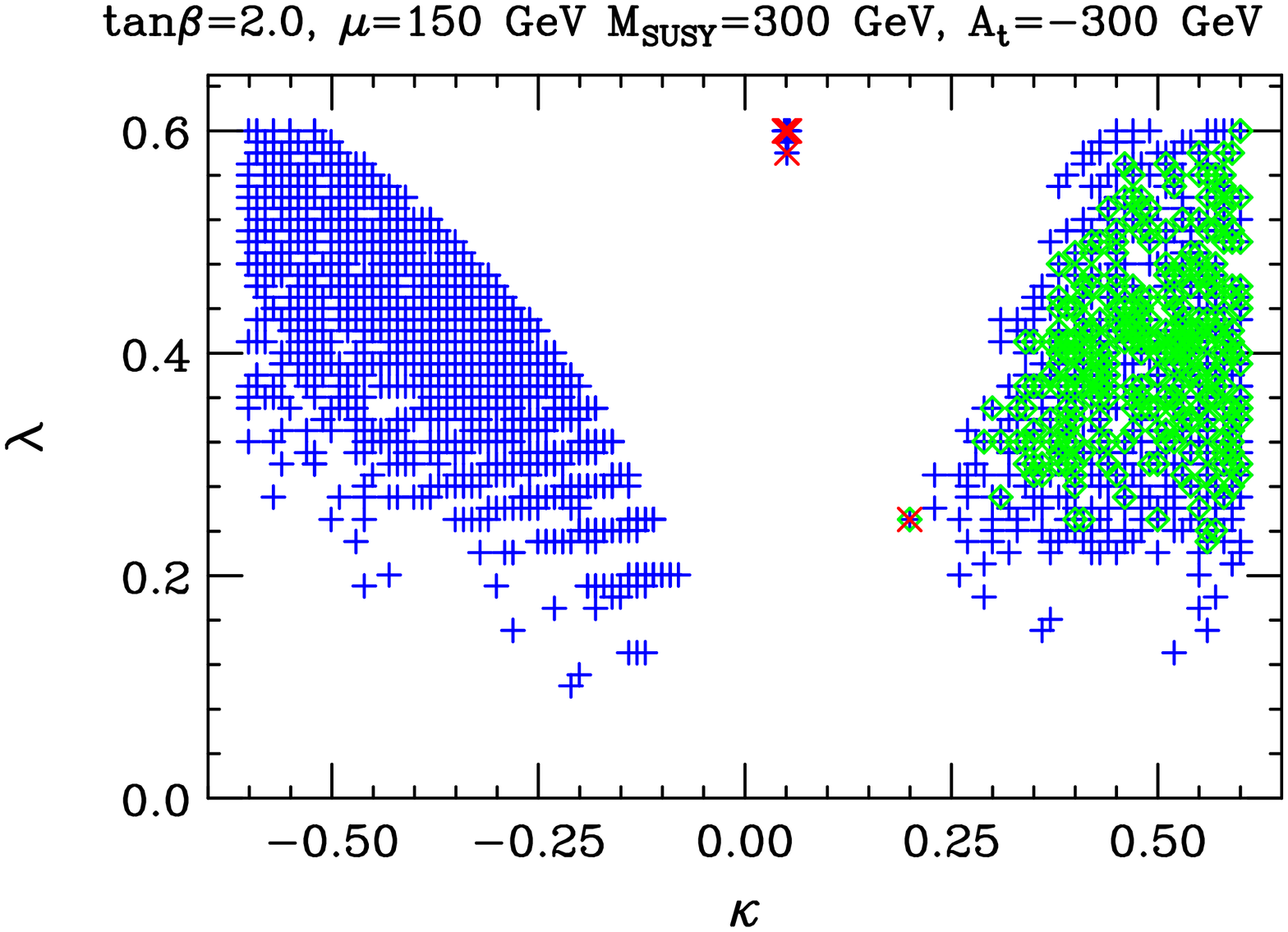}
\caption{$\lam$ is plotted vs. $\kap$ for the $\tanb=2$,
  $\msusy=300\gev$, $A=-300\gev$ scenario. }
\label{lamvskap_tb2}
\end{figure}

\begin{figure}
\includegraphics[height=0.4\textwidth,angle=90]{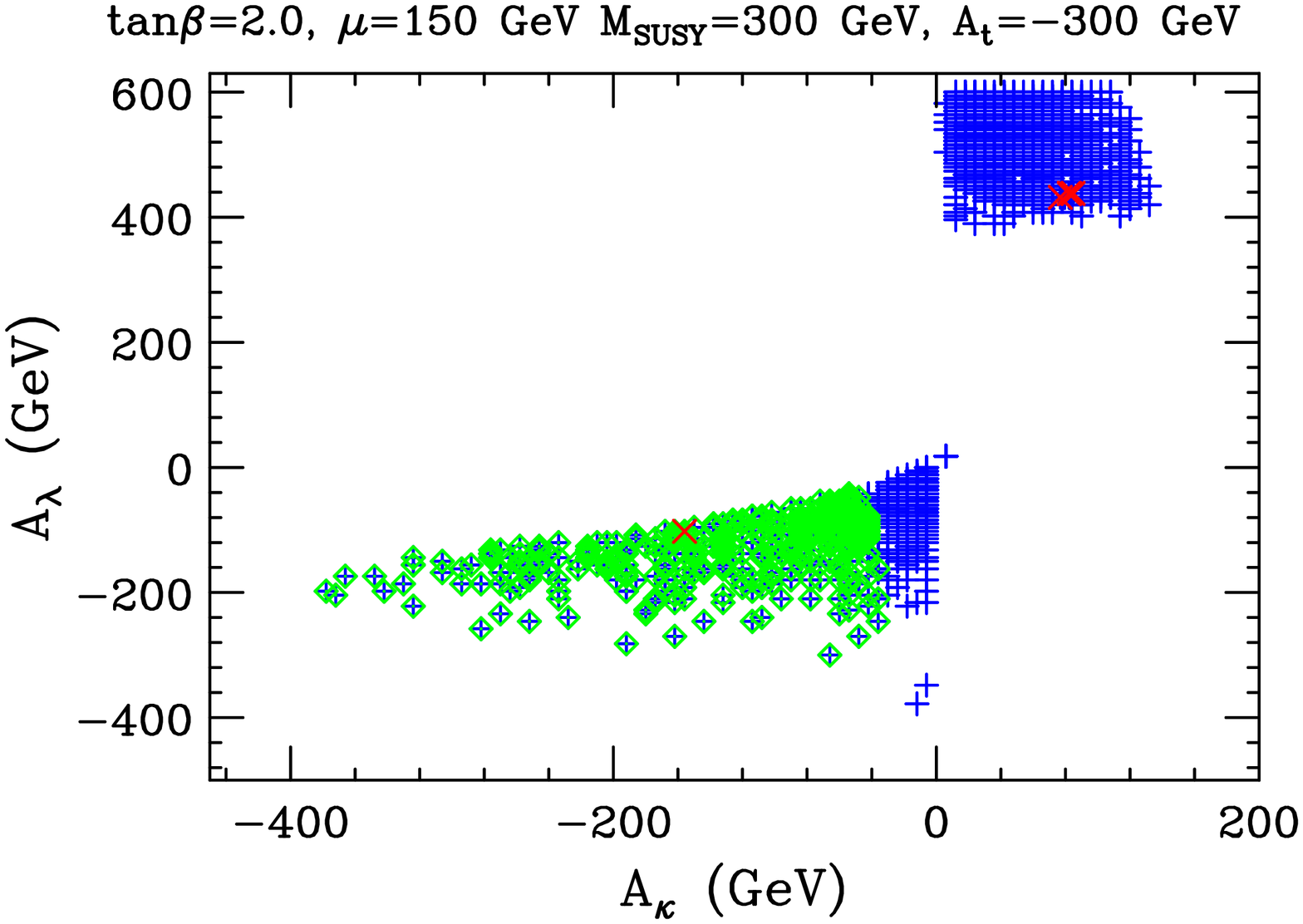}
\caption{$\alam$ is plotted vs. $\akap$ for the $\tanb=2$,
  $\msusy=300\gev$, $A=-300\gev$ scenario. }
\label{alamvsakap_tb2}
\end{figure}

A convenient reference scenario is the case of  $\tan \beta =2$ with
$M_{SUSY} = 300 $ GeV and $A_t=A_b=A_\tau=-300\gev$. 
The plots for this case are Figs.~\ref{ctasqvsmai_tb2} -- \ref{alamvsakap_tb2}.
In our plots, the blue $+$'s are all points that
satisfy the NMHDECAY constraints, while green diamonds are
those which in addition have a light CP-odd Higgs which is
doublet-like, $\cos^2 \theta_A > 0.5$. The red crosses single out those
points for which $\mhi<65\gev$. 
Because of the limits on $|\cta|$ discussed above, $\mai$ values below
about $7.5\gev$ are disallowed for $\cos^2\theta_A>0.5$, as are many
points with  $\cos^2\theta_A<0.5$.  This is illustrated in
Fig.~\ref{ctasqvsmai_tb2}. The jagged shape of the boundary 
in the $\mai<7.5\gev$ region for the
$\cos^2\theta_A<0.5$ points simply reflects the rather rapid
variations in the limits from $\brups$ decays.


The first plot, Fig.~\ref{maivsmhi_tb2} shows the $\mhi$ masses that
are obtained in our scan and the correlation with $\mai$. Higgs with
$\mhi<114\gev$ are not excluded by LEP data. The reason is apparent
from Figs.~\ref{cvisqvsmhi_tb2}, \ref{brhiaiaivsmhi_tb2} and
\ref{gzzhisqbrhibbvsmhi_tb2}. There, we
plot $\cvisq\equiv g_{ZZ\hi}^2/g_{ZZ\hsm}^2$, $\br(\hi\to\ai\ai)$ and
$\cvisq\br(\hi\to b\anti b)$
vs. $\mhi$. We see that the light $\hi$ escapes LEP constraints mainly
because of large $\br(\hi\to\ai\ai)$ (where $\ai\to\tau^+\tau^-$
yields a $4\tau$ final state that is weakly constrained by LEP data)
although there are a significant number of points for which the $\ai$
is mainly singlet with small $\cvisq$. The plot of
Fig.~\ref{gzzhisqbrhibbvsmhi_tb2} shows the net rate for 
$\epem\to Z b\anti b$ relative to the SM prediction.  We observe that
away from the $90\gev$ to $105\gev$ window in $\mhi=m_{b\anti b}$, in
which there is an excess of LEP events relative to background, this net
rate must be quite small.  In the $90\gev$ to $105\gev$ window, the
best fit to the experimental data corresponds to an an excess of order $0.1$
times the expected SM rate is allowed.  However, in this window, an excess as
large as $0.2$ times the SM rate is still allowed at 90\% CL, as reflected in the
plot. Note that it is mainly the points with $\cos^2\theta_A\lsim 0.5$
that best explain the observed $0.1\times$ SM excess in this region.

The next interesting feature of these small $\mai$, small $\tanb$
scenarios is the very substantial probability that the $\hp$ will also
be quite light.  As shown in Fig.~\ref{mhpvsmhi_tb2}, this is
particularly the case for parameters such that the $\ai$ is mainly
doublet. For these $\ai$-doublet-like scenarios, we observe that there
are cases for which $\mhi$ is well below $100\gev$ while $\mhp$ is of
order $100\gev$, and in the vast majority of these $\ai$-doublet-like
scenarios $\mhp<170\gev$ so that the $\hp$ would have been produced in
top decays. At the same time, as shown in Fig.~\ref{mhiivsmhi_tb2},
for the $\ai$-doublet-like scenarios $\mhii$ can also be of order
$100\gev$, and in nearly all cases $\mhii<200\gev$ so that $\epem\to
Z\hii$ production events would be present in LEP data.  In the case of
$\mhii\sim 100\gev$ it is the reduced $\cviisq$
(Fig.~\ref{cviisqvsmhii_tb2}) coupled with large $\br(\hii\to \ai\ai)$
(Fig.~\ref{brhiiaiaivsmhii_tb2}) that makes LEP sensitivity in the
$Z\hii\to Z b\anti b$ channel small. Indeed,
Fig.~\ref{gzzhiisqbrhiibbvsmhii_tb2} shows that the $\hii$
contribution to the $Z\hii\to Z b\anti b$ channel can only be
significant for $\mhii\sim 125\gev$, well above the LEP kinematic
reach.  (However, as we shall see, this conclusion does not apply to
all choices of $\tanb$ and soft-SUSY-breaking parameters.)  For these
same scenarios with large $\ctasq$, the $\ai$ and $\hii$ both have
substantial doublet component, and the $Z\to \hii\ai$ rate at LEP
would also have been significant. For $\mhii$ near $100\gev$, the
$\hii\ai$ final states would have escaped LEP detection because of
large $\br(\hii\to\ai\ai)$. For larger $\mhii$ up near $200\gev$,
$\br(\hii\to Z\ai)$ would have been large, see
Fig.~\ref{brhiizaivsmhii_tb2}, and LEP did not analyze their data in
such a way as to be sensitive to $\hii\ai\to Z\ai\ai$ final states,
especially given that $\ai$ decays to either two taus or two jets.

Turning to the charged Higgs boson, 
most LEP searches for the $\hp$ were based on the dominant $\epem\to \hp\hm$
production mechanism assuming that $\hp\to\tau^+\nu_\tau$ and $\hp\to
c\anti s$ were the only two decay modes. However,
Fig.~\ref{brhpwai_tb2} shows that $\hp\to W^{+\,(*)}\ai$ is dominant for
the $\ai$-doublet-like scenarios. Limits on $\mhp$ weaken as
$\br(\hp\to\tau^+\nu_\tau)$ declines --- if $\br(\hp\to
\tau^+\nu_\tau)\sim 1,0.5,0$ the limits are roughly
$\mhp>90,80,80\gev$, respectively~\cite{:2001xy}.  DELPHI considered
the possibility of $\hp\to W^{+\,(*)} a$ assuming $a\to b\anti b$ is
dominant~\cite{Abdallah:2003wd}. However, their limits on $\mhp$  do not
apply to the case of $m_a<2m_b$ of interest here.

Overall, we have the remarkable result that for the chosen $\tanb=2$
and $\msusy=300\gev,A=-300\gev$ parameters there are a large number of
model points (the $\ai$-doublet-like points) for which the $\hi$ and
$\hpm$ have mass at or below $100\gev$ and the $\hii$ has mass in the
range $100-190\gev$. All would have been copiously produced at LEP,
and yet all would have escaped LEP detection.

The primary sensitivity of the Tevatron to the $\ai$-doublet-like
scenarios with a light $\hp$ is through searches for $t\anti t$
production with one $t$ decaying via $t\to \hp \anti
b$~\cite{Grenier:2007xj,newcdf,newd0}.  The recent preliminary
Tevatron analyses~\cite{newcdf,newd0} set limits on the $\br(t\to \hp
b)\br(\hp\to\tau^+\nu_\tau)$ as a function of $\mhp$.  Fitting
simultaneously the branching ratio product and $\sigma(p\bar p\to
t\anti t)$, the limit for $\mhp=80\gev$ is $\br(t\to \hp
b)\br(\hp\to\tau^+\nu_\tau)<0.12$.  In the present scenario, all these
searches would have suppressed sensitivity for the cases where
$\mhp\sim 100\gev$. The reason is that $\hp\to \wstarp \ai$ always has
branching ratio $>0.5$, and the $\ai$ decays primarily to
$\tau^+\tau^-$ for $\mai>2\mtau$ and to various lighter final states
if $\mai<2\mtau$. $\br(\hp\to \wstarp\ai)$ is shown in
Fig.~\ref{brhpwai_tb2}.  From Fig.~\ref{brhptaunu_tb2}, we see that
$\br(\hp\to\tau^+\nu_\tau)\sim 1-\br(\hp\to \wstarp\ai)$ for
$\mhp<\mt+\mb$.

Of course, $\br(t\to\hp \anti b)$ is $\tanb$ dependent. For
$\tanb=1.2,1.7,2$ it is of order $0.3,0.173,0.126$ for $\mhp\sim 90\gev$. For
$\mhp>\mt+\mb$ one finds that $\hp\to t\anti b$ is the dominant
decay. At $\tanb=2$, $\br(\hp\to \wstarp\ai)$ can be of order $0.5$ out to
relatively large $\mhp$. LHC search strategies sensitive to all these
unusual scenarios need to be developed. Some discussion of the
possibilities appears in Ref.~\cite{Akeroyd:2007yj}.


Given the fact that the $\ai$ appears in so many decays, it is useful
to review its branching ratios.  In NMHDECAY, these are computed using
partonic final states and masses.  This implies a few inaccuracies.
In particular, $\ai\to s\anti s$ is non-zero even when $\mai<2m_K$
since NMHDECAY employs $m_s=0.19\gev$. The $\ai$ branching ratios appear
in Figs.~\ref{braitautau_tb2}, \ref{braicc_tb2}, \ref{braiss_tb2} and
\ref{braigg_tb2}.
As expected, if $\mai<2m_b$ but well above $2\mtau$ (as is the case for all
$\ai$-doublet-like scenarios), $\ai\to \tauptaum$ is the dominant
decay, with the remainder being in the $\ai\to gg$ and $\ai\to c\anti
c$ modes (in that order). For cases where the $\ai$ is approaching $2\mtau$,
$\br(\ai\to \tauptaum)$
declines, but is always bigger than $0.5$ if $\mai>2\mtau$ with the
residual mainly taken up by $\br(\ai\to c\anti c)$.  For the few
$\mai<2m_c$ points, $\ai\to s\anti s$ is dominant.


As noted earlier, an important constraint on scenarios with a light
$\ai$ is the branching ratio $\brups$ which is strongly constrained by
data from CLEO-III~\cite{cleoiii} for $\mai>2\mtau$. In particular,
depending upon the precise value of $\mai$ in the range between
$2\mtau$ and $7.5\gev$, the 95\% CL upper limit on $\brups$ is between
$6\times 10^{-5}$ and $1.2\times 10^{-5}$.  In
Fig.~\ref{brupsvsmai_tb2} we plot $\brups$ as a function of $\mai$
after imposing $\brups$ constraints.  We see that points with
$\cos^2\theta_A>0.5$ (the green diamonds) are only allowed at
relatively large $\mai$ and that even some points with
$\cos^2\theta_A<0.5$ have been eliminated in the $2\mtau<\mai<7.5\gev$
range. Thus, it is the $\Upsilon\to \gam \ai\to \gam\tauptaum$ decay limits
that rule out $\ai$-doublet-like scenarios with $\mai\lsim
7.5\gev$. The underlying reason for the $\ai$-doublet-like points to
be more strongly excluded by $\Upsilon$ decays is that, as discussed
earlier, the $\ai b\anti b$ coupling is given by $\caibb=\cta\tanb$,
which is, of course, largest for large $|\cta|$.

Let us turn for a moment to a discussion of whether or not these
scenarios are perturbative after evolution to the GUT
scale.~\footnote{For small $\tan \beta$ the the top Yukawa coupling becomes
  non-perturbative close to the grand unification (GUT) scale. The
  exact value of $\tan \beta$ consistent with perturbativity all the
  way to the GUT scale depends on superpartner masses through SUSY
  threshold corrections to the top Yukawa coupling, and in the NMSSM
  it is about $\tan \beta \gtrsim 1.6$. However, adding extra
  vector-like complete SU(5) matter multiplets at the TeV scale, \eg\ the
  parts of the sector that mediate SUSY breaking (messengers) or are 
  present for no particular reason, does not affect the unification of
  gauge couplings while it slows down the running of the top Yukawa
  coupling~\cite{Masip:1998jc, Barbieri:2007tu} and even $\tan \beta
  \simeq 1$ can be consistent with perturbative unification of gauge
  couplings.}  The couplings of interest are $\lam$, $\kap$, $h_t$ and
$h_b$.  At low $\tanb$, $h_b$ always remains perturbative but $\lam$,
$\kap$ and $h_t$ can become large.  In Fig.~\ref{ymaxvslam_tb2} we
plot the value of
\beq
\ymax\equiv {{\rm max}\{\lam,\kap,h_t,h_b\}\over 4\pi}
\eeq
at the GUT scale as a function of $\lam$ for the various scenarios in
our $\tanb=2$ scan. A value of $\ymax=0.5$ indicates that the
evolution has gone non perturbative. In Fig.~\ref{ymaxvsnmax_tb2}, we
show which of the couplings is largest or has gone non-perturbative
first using the code
$4\equiv \lam$, $5\equiv \kap$, $6\equiv h_t$ and $7\equiv h_b$. We
observe that it is most often $\kap$ that has the largest coupling at
the GUT scale, especially for the $\cos^2\theta_A>0.5$ scenarios.

Another issue of interest is whether finetuning of the NMSSM
parameters (in particular $\alam$ and $\akap$) is required (either at
scale $\mz$ or at the GUT scale) in order to obtain $\mai<10\gev$ and
scenarios that escape LEP and other limits.  In~\cite{Dermisek:2006wr}
we developed a measure $G$ of this fine tuning.  In
Fig.~\ref{gvscta_tb2}, we plot $G$ as a function of $\cta$.  We see
that small values of $G$ arise for quite specific values of $\cta$,
namely $-0.6\lsim \cta \lsim -0.4$ and $0.15\lsim\cta\lsim 0.22$. Note
that the $\ai$-doublet-like scenarios typically have moderately large
$G$ values --- only if the $\ai$ is singlet like is it possible for there
to be no need for tuning $\alam$ and $\akap$ in order to achieve
$\mai<2m_b$ {\it and} large $\br(\hi\to\ai\ai)$ (to escape LEP limits)
simultaneously.
 
In our scans, we did not specifically exclude scenarios because of
difficulties with precision electroweak constraints (mainly the
parameter $T$) or the anomalous magnetic moment of the muon,
$a_\mu$.  In fact, for all the
points plotted, $-0.002<\Delta T<0.011$, where $\Delta T$ is defined
relative to a SM-like Higgs with mass $100\gev$, and $-2.2\times
10^{-10}<\delta a_\mu<-1.4\times 10^{-10}$ where $\delta a_\mu$ is the
net contribution of the entire Higgs sector. Clearly, the size of
$\Delta T$ is such that the Higgs sector of the NMSSM models being
considered makes a very small contribution to $T$ while $\delta a_\mu$
is also so small as to have little impact on the current discrepancy
between the SM prediction for $a_\mu$ and the experimental
observation, which difference is of order $30\times 10^{-10}$.


Finally, we show in Figs.~\ref{lamvskap_tb2} and
$\ref{alamvsakap_tb2}$ the $\lam$, $\kap$, $\alam$ and $\akap$ values
which yield the points plotted in the preceding figures. The main
observation is that the $\cos^2\theta_A>0.5$ points require $\kap>0$
and $\akap,\alam<0$.  Note also the small number of points with $\kap$
close to zero.  Many, but not all, of the very small $\mhi<65\gev$
scenarios arise from these points.

\subsection{Results for $\tan \beta =1.7$}

\begin{figure}
\includegraphics[height=0.4\textwidth,angle=90]{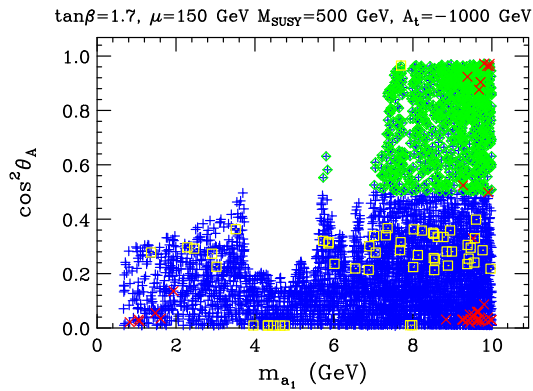}
\caption{$\cos^2\theta_A$ is plotted vs. $\mai$ for the $\tanb=1.7$,
  $\msusy=500\gev$, $A=-1000\gev$ scenario. }
\label{ctasqvsmai_tb1pt7}
\end{figure}

\begin{figure}
\includegraphics[height=0.4\textwidth,angle=90]{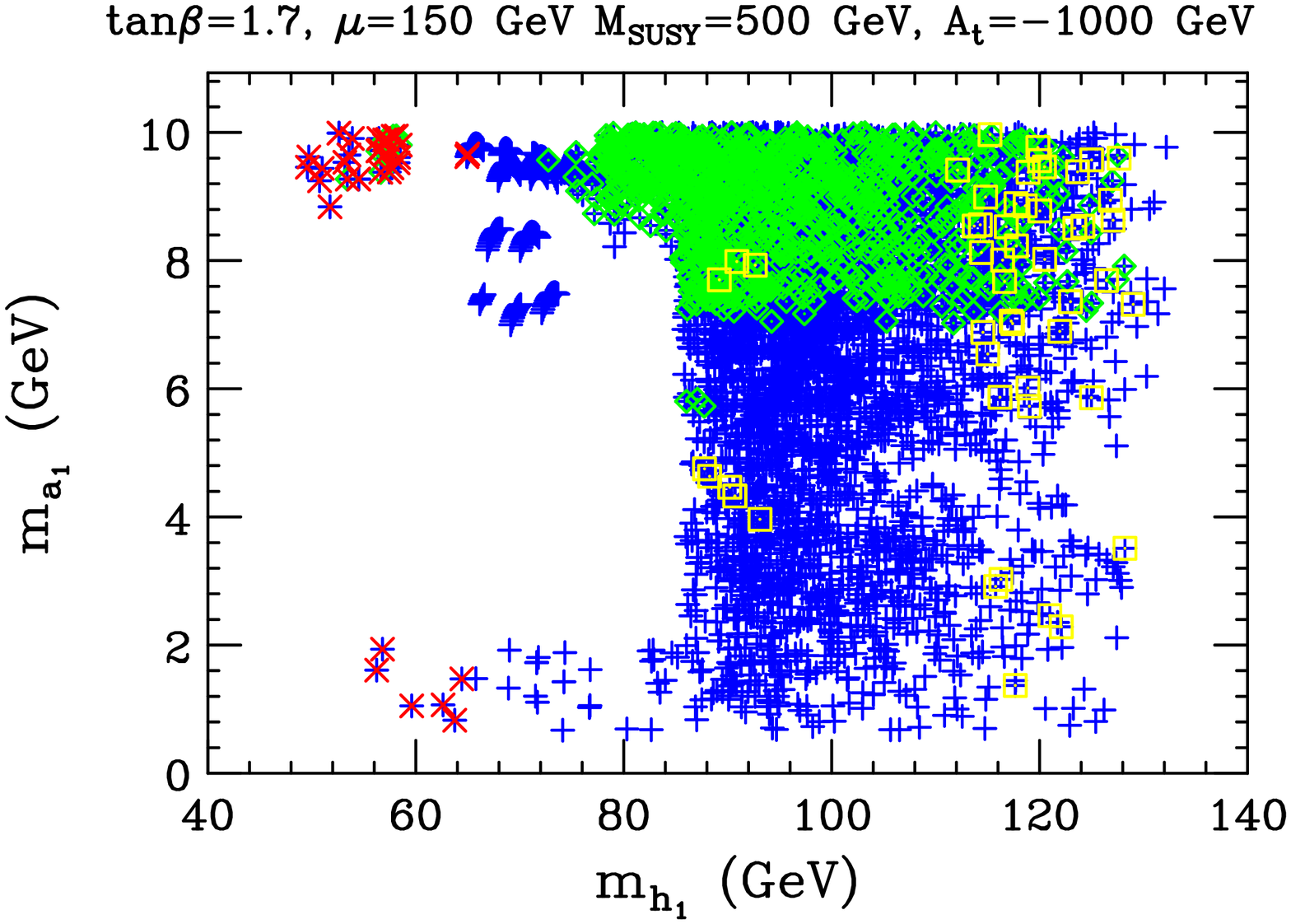}
\caption{$\mai$ is plotted vs. $\mhi$ for the $\tanb=1.7$,
  $\msusy=500\gev$, $A=-1000\gev$ scenario. }
\label{maivsmhi_tb1pt7}
\end{figure}

\begin{figure}
\includegraphics[height=0.4\textwidth,angle=90]{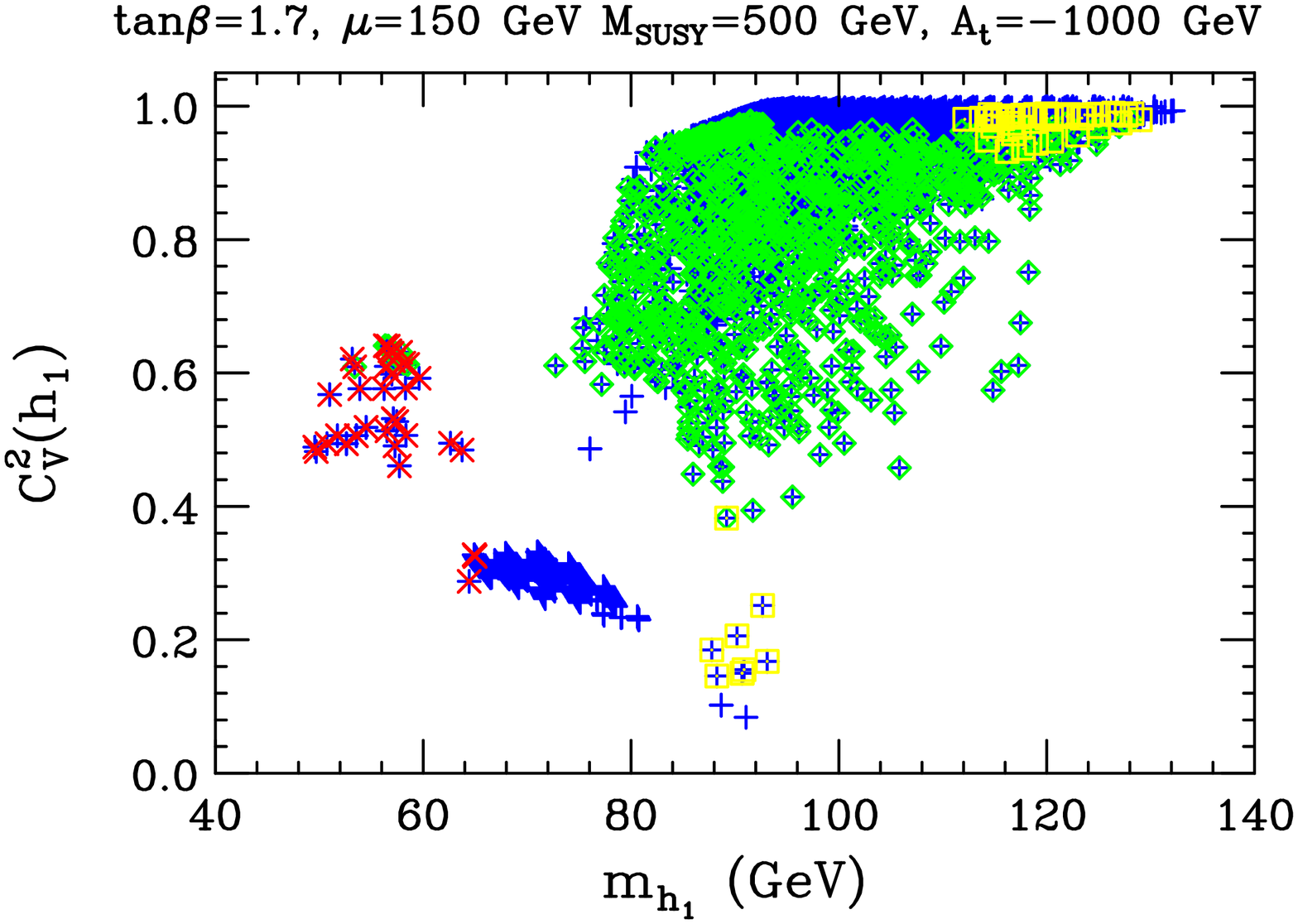}
\caption{$\cvisq$ is plotted vs. $\mhi$ for the $\tanb=1.7$,
  $\msusy=500\gev$, $A=-1000\gev$ scenario.}
\label{cvisqvsmhi_tb1pt7}
\end{figure}

\begin{figure}
\includegraphics[height=0.4\textwidth,angle=90]{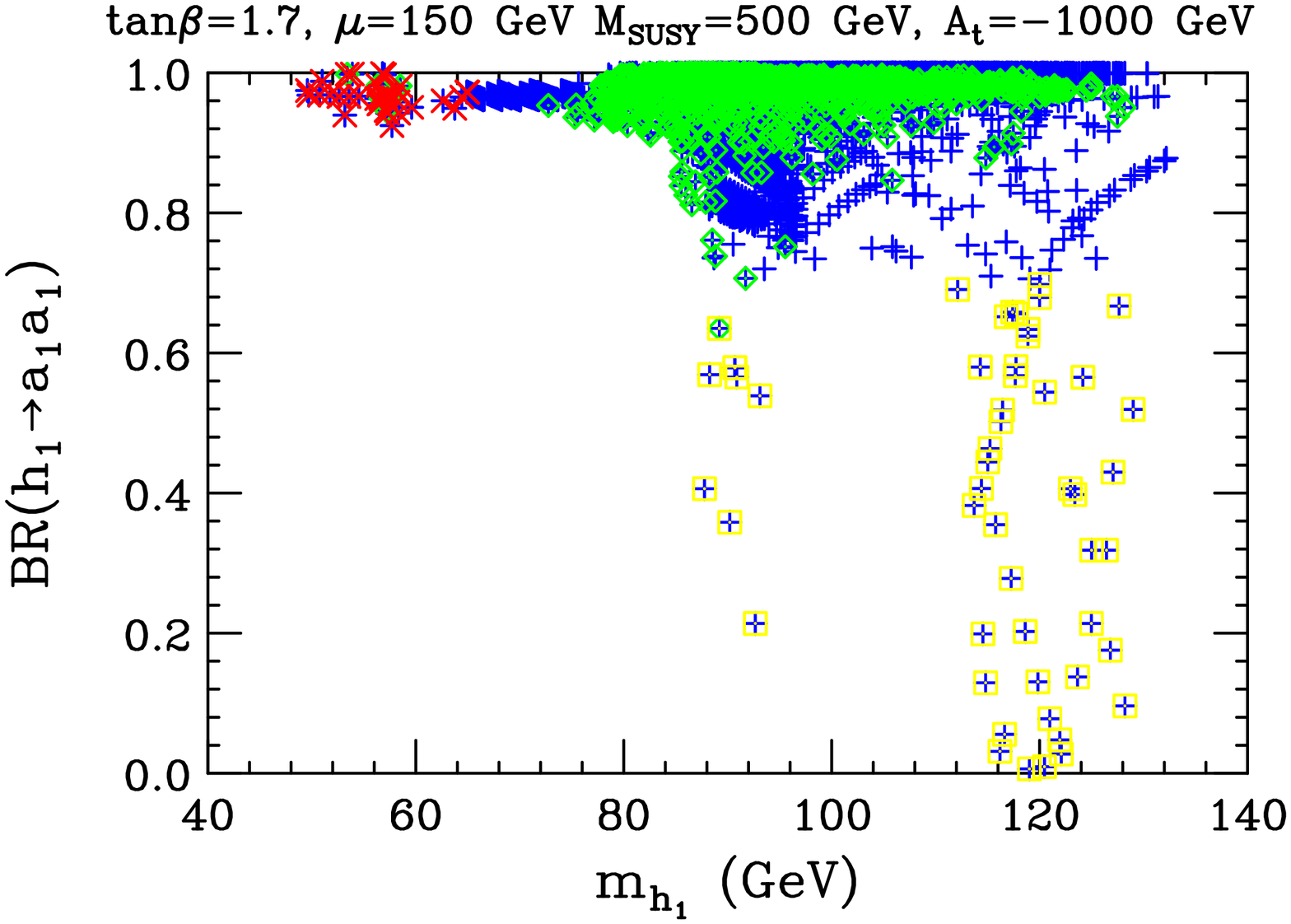}
\caption{$\br(\hi\to\ai\ai)$ is plotted vs. $\mhi$ for the $\tanb=1.7$,
  $\msusy=500\gev$, $A=-1000\gev$ scenario.}
\label{brhiaiaivsmhi_tb1pt7}
\end{figure}

\begin{figure}
\includegraphics[height=0.4\textwidth,angle=90]{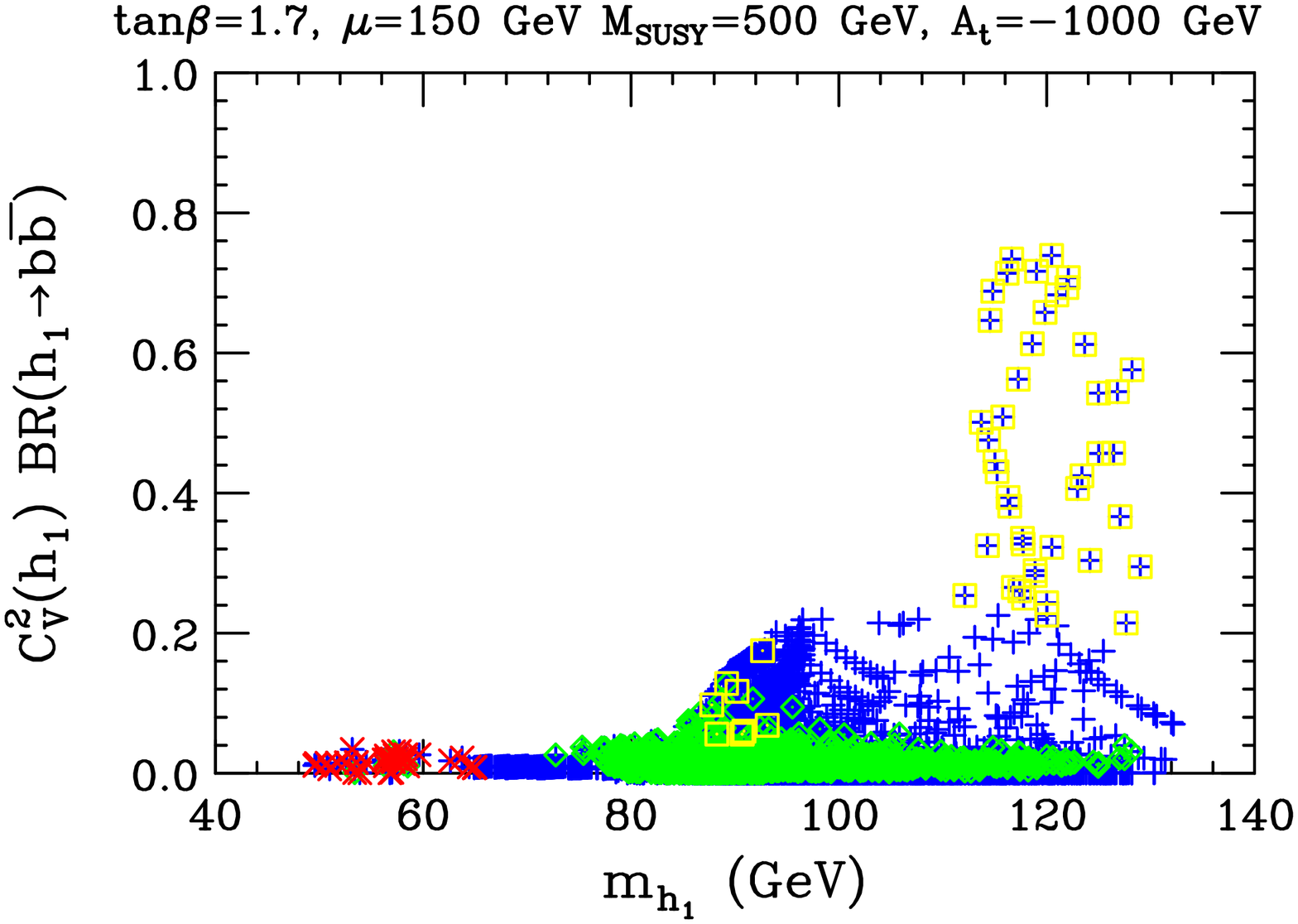}
\caption{$\cvisq\br(\hi\to b\anti b)$ is plotted vs. $\mhi$ for the $\tanb=1.7$,
  $\msusy=500\gev$, $A=-1000\gev$ scenario.}
\label{gzzhisqbrhibbvsmhi_tb1pt7}
\end{figure}

\begin{figure}
\includegraphics[height=0.4\textwidth,angle=90]{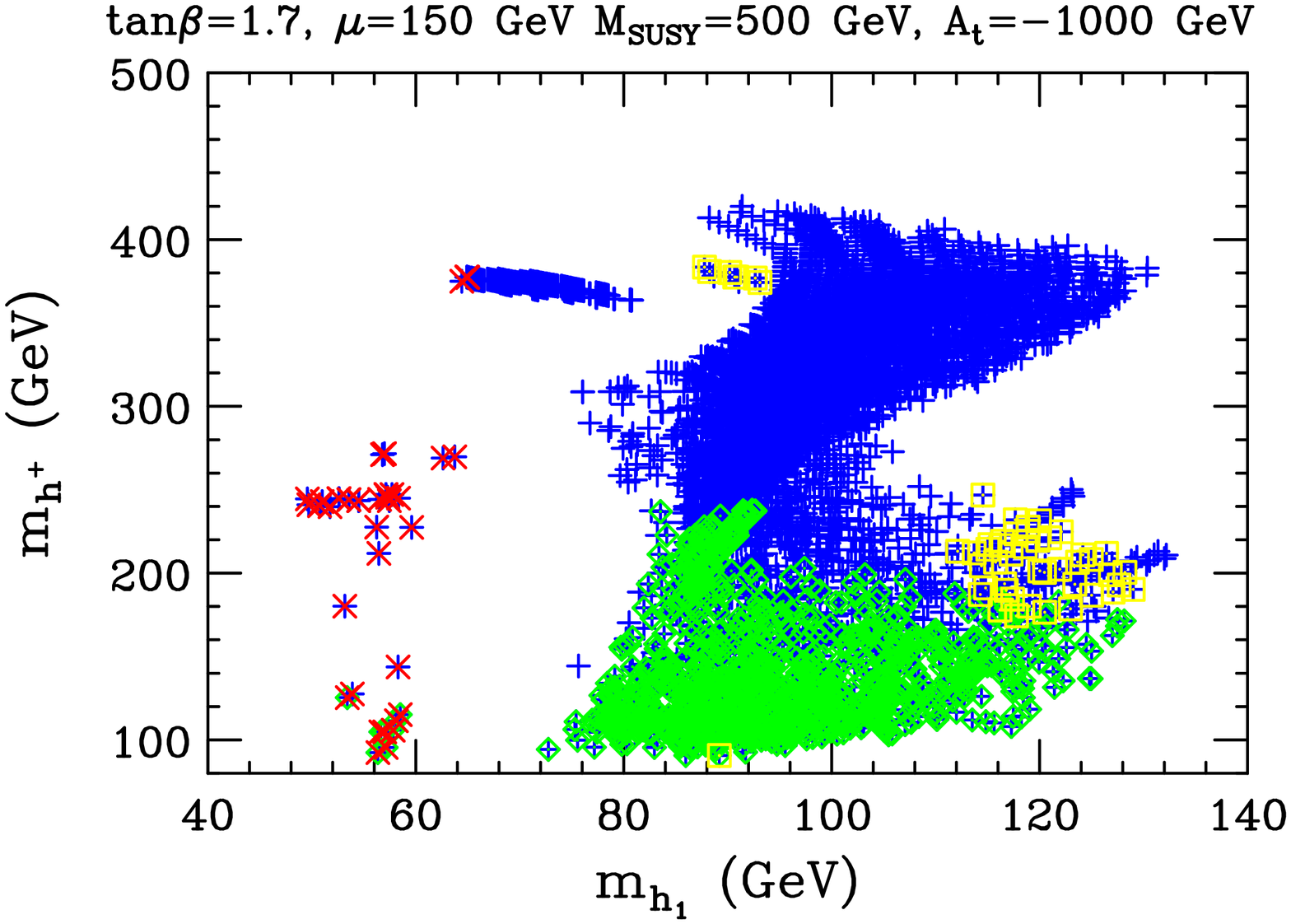}
\caption{$\mhp$ is plotted vs. $\mhi$ for the $\tanb=1.7$,
  $\msusy=500\gev$, $A=-1000\gev$ scenario.}
\label{mhpvsmhi_tb1pt7}
\end{figure}

\begin{figure}
\includegraphics[height=0.4\textwidth,angle=90]{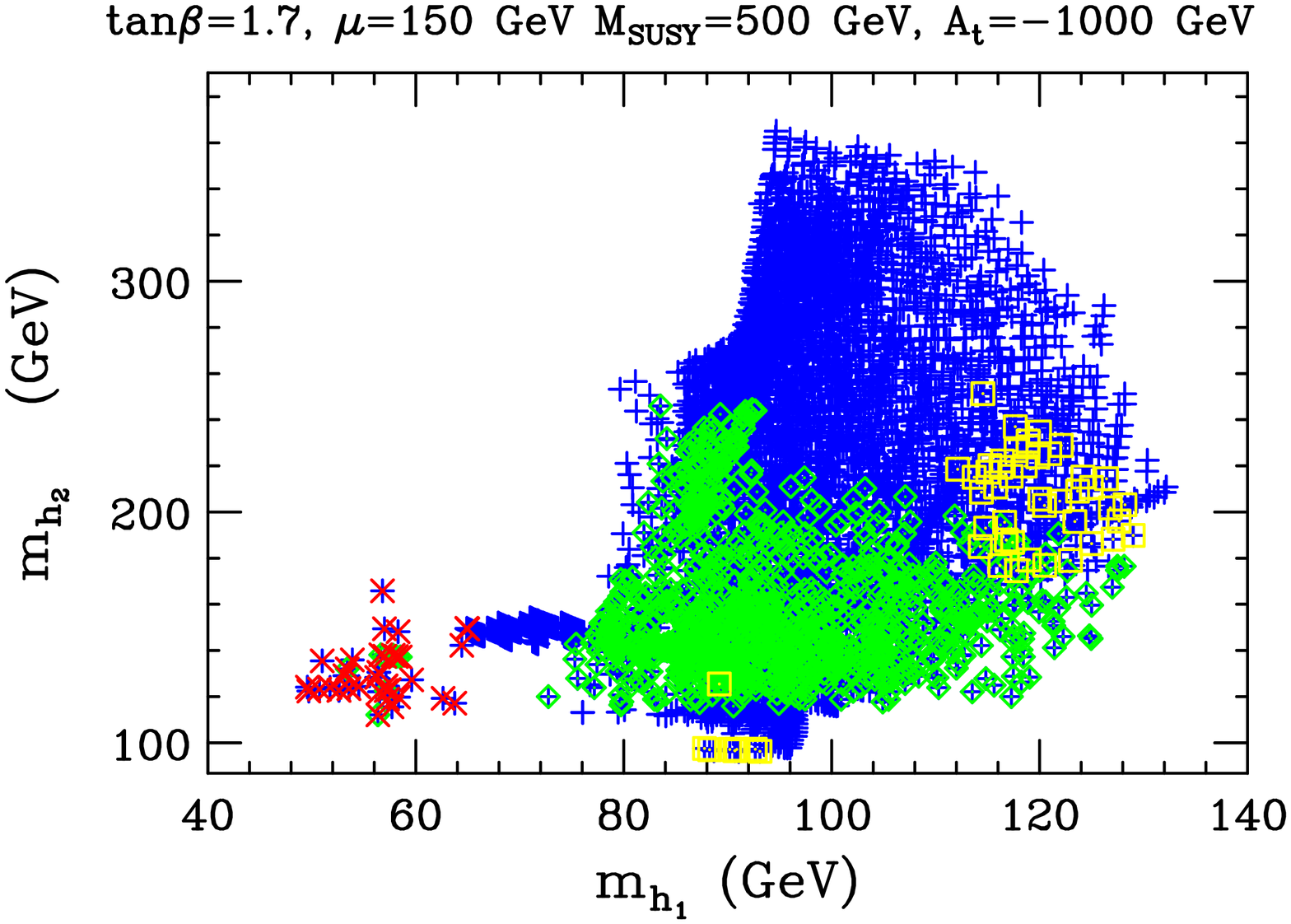}
\caption{$\mhii$ is plotted vs. $\mhi$ for the $\tanb=1.7$,
  $\msusy=500\gev$, $A=-1000\gev$ scenario.}
\label{mhiivsmhi_tb1pt7}
\end{figure}

\begin{figure}
\includegraphics[height=0.4\textwidth,angle=90]{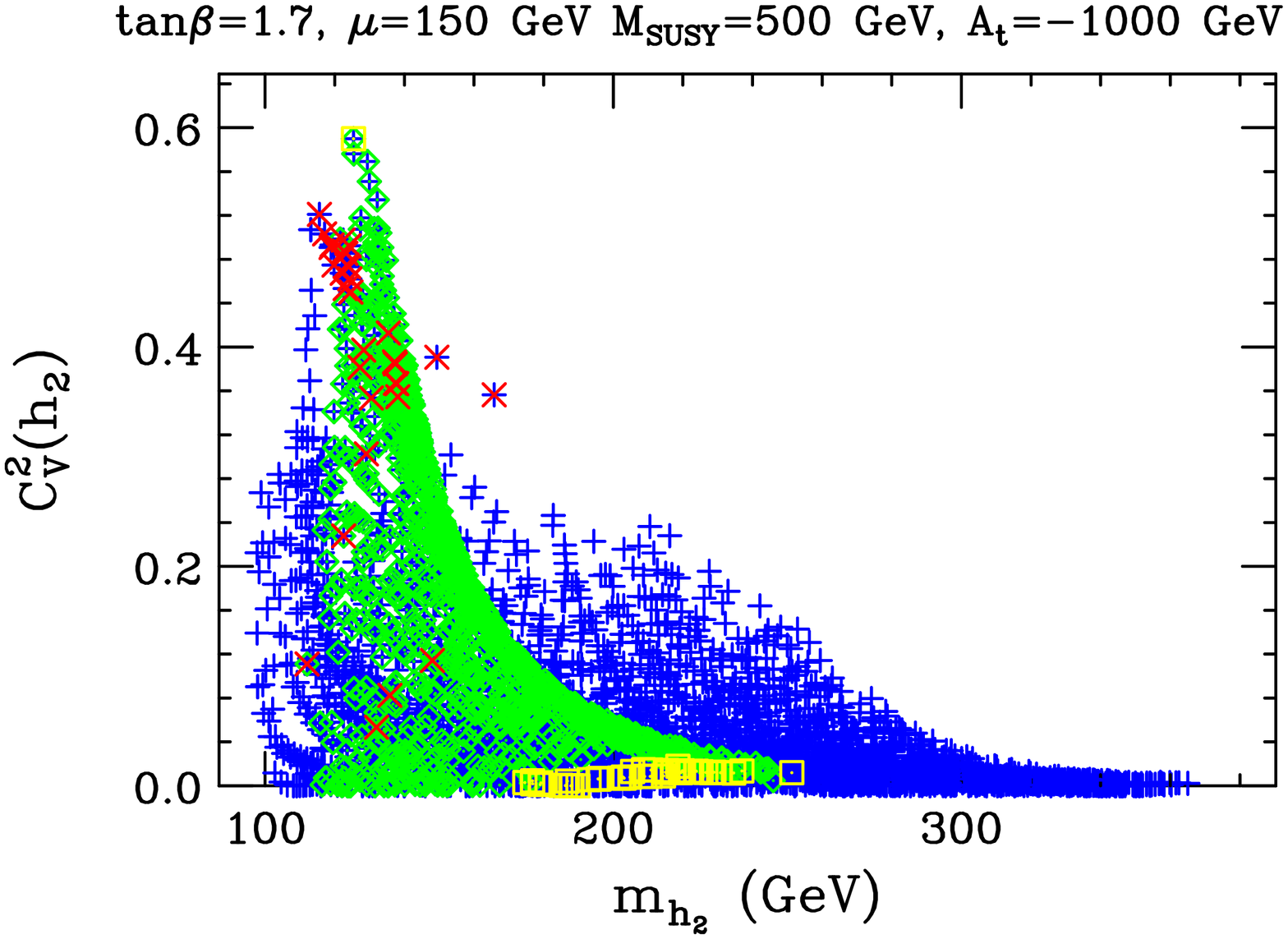}
\caption{$\cviisq$ is plotted vs. $\mhii$ for the $\tanb=1.7$,
  $\msusy=500\gev$, $A=-1000\gev$ scenario.}
\label{cviisqvsmhii_tb1pt7}
\end{figure}

\begin{figure}
\includegraphics[height=0.4\textwidth,angle=90]{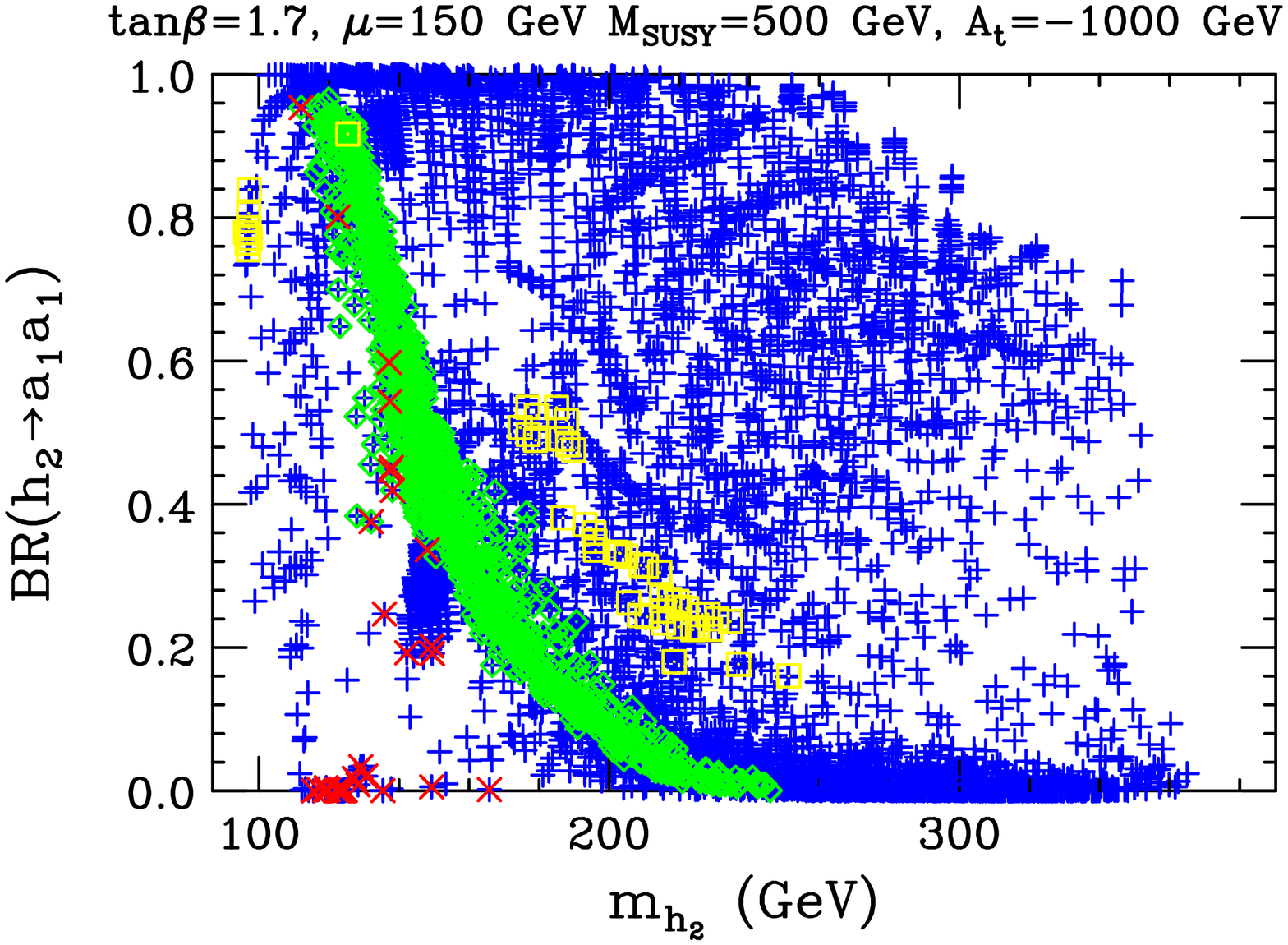}
\caption{$\br(\hii\to\ai\ai)$ is plotted vs. $\mhii$ for the $\tanb=1.7$,
  $\msusy=500\gev$, $A=-1000\gev$ scenario.}
\label{brhiiaiaivsmhii_tb1pt7}
\end{figure}

\begin{figure}
\includegraphics[height=0.4\textwidth,angle=90]{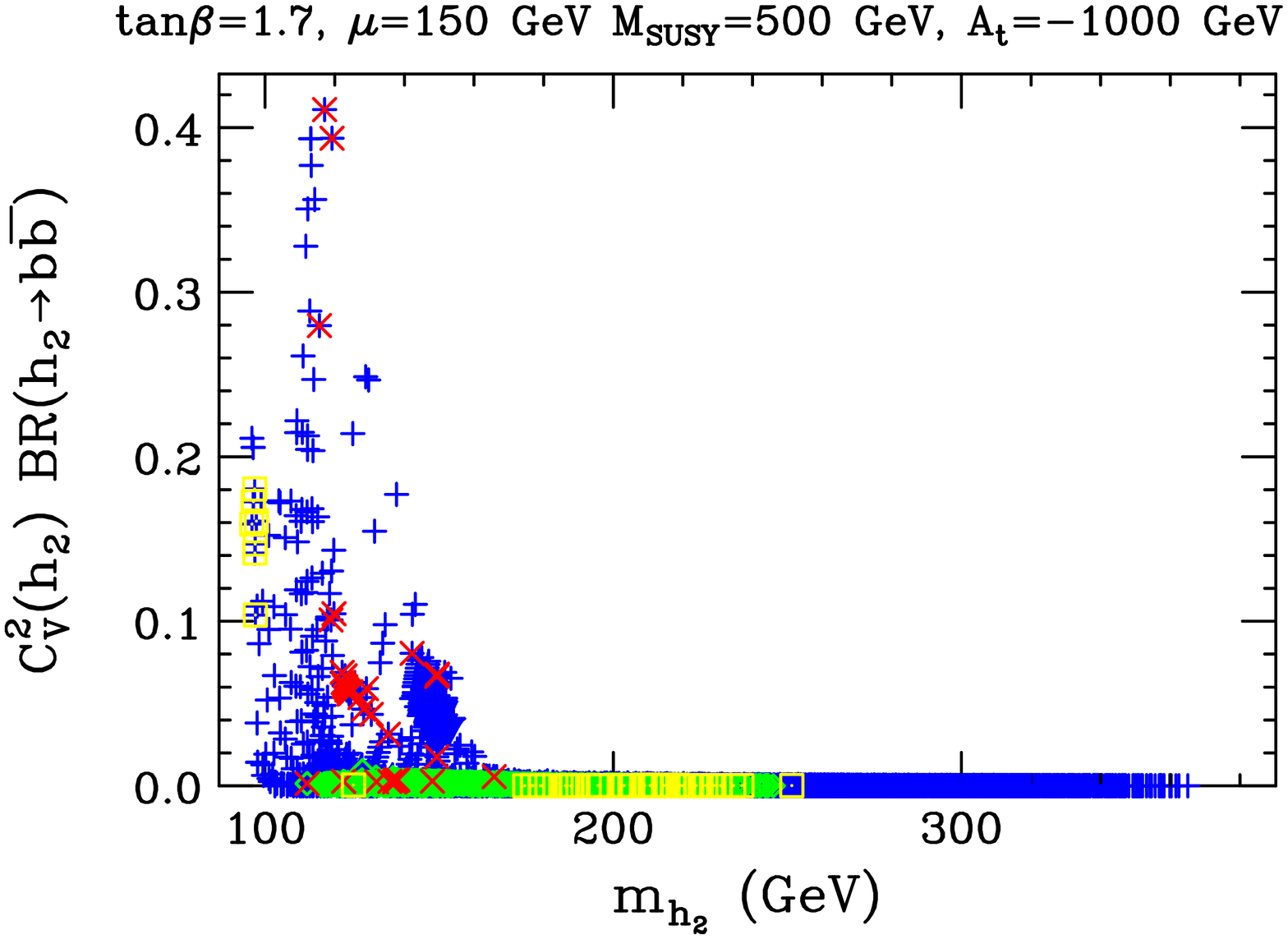}
\caption{$\cviisq\br(\hii\to b\anti b)$ is plotted vs. $\mhii$ for the $\tanb=1.7$,
  $\msusy=500\gev$, $A=-1000\gev$ scenario.}
\label{gzzhiisqbrhiibbvsmhii_tb1pt7}
\end{figure}

\begin{figure}
\includegraphics[height=0.4\textwidth,angle=90]{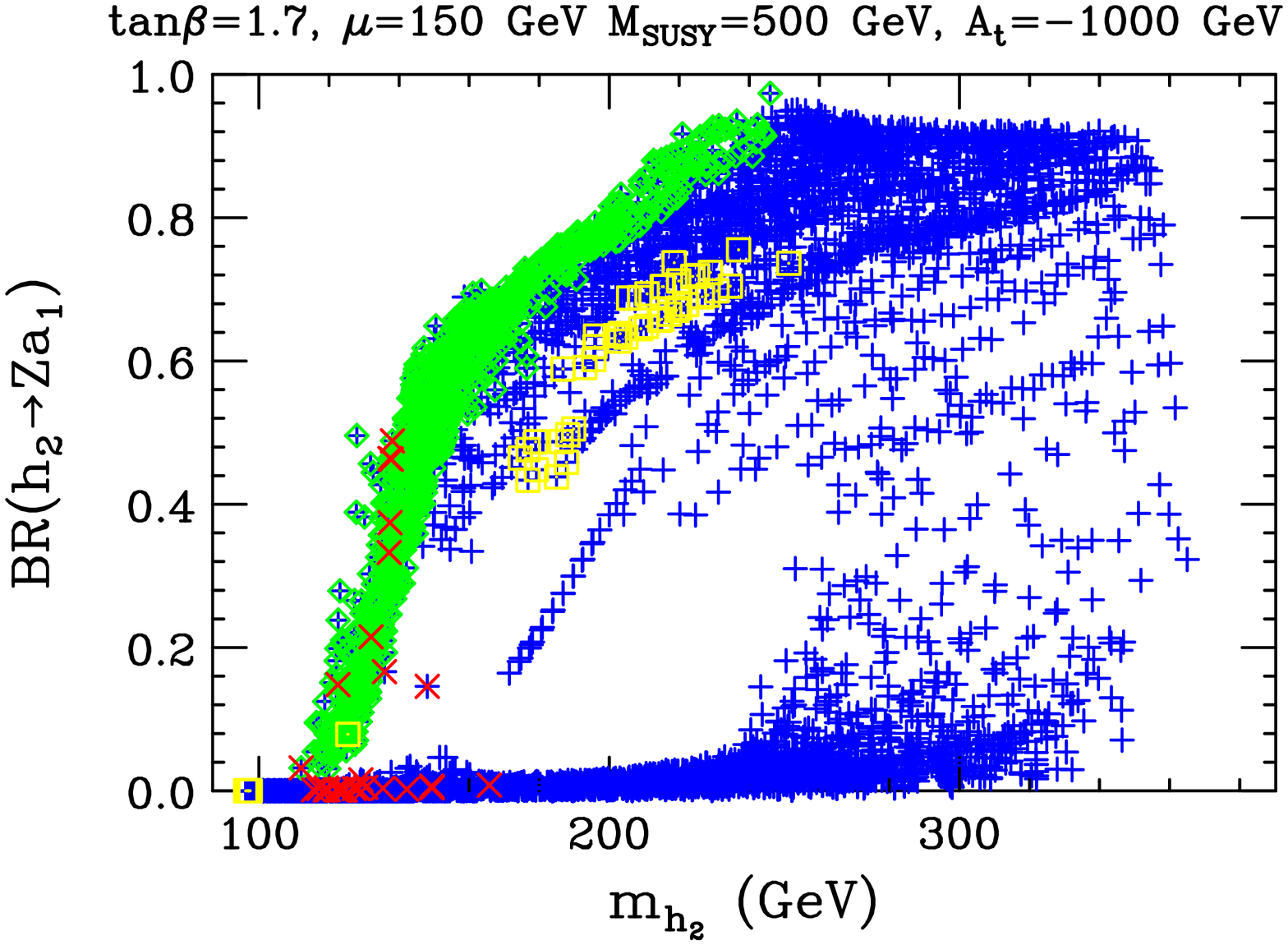}
\caption{$\br(\hii\to Z\ai)$ is plotted vs. $\mhii$ for the $\tanb=1.7$,
  $\msusy=500\gev$, $A=-1000\gev$ scenario.}
\label{brhiizaivsmhii_tb1pt7}
\end{figure}

\begin{figure}
\includegraphics[height=0.4\textwidth,angle=90]{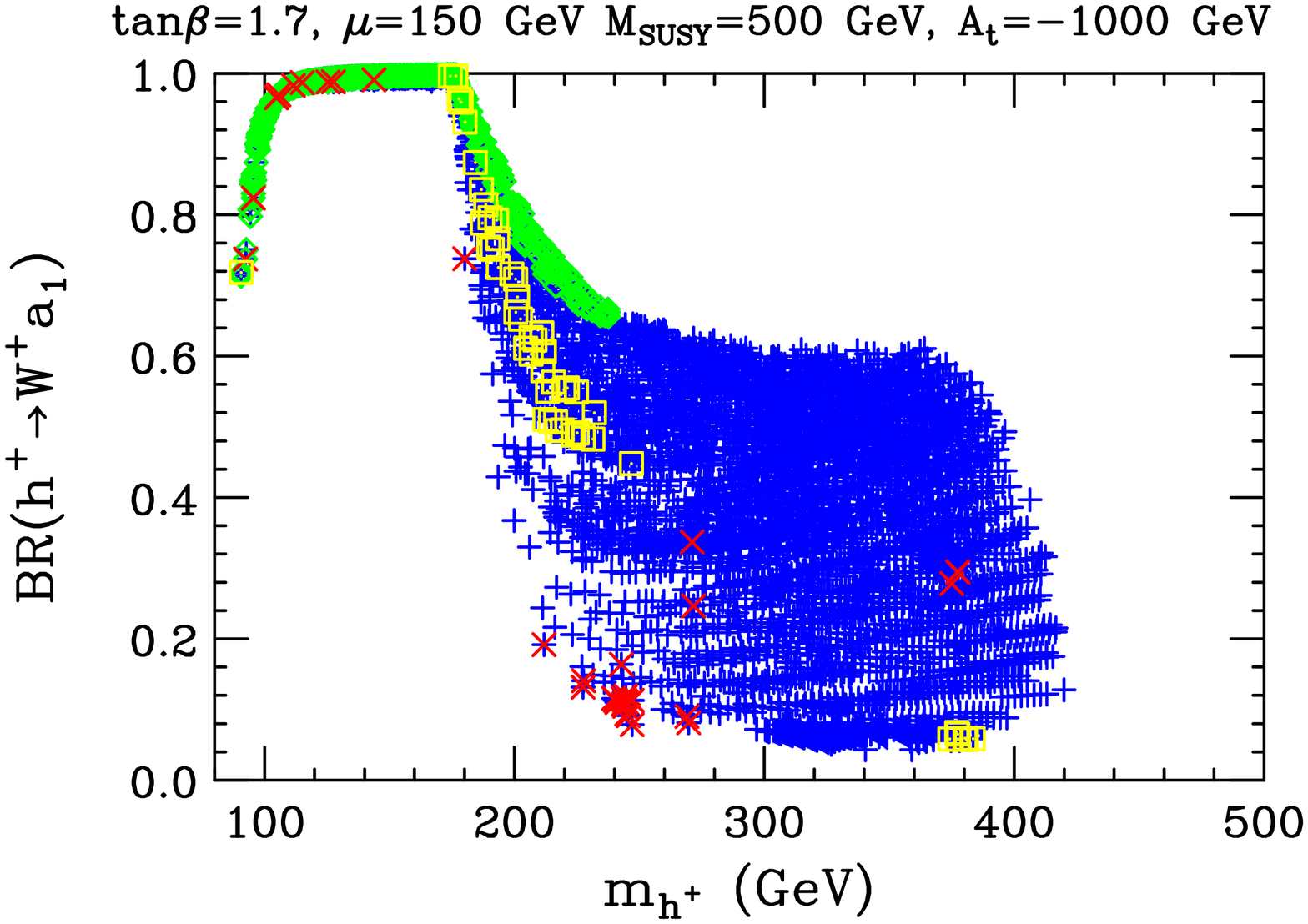}
\caption{$\br(\hp\to \wstarp \ai)$ is plotted vs. $\mhp$ for the $\tanb=1.7$,
  $\msusy=500\gev$, $A=-1000\gev$ scenario.}
\label{brhpwai_tb1pt7}
\end{figure}

\begin{figure}
\includegraphics[height=0.4\textwidth,angle=90]{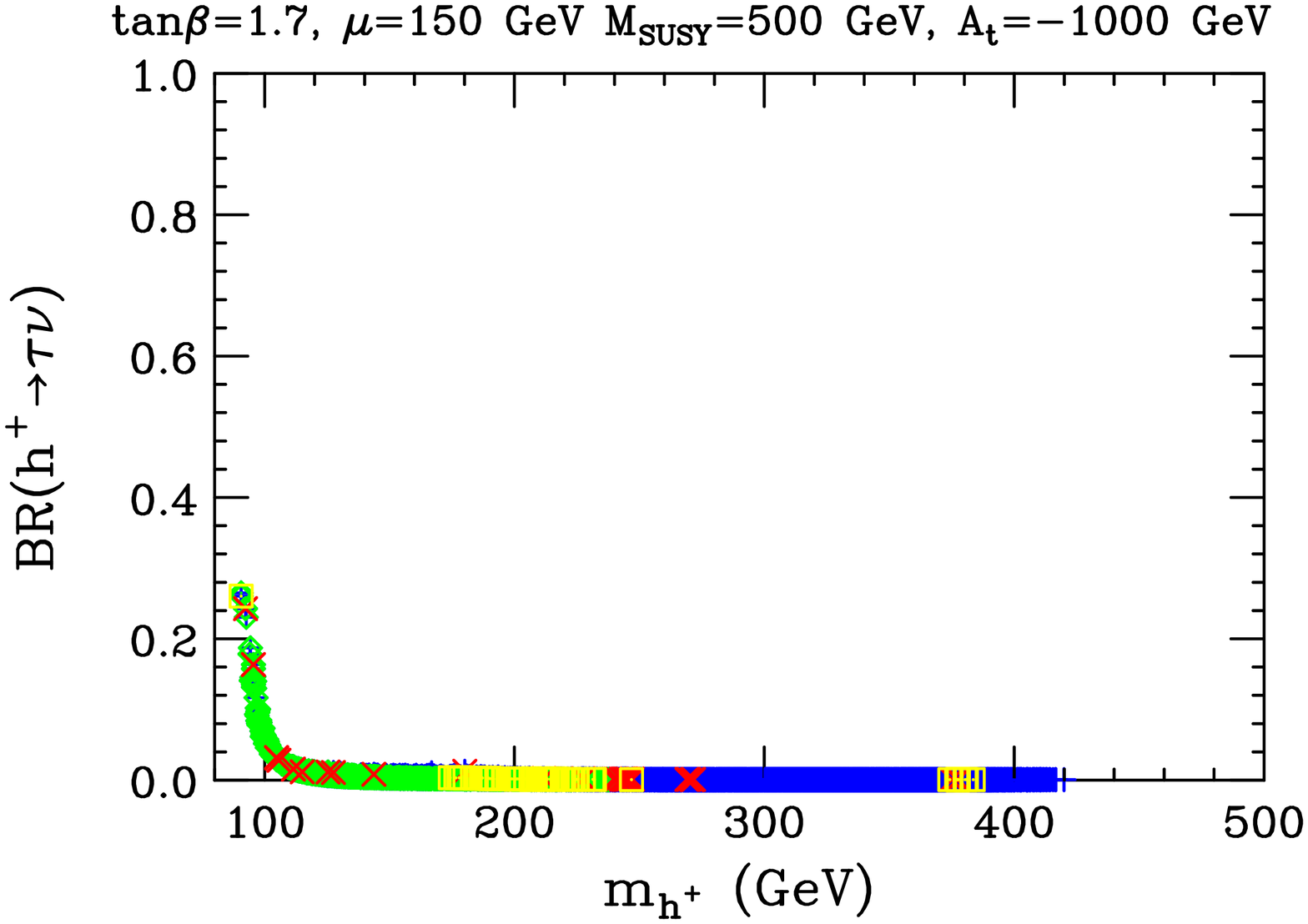}
\caption{$\br(\hp\to \tau^+\nu_\tau)$ is plotted vs. $\mhp$ for the $\tanb=1.7$,
  $\msusy=500\gev$, $A=-1000\gev$ scenario.}
\label{brhptaunu_tb1pt7}
\end{figure}

\begin{figure}
\includegraphics[height=0.4\textwidth,angle=90]{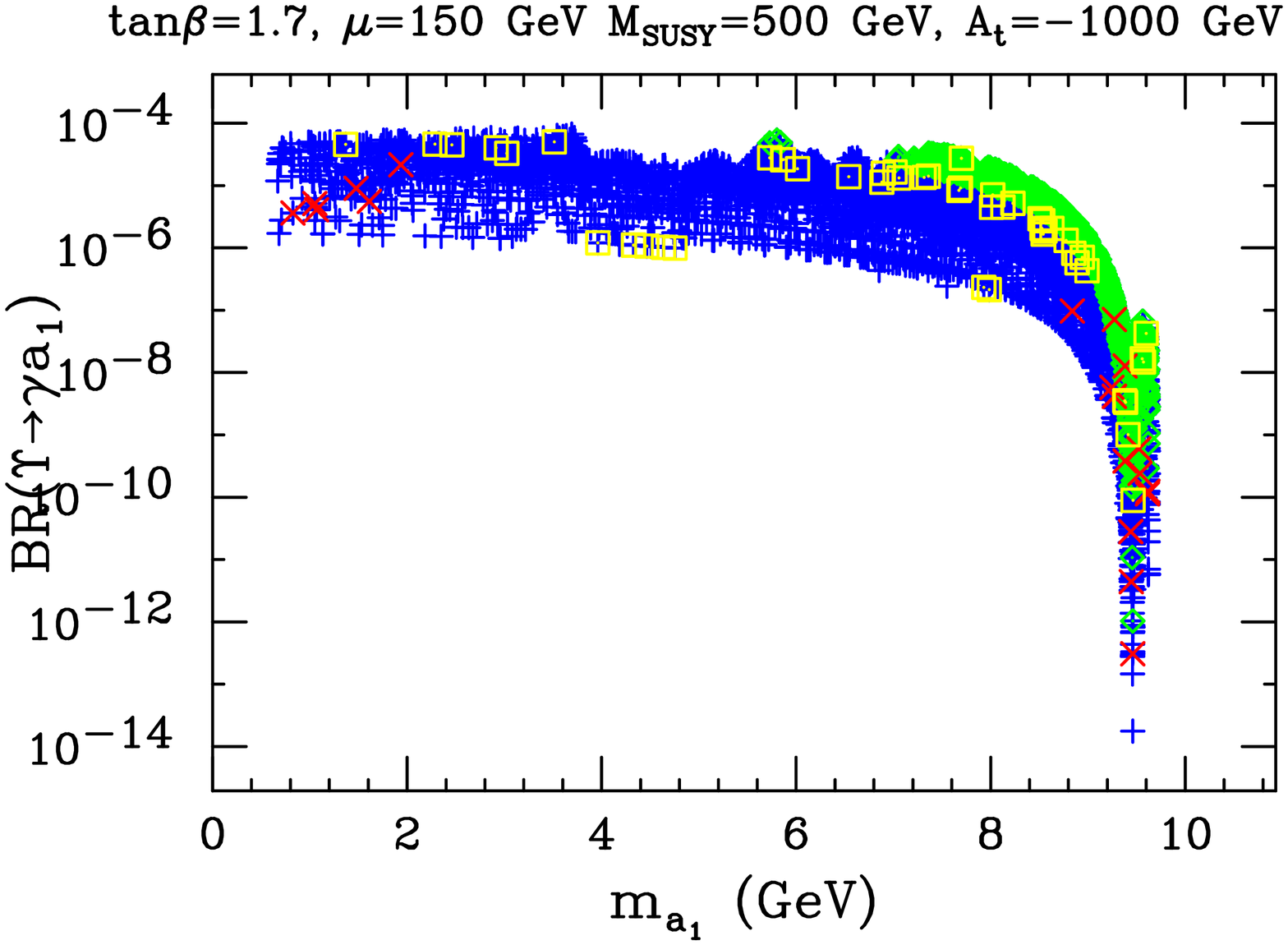}
\caption{$\brups$ is plotted vs. $\mai$ for the $\tanb=1.7$,
  $\msusy=500\gev$, $A=-1000\gev$ scenario.}
\label{brupsvsmai_tb1pt7}
\end{figure}

\begin{figure}
\includegraphics[height=0.4\textwidth,angle=90]{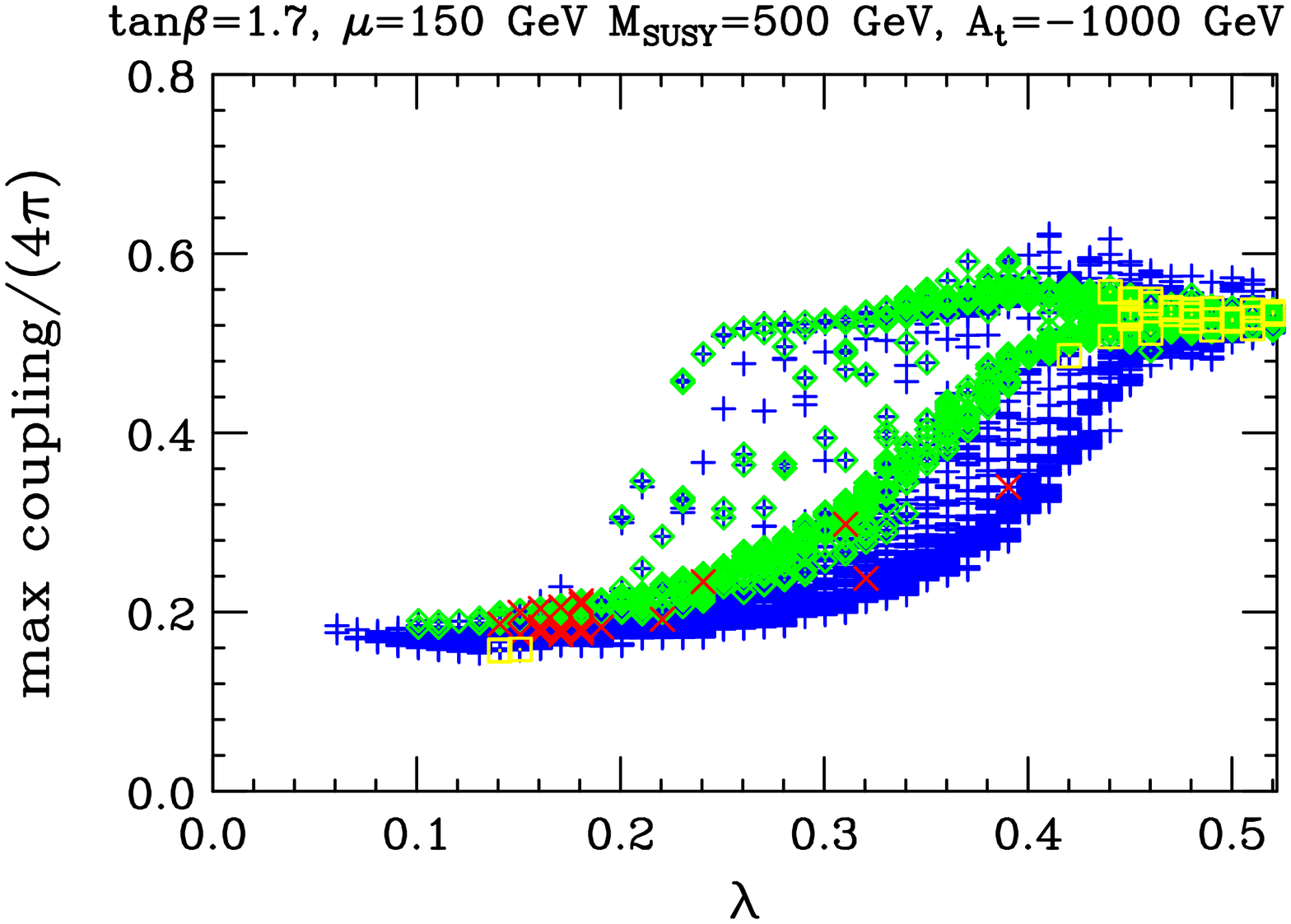}
\caption{$\ymax$ is plotted vs. $\lam$ for the $\tanb=1.7$,
  $\msusy=500\gev$, $A=-1000\gev$ scenario.}
\label{ymaxvslam_tb1pt7}
\end{figure}

\begin{figure}
\includegraphics[height=0.4\textwidth,angle=90]{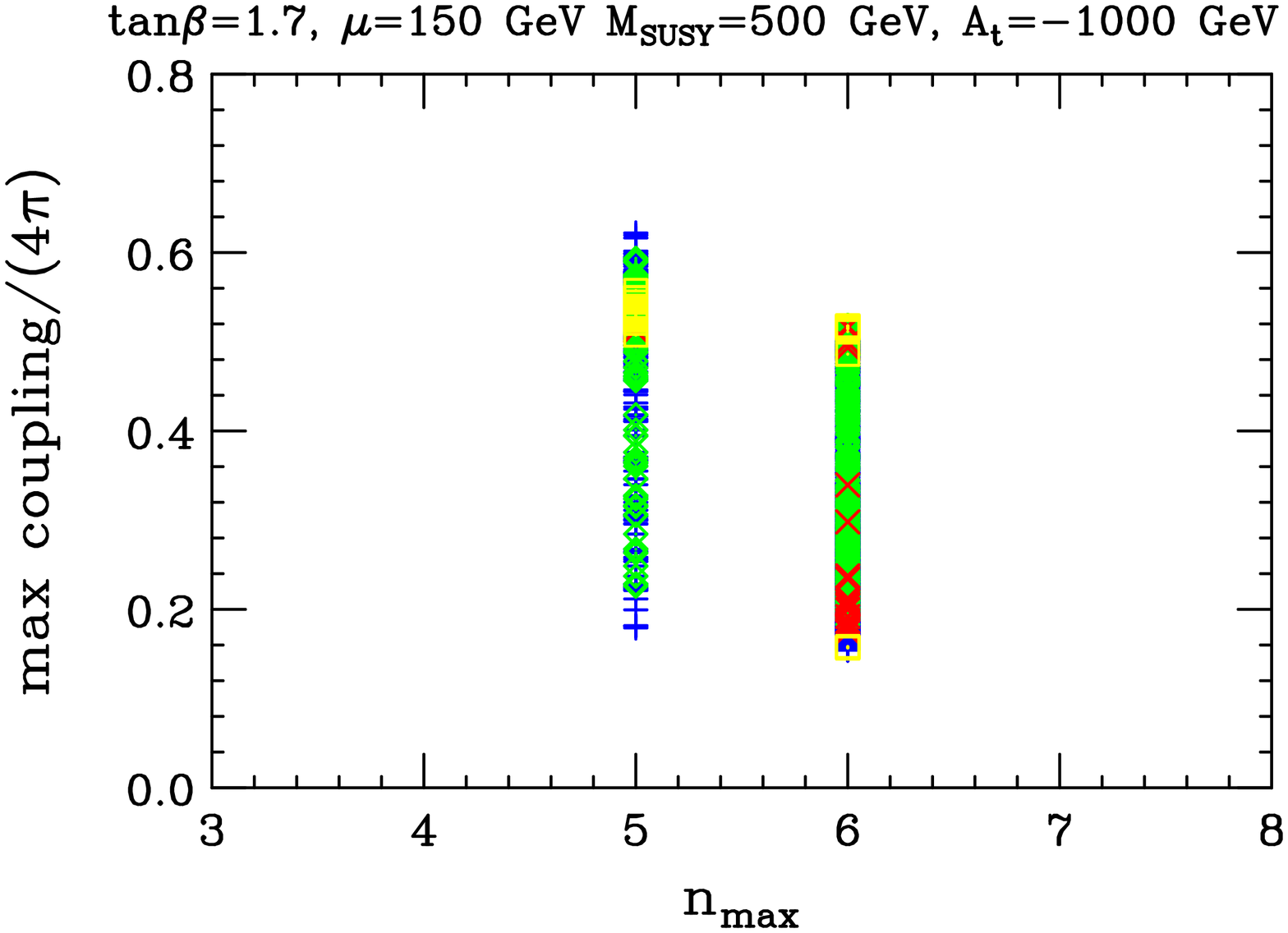}
\caption{$\ymax$ is plotted vs. $\nmax$ for the $\tanb=1.7$,
  $\msusy=500\gev$, $A=-1000\gev$ scenario.}
\label{ymaxvsnmax_tb1pt7}
\end{figure}

\begin{figure}
\includegraphics[height=0.4\textwidth,angle=90]{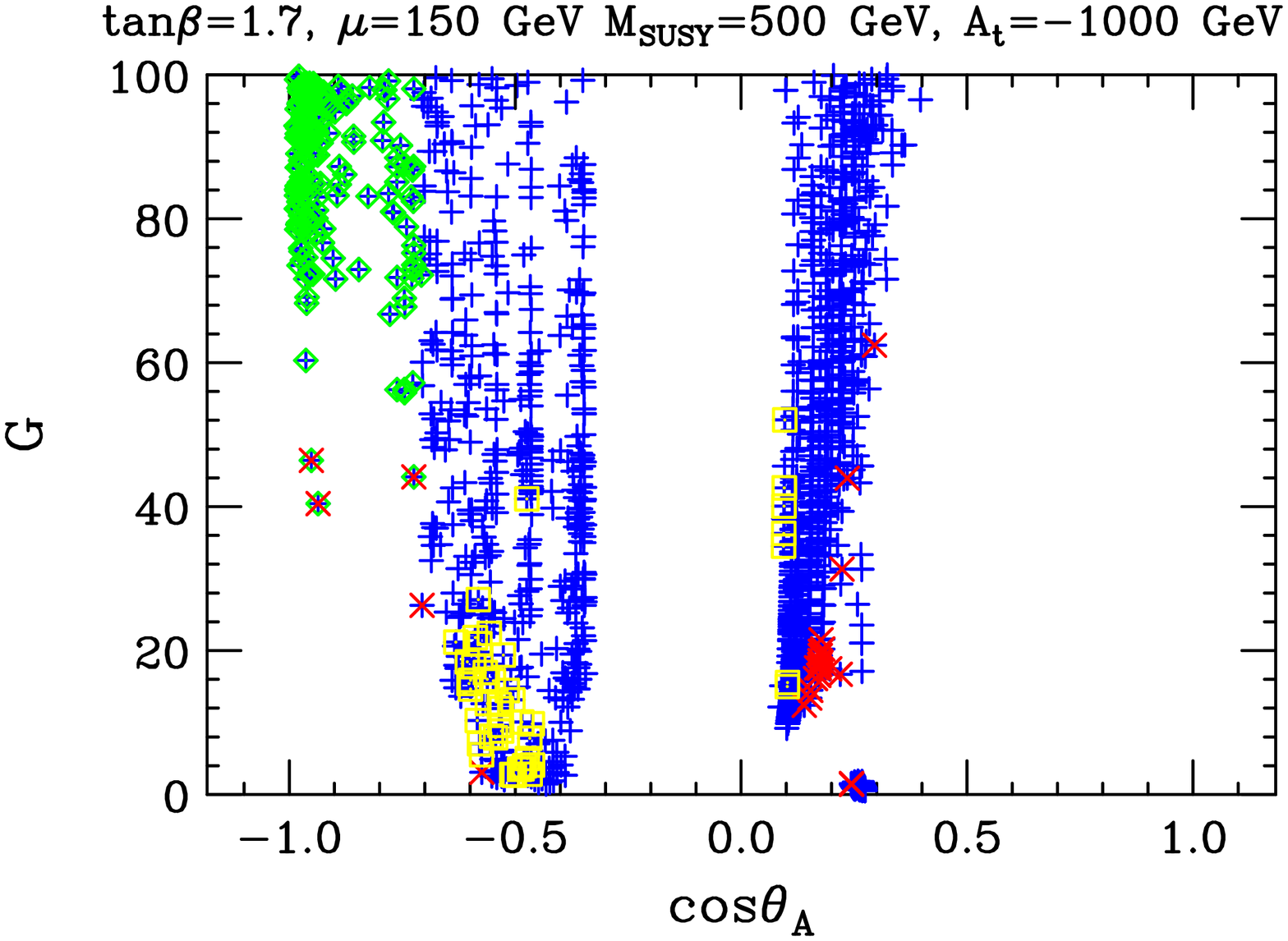}
\caption{$G$ is plotted vs. $\cta$ for the $\tanb=1.7$,
  $\msusy=500\gev$, $A=-1000\gev$ scenario. The displayed points
  comprise only a small fraction of the total number of points
  appearing in previous figures.}
\label{gvscta_tb1pt7}
\end{figure}

\begin{figure}
\includegraphics[height=0.4\textwidth,angle=90]{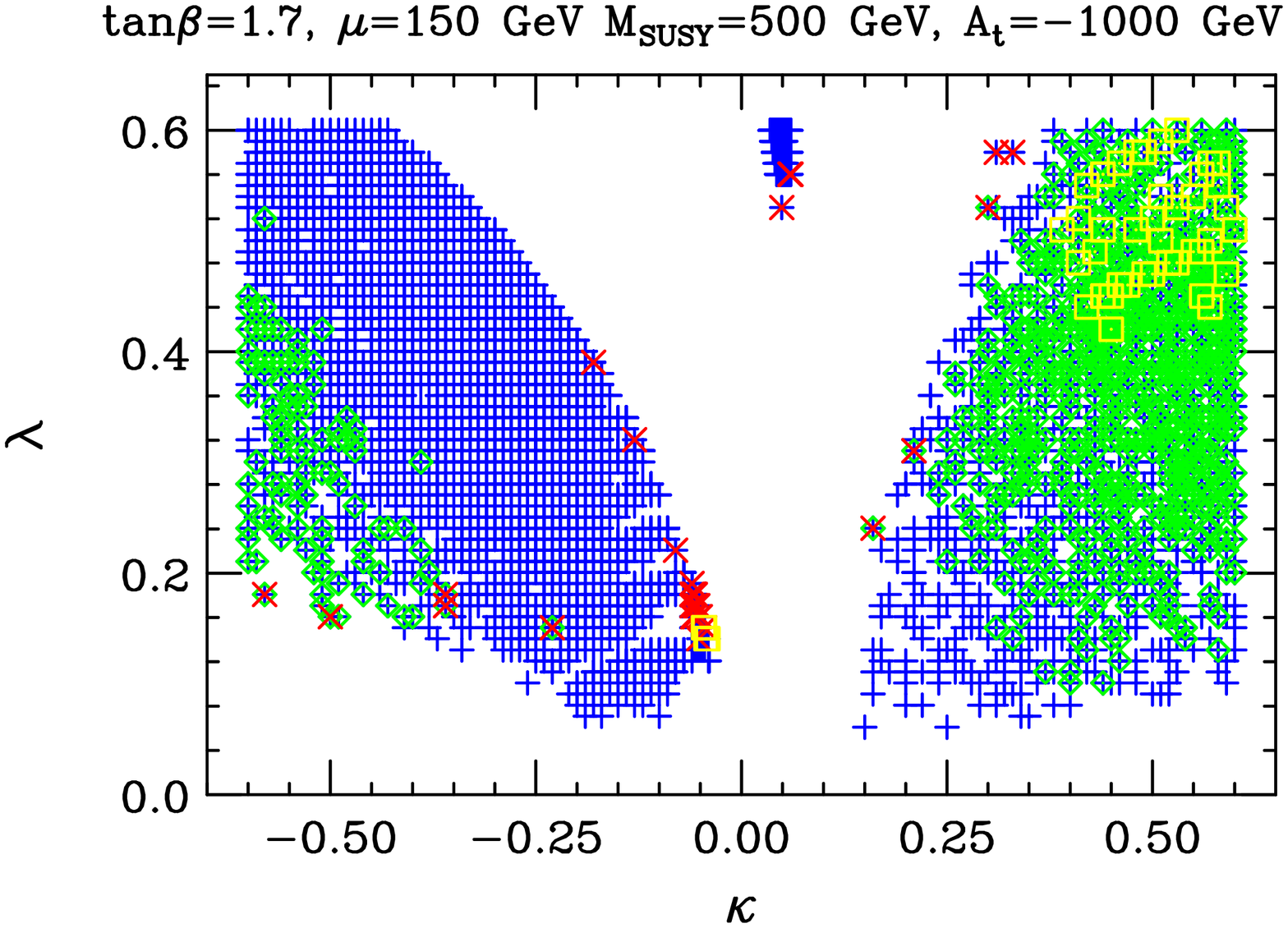}
\caption{$\lam$ is plotted vs. $\kap$ for the $\tanb=1.7$,
  $\msusy=500\gev$, $A=-1000\gev$ scenario. }
\label{lamvskap_tb1pt7}
\end{figure}

\begin{figure}
\vspace*{.2in}
\includegraphics[height=0.4\textwidth,angle=90]{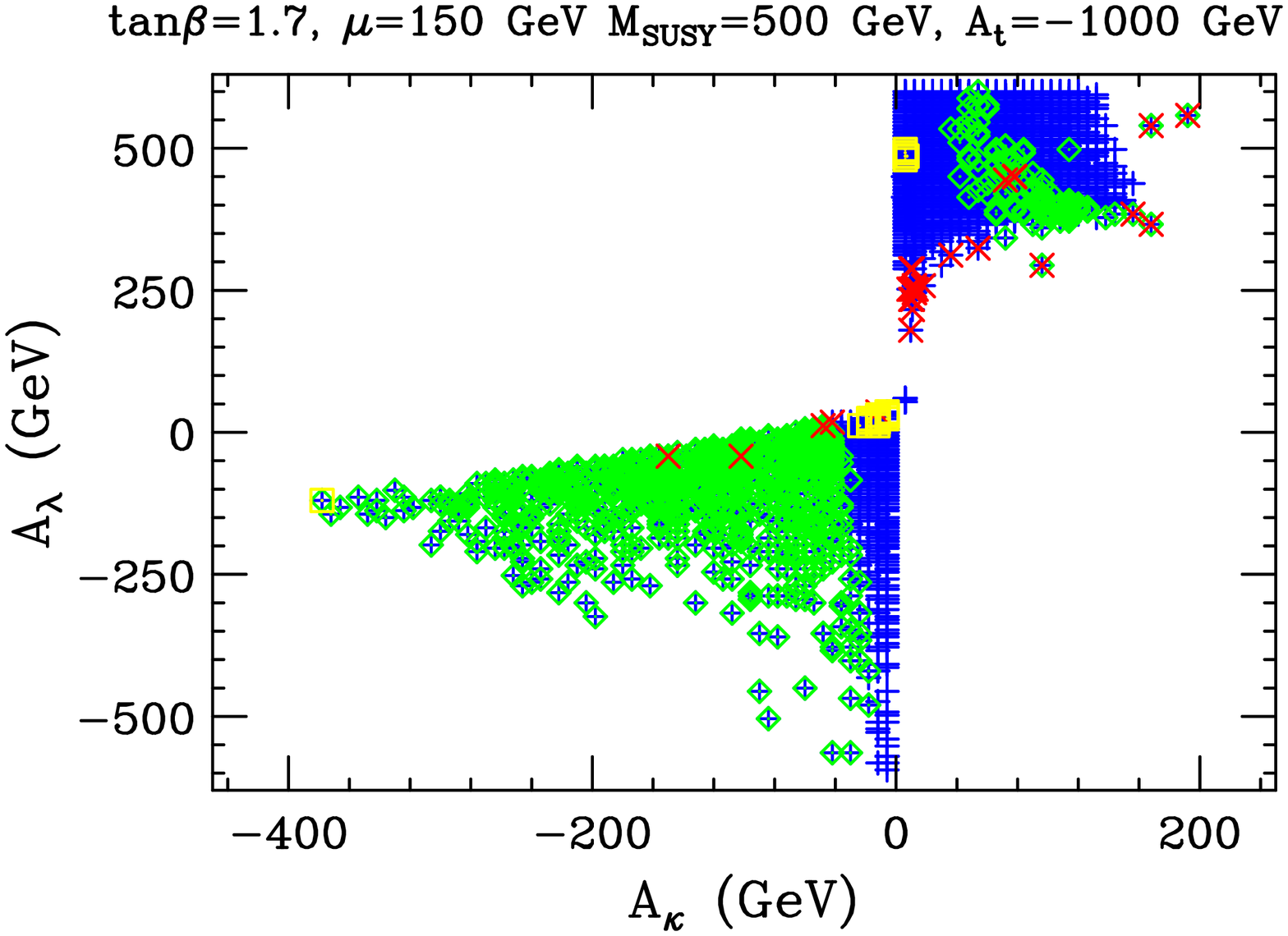}
\caption{$\alam$ is plotted vs. $\akap$ for the $\tanb=1.7$,
  $\msusy=500\gev$, $A=-1000\gev$ scenario. }
\label{alamvsakap_tb1pt7}
\end{figure}

One can avoid non-perturbative couplings for a large number of allowed
points for lower $\tanb$ if $\msusy$ and $A_t$ are somewhat larger. As
an example, we present results for the case of $\tanb=1.7$,
$\msusy=500\gev$ and $A=-1000\gev$ in
Figs.~\ref{ctasqvsmai_tb1pt7}-\ref{alamvsakap_tb1pt7}. The point
notation is as for $\tanb=2$, except that in this case there are
points for which $\br(\hi\to \ai\ai)<0.7$.  These points are indicated
by the yellow squares.

As in the previous case, significant restrictions are placed on
$|\cta|$ due to limits on the $\caibb$ coupling.
Fig.~\ref{ctasqvsmai_tb1pt7} shows that once again these restrictions
basically imply a limit on $\cos^2\theta_A$ that is significantly
below $0.5$ if $\mai\lsim 7.5\gev$.

We now repeat the same set of figures as in the $\tanb=2$ case.  Many
of the same comments apply.  Where appropriate we shall comment on
differences.

We note that we have found many more points with quite low $\mhi$ for
this case as compared to the $\tanb=2$ scenario previously considered.
Of particular interest is the fact that there are a significant number
of model points for which $\mhi$ is near $90\gev$ and $\mhii$ is just
below $100\gev$ with $\cvisq\br(\hi\to b\anti b)\sim 0.1-0.2$ and
$\cviisq\br(\hii\to b\anti b)\sim 0.1-0.2$. A particular subset of these can be identified
in Figs.~\ref{gzzhisqbrhibbvsmhi_tb1pt7} and
\ref{gzzhiisqbrhiibbvsmhii_tb1pt7} as the yellow squares with the above
attributes. (However, there are quite a few blue points that also
satisfy these criteria.) The precise masses and $C_V^2\br(h\to b\anti
b)$ values of the yellow-square points
are tabulated in Table~\ref{cvsqbrhbb_tb1pt7}.
These points appear in Fig.~\ref{ctasqvsmai_tb1pt7} as the yellow
squares with $\ctasq\sim 0$ (more precisely $\cta\sim 0.1$) and
$\mai\sim 4\gev$ or $8\gev$. They appear in Figs.~\ref{lamvskap_tb1pt7} and
\ref{alamvsakap_tb1pt7} as the yellow square points with $\kap\in [-0.046,-0.041]$, $\lam\sim 0.14-0.15$, $\akap\sim 6-7\gev$ and $\alam\sim
486-492\gev$. The $\cta$, $\kap$ and $\akap$ values indicate that
these points are ones that are close to the Peccei-Quinn symmetry
limit of the NMSSM. 

\begin{table}
\label{cvsqbrhbb_tb1pt7}
\caption{Selected points for which $\mhi$ and corresponding $\mhii$
lie within the LEP excess region and the corresponding $\cvisq\br(\hi
\to b\anti b)$ and $\cviisq\br(\hii\to b\anti b)$ values.}
\begin{tabular} {|c|c|c|c|}
\hline
$\mhi$ & $\cvisq\br(\hi\to b\anti b)$ & $\mhii$ & $\cviisq\br(\hii\to
b\anti b)$ \cr
\hline
  93.1 & 0.0684  & 96.2 & 0.1590\cr
  90.7 & 0.0560  & 96.6 & 0.1726 \cr
  90.2 & 0.1171  & 97.2 & 0.1468\cr
  88.3 & 0.0557  & 97.0 & 0.1803\cr
  87.8 & 0.0974  & 97.5 & 0.1609\cr
  90.7 & 0.0560  & 96.6 & 0.1727\cr
  92.7 & 0.1748  & 97.2 & 0.1037\cr
  90.9 & 0.0599  & 97.1 & 0.1416\cr
\hline
\end{tabular}
\end{table}

The reason that these points are of particular interest is that the
two Higgs bosons combine to nicely explain the LEP excess seen
throughout the entire $m_{b\anti b}\in[88\gev,100\gev]$ mass region in
the $\epem\to Zb\anti b$ channel. The level of this excess corresponds
to $C_V^2\br(h\to b\anti b)\sim 0.1-0.2$ for any single $h$ mass in
this region.  The masses of the two Higgs bosons are typically
separated by about $2\sigma_{res}$ where $\sigma_{res}\sim 3\gev$ was the LEP mass
resolution. As a result, the combination of the two Higgses would
give the broad excess observed.  The manner in which $C_V^2\br(h\to
b\anti b)\sim 0.1-0.2$ is achieved is quite different for $\hi$ vs.
$\hii$. In the case of the $\hi$, $\cvisq$ (see
Fig.~\ref{cvisqvsmhi_tb1pt7}) is small and $\br(\hi\to b\anti b)$ is
fairly large (because $\br(\hi\to \ai\ai)$ is relatively small (see
Fig.~\ref{brhiaiaivsmhi_tb1pt7}).  In the case of the $\hii$, $\cviisq$ (see
Fig.~\ref{cviisqvsmhii_tb1pt7}) is large and $\br(\hii\to b\anti b)$ is
fairly small (because $\br(\hii\to \ai\ai)$ is relatively large (see
Fig.~\ref{brhiiaiaivsmhii_tb1pt7}).

These special points are also rather attractive in that they are ones
for which the couplings remain perturbative after evolution evolution
to the GUT scale (see the yellow-square points with small $\lam$ in
Fig.~\ref{ymaxvslam_tb1pt7}).  However, for these points the 
$\akap$-$\alam$ fine-tuning measure $G$ (see the yellow-square points
in Fig.~\ref{gvscta_tb1pt7} with $\cta\sim 0.1$) is somewhat large.



As regards the $\ai$-doublet-like (green diamond) points, we observe
from Fig.~\ref{mhiivsmhi_tb1pt7} that the lower bound on $\mhii$ has
been pushed to about $110\gev$ vs. the $\sim 100\gev$ value obtained
for $\tanb=2$.  This means that $Z\hii$ production at LEP would have been
minimal or absent for such cases, but there would still have been a
significant rate for $\hii\ai$ production.  However, to repeat, LEP
did not look for the relevant $\hii\ai\to \ai\ai\ai$ or $\hii\ai\to
Z\ai\ai$ final states that would have been dominant
(Figs.~\ref{brhiiaiaivsmhii_tb1pt7} and \ref{brhiizaivsmhii_tb1pt7},
respectively).

\subsection{Results for $\tan \beta =1.2$}

\begin{figure}
\vspace*{.2in}
\includegraphics[height=0.4\textwidth,angle=90]{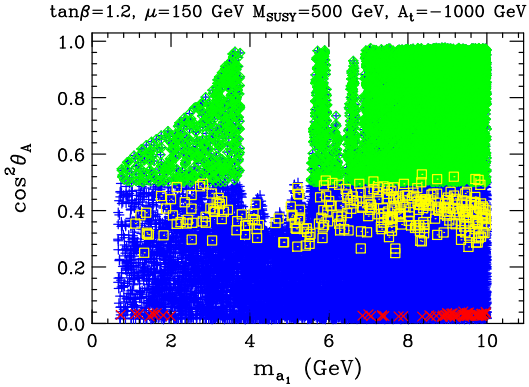}
\caption{$\cos^2\theta_A$  is plotted vs. $\mai$ for the $\tanb=1.2$,
  $\msusy=500\gev$, $A=-1000\gev$ scenario. }
\label{ctasqvsmai_tb1pt2}
\end{figure}

We have also performed a scan for the case of $\tanb=1.2$,
$\msusy=500\gev$ and $A=-1000\gev$. For the most part, results are
very similar to those for $\tanb=1.7$, $\msusy=500\gev$ and
$A=-1000\gev$.  One difference arises because the coupling of the
$\ai$ to $b\anti b$, proportional to $\cta\tanb$ is weaker for the
lower $\tanb$ value. This implies that the experimental upper limits
on this coupling are less restrictive at a given value of $\cta$. The
result is that $|\cta|^2>0.5$ is possible for $\mai<2\mtau$ (\ie\ 
below the $\mai$ values for which the $\ai b\anti b$ coupling is so
strongly limited by CLEO-III results). This is made apparent by
comparing Fig.~\ref{ctasqvsmai_tb1pt2} to
Fig.~\ref{ctasqvsmai_tb1pt7}.  Another difference is that for all but
a special class (to be described later) of the $\tanb=1.2$ scenarios, one or more of the
couplings, $\lam$, $\kap$, $\alam$ or $\akap$ becomes non-perturbative
in evolving to the GUT scale.

\begin{figure}
\vspace*{.2in}
\includegraphics[height=0.4\textwidth,angle=90]{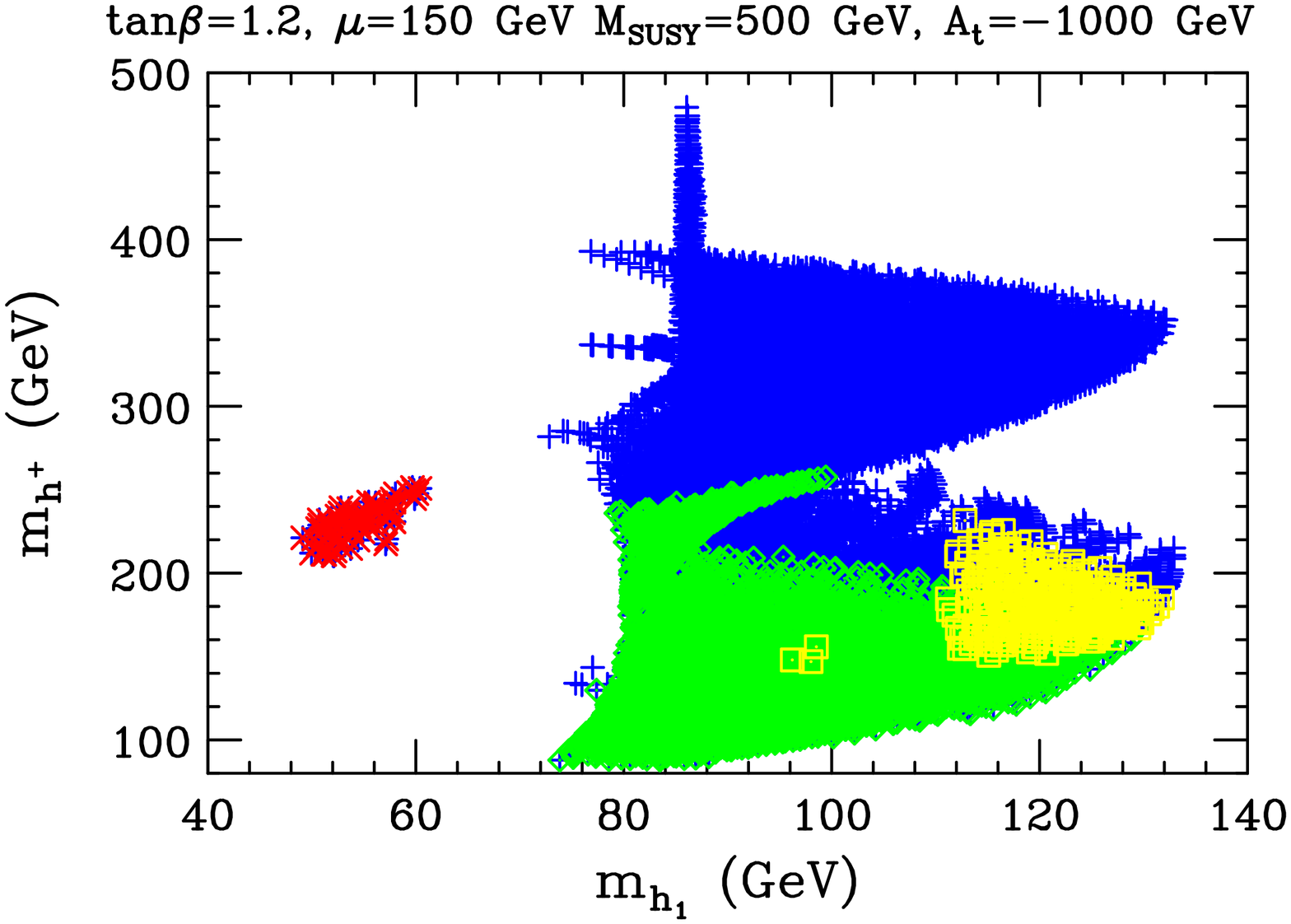}
\caption{$\mhp$ is plotted vs. $\mhi$ for the $\tanb=1.2$,
  $\msusy=500\gev$, $A=-1000\gev$ scenario. }
\label{mhpvsmhi_tb1pt2}
\end{figure}

\begin{figure}
\vspace*{.2in}
\includegraphics[height=0.4\textwidth,angle=90]{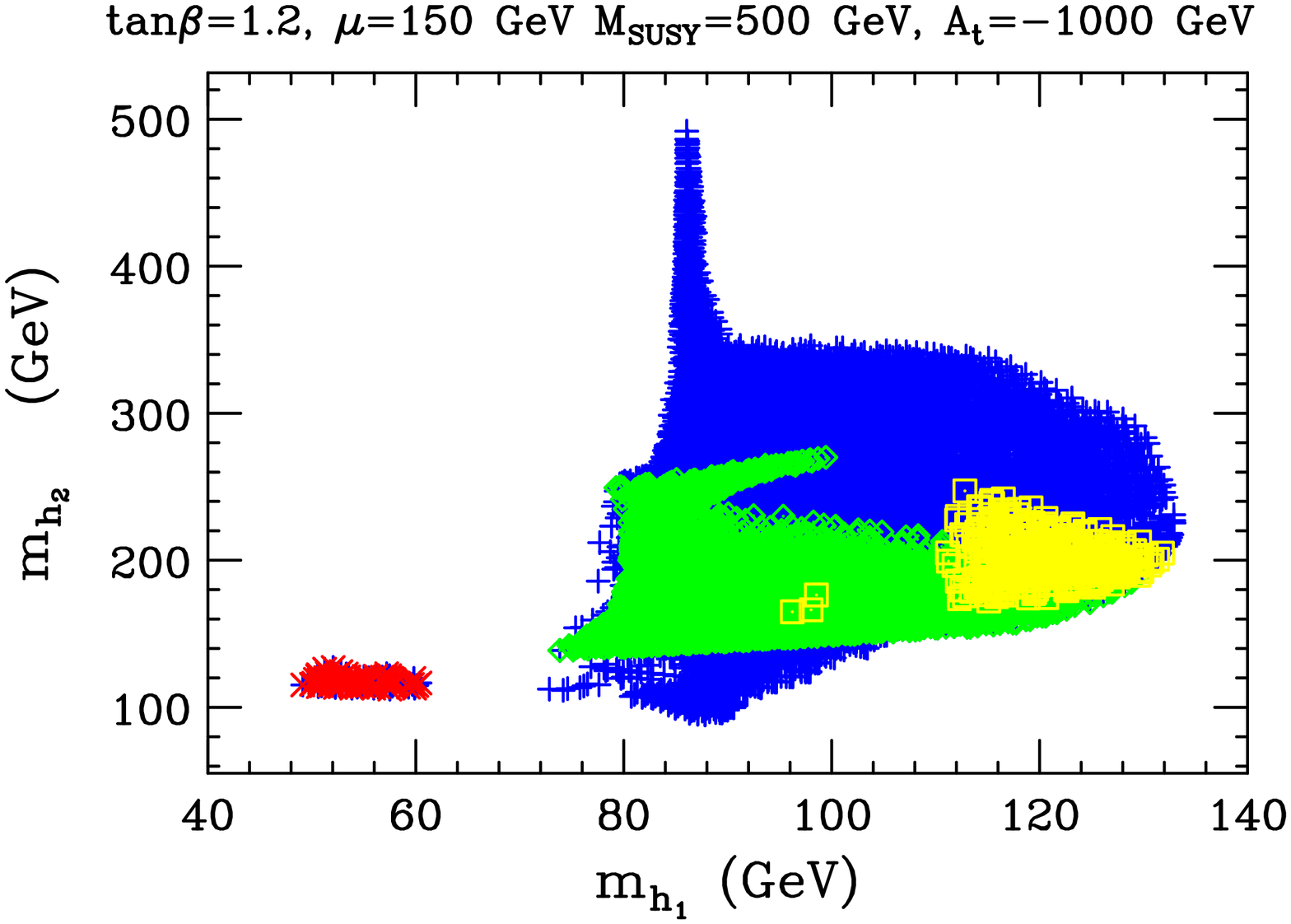}
\caption{$\mhii$ is plotted vs. $\mhi$ for the $\tanb=1.2$,
  $\msusy=500\gev$, $A=-1000\gev$ scenario. }
\label{mhiivsmhi_tb1pt2}
\end{figure}

In Fig.~\ref{mhpvsmhi_tb1pt2} we plot $\mhp$ vs. $\mhi$.  We see that
the lowest value of $\mhp$ is about $90\gev$ and arises for the
$\ai$-doublet-like scenarios.  There are a significant number of
points with $\mhi<65\gev$, all of which have a very singlet-like
$\ai$, as is most apparent from the red crosses in
Fig.~\ref{ctasqvsmai_tb1pt2}.
Fig.~\ref{mhiivsmhi_tb1pt2} shows $\mhii$ vs. $\mhi$. We see that
$\mhii$ values as low as $90\gev$ are possible for singlet-like $\ai$,
whereas the lower limit on $\mhii$ for $\ai$-doublet-like scenarios
has risen to about $140\gev$ as compared to the lower values found for
$\tanb=1.7$ and $2$. 

For this $\tanb=1.2$ case, there are many points with $\mhii$ in the
$95-100\gev$ interval and $\mhi$ in the $90-96\gev$ interval with both
$\cvisq\br(\hi\to b\anti b)$ and $\cviisq\br(\hii\to b\anti b)$
between 0.05 and 0.15 that would explain the broad excess in this
region seen at LEP.  As for $\tanb=1.7$, all these points have very
small $\kap$ and $\akap$ and are therefore close to the PQ symmetry
limit of the model. Most of these points are such that the couplings
do not quite reach the non-perturbative value of coupling/$(4\pi)=0.5$
at the GUT scale. Rather coupling/$(4\pi)\sim 0.4$ is a typical
maximum value. In this sense they are the most attractive of the
$\tanb=1.2$ scenarios.  As for the blue points of this type in the
$\tanb=1.7$ case, the points in this special class at $\tanb=1.2$
typically also have very modest $\alam$, $\akap$ finetuning measure
$G$, with $G$ values between 10 and 30 being typical. One point of
difference with $\tanb=1.7$ is that the $\tanb=1.2$ special points all
have $\br(\hi\to \ai\ai)>0.75$.

\begin{figure}
\vspace*{.2in}
\includegraphics[height=0.4\textwidth,angle=90]{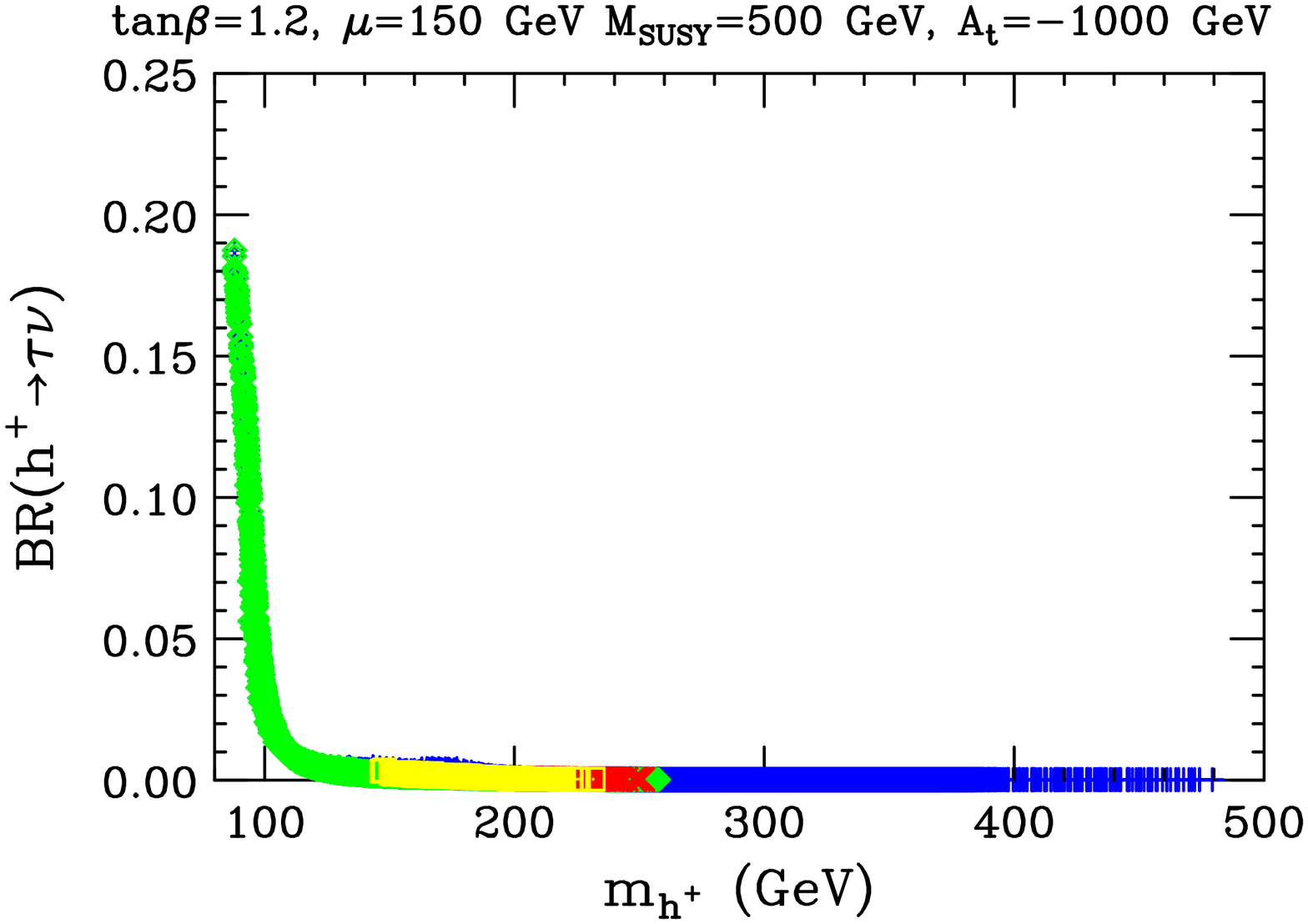}
\caption{$\br(\hp\to \wstarp \ai)$ is plotted vs. $\mhp$ for the $\tanb=1.2$,
  $\msusy=500\gev$, $A=-1000\gev$ scenario. }
\label{brhptaunuvsmhp_tb1pt2}
\end{figure}

\begin{figure}
\vspace*{.2in}
\includegraphics[height=0.4\textwidth,angle=90]{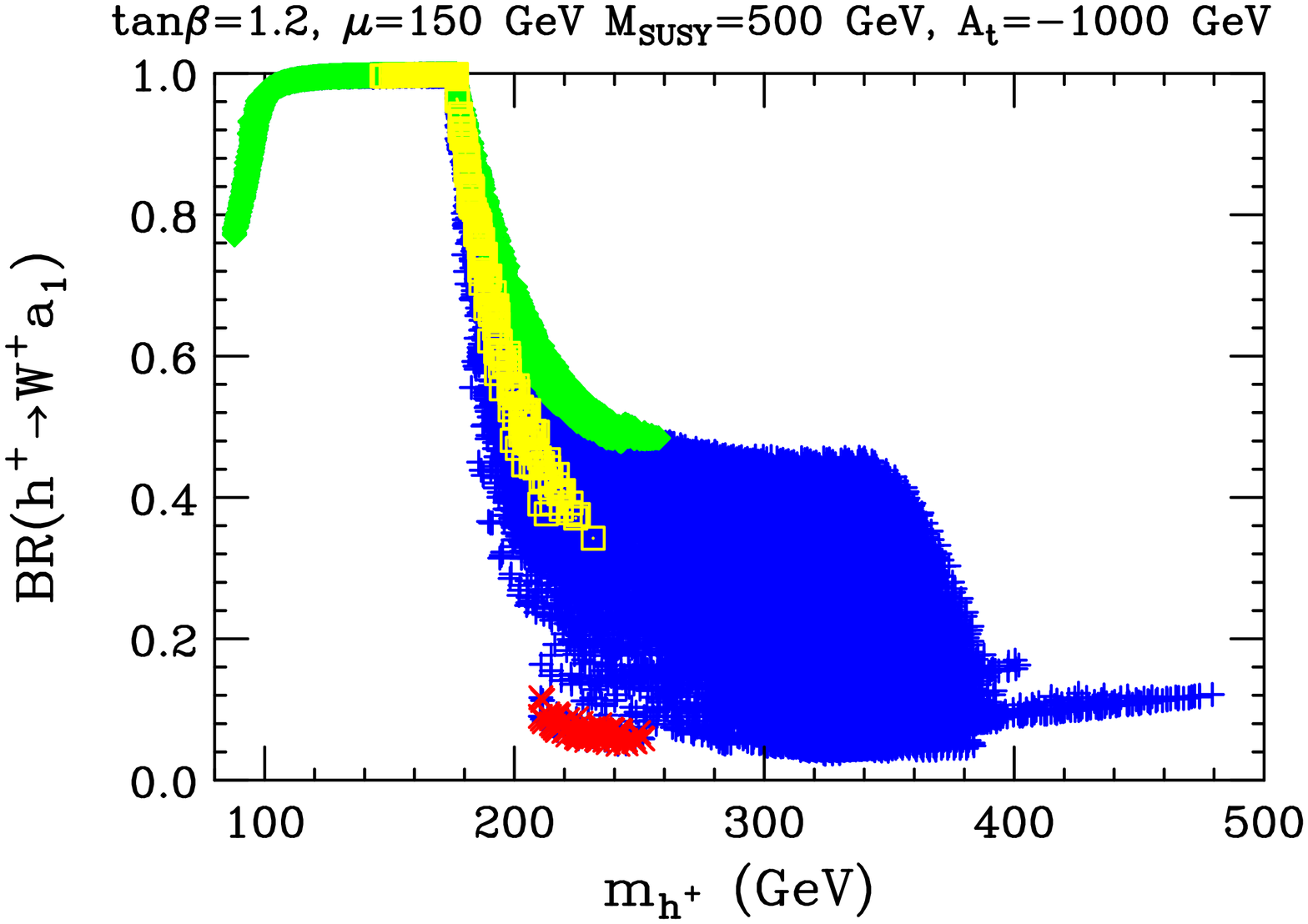}
\caption{$\br(\hp\to \wstarp \ai)$ is plotted vs. $\mhp$ for the $\tanb=1.2$,
  $\msusy=500\gev$, $A=-1000\gev$ scenario. }
\label{brhpwaivsmhp_tb1pt2}
\end{figure}

For this smaller $\tanb=1.2$ value, $\br(t\to\hp b)$ is larger
($\sim 0.3$) at the lowest $\mhp\sim 90\gev$ mass than for 
$\tanb=1.7$. Nonetheless, the Tevatron is still unable to limit these
scenarios since $\br(\hp\to\tau^+\nu_\tau)<0.2$ 
(see Fig.~\ref{brhptaunuvsmhp_tb1pt2}) given the dominance of $\hp\to
\wstarp b$ decays (Fig.~\ref{brhpwaivsmhp_tb1pt2}).

\section{Conclusions}

For low $\tanb$ values in the NMSSM, we have found many interesting
new Higgs scenarios with a light CP-odd scalar with mass below
$10\gev$.  For many of the experimentally allowed parameter choices,
the $\hi$, $\hp$ and $\hii$ are all sufficiently light as to be
kinematically accessible at LEP and the Tevatron, but they decay into
final states containing the light CP-odd scalar and therefore escaped
detection. One particularly interesting parameter space region is that
associated with the PQ symmetry limit (small $|\kap|$ and small
$\akap$) in which the $\hi$ has mass near $90\gev$ and the $\hii$ has
mass near $100\gev$ (\ie\ basically non-overlapping within
experimental resolution) and $g_{ZZ\hi}^2\br(\hi\to b\anti b)$ and
$g_{ZZ\hii}^2\br(\hii\to b\anti b)$ are such as to explain the
observed LEP excess throughout this region. These points, such that
both $\hi$ and $\hii$ contribute to the LEP excess, are present
for the $\tanb=1.2$ and $\tanb=1.7$ cases, but not for $\tanb=2$.

Another important common feature of all these low-$\tanb$ scenarios
that is also shared with the high-$\tanb$ scenarios explored in
earlier papers is that for any given $\mai$ there is always a lower
limit on $\br(\Upsilon\to \gam\ai)$.  This lower limit is above about
$5\times 10^{-7}$ for $\mai<7.5\gev$. We are hopeful that this is a
level that can eventually be probed by BaBar and Belle.  This lower
limit arises because there is a lower limit on $|\cta|$, 
and therefore on $\caibb=\cta\tanb$, below which
$\br(\hi\to\ai\ai)$ is not large enough for a light $\hi$ to have
escaped LEP limits. 

We should further comment that all the scenarios of the present paper,
as well as previous papers that focused on higher-$\tanb$, are such
that the scenarios that survive all experimental constraints are ones for
which the contributions of the Higgs sector to both the precision
electroweak observables, $\Delta T$ and $\Delta S$ (relative to the
SM-Higgs contribution for $\mhsm=110\gev$), and to the muon
anomalous magnetic moment, $a_\mu$ (relative to the observed
experimental discrepancy) are very small. In the case of $\Delta T$
and $\Delta S$ the small extra $\Delta T$ can be understood as a
natural result of either $\hiii$ and $\aii$ decoupling or of $\hii$
and $\aii$ decoupling.

Overall, the NMSSM provides a huge opportunity to have an ``Ideal
Higgs'' boson scenario in which there is one or two light Higgs bosons
(masses at or below $100\gev$) that in combination have all the
$ZZ$-Higgs coupling squared and therefore give values for the
precision electroweak observables $S$ and $T$ that are in excellent
agreement with data.  These Higgs bosons escape LEP limits because of
unusual decays involving the light $\ai$ with $\mai<2m_b$ that is the
common future of all these ``Ideal'' models. They also provide an
excellent possibility for describing the broad excess in the $\epem\to
Zb\anti b$ channel in the region $m_{b\anti b}\in [90\gev,105\gev$]
seen at LEP.  We look forward to possibly discovering the $\ai$ in
$\Upsilon$ decays at Babar or Belle, or direct detection of 
$\hi\to \ai\ai$ at the LHC, if not the Tevatron.

\acknowledgments

\vspace{0.2cm} We would like to thank A. Akeroyd and Ricardo Eusebi
for discussions. JFG is supported by U.S. Department of Energy grant
DE-FG02-91ER40674.   JFG would like to thank the Aspen Center for
Physics and KITP at U.C. Santa Barbara for support and hospitality
during various phases of this work. The part of this research
performed at KITP was supported by the National Science
Foundation under Grant No. PHY05-51164.

\vspace{0.2cm}
\vfill


\end{document}